\documentclass[11pt,a4paper]{article}

\usepackage{epsfig}
\usepackage{amsmath}
\usepackage{amssymb}
\usepackage{verbatim}
\usepackage{bm}
\usepackage{dsfont}
\usepackage[footnotesize]{caption}
\usepackage{subfigure}
\usepackage{cite}
\usepackage{textcomp}
\usepackage{calc}
\usepackage{geometry}
\usepackage{bbm}
\usepackage{xcolor}
\usepackage[utf8]{inputenc}

\usepackage{hyperref}
\hypersetup{colorlinks=true,urlcolor=blue,linkcolor=red,citecolor=green!60!black}

\renewcommand{\baselinestretch}{1.3}

\begin{document}


\noindent December 2017
\hfill DESY 17-223

\vskip 1.5cm

\begin{center}

\bigskip
{\huge\bf Inflation from High-Scale\\\medskip Supersymmetry Breaking}

\vskip 2cm

\renewcommand*{\thefootnote}{\fnsymbol{footnote}}

{\large
Valerie Domcke\,$^{a,\,\hspace{-0.25mm}}$%
\footnote{\href{mailto:valerie.domcke@desy.de}{valerie.domcke@desy.de}}
and Kai~Schmitz\,$^{b,\,\hspace{-0.25mm}}$%
\footnote{\href{mailto:kai.schmitz@mpi-hd.mpg.de}{kai.schmitz@mpi-hd.mpg.de}}}\\[3mm]
{\it{
$^{a}$ Deutsches Elektronen-Synchrotron (DESY), 22607 Hamburg, Germany\\
$^{b}$ Max-Planck-Institut f\"ur Kernphysik (MPIK), 69117 Heidelberg, Germany}}

\end{center}

\vskip 1cm

\renewcommand*{\thefootnote}{\arabic{footnote}}
\setcounter{footnote}{0}



\begin{abstract}


Supersymmetry breaking close to the scale of grand unification
can explain cosmic inflation.
As we demonstrate in this paper, this can be achieved in strongly coupled supersymmetric
gauge theories, such that the energy scales of inflation and supersymmetry breaking
are generated dynamically.
As a consequence, both scales are related to each other
and exponentially suppressed compared to the Planck scale.
As an example, we consider a dynamical model in which gauging a global flavor symmetry
in the supersymmetry-breaking sector gives rise to a Fayet-Iliopoulos D term.
This results in successful D-term hybrid inflation in agreement 
with all theoretical and phenomenological constraints.
The gauged flavor symmetry can be identified with $U(1)_{B-L}$, where $B$ and $L$
denote baryon and lepton number, respectively.
In the end, we arrive at a consistent cosmological scenario that provides a unified
picture of high-scale supersymmetry breaking, viable D-term hybrid inflation, spontaneous
$B$$-$$L$ breaking at the scale of grand unification, baryogenesis via leptogenesis,
and standard model neutrino masses due to the type-I seesaw mechanism.


\end{abstract}


\thispagestyle{empty}


\newpage

{\hypersetup{linkcolor=black}\renewcommand{\baselinestretch}{1}\tableofcontents}



\section{Introduction: GUT-scale SUSY breaking as the origin of inflation}
\label{sec:introduction}


\textit{Supersymmetry} (SUSY) is an elegant and well-motivated extension
of the \textit{standard model} (SM).
In recent years, the experimental data collected at the \textit{Large Hadron
Collider} (LHC) has, however, put increasing pressure on conventional
SUSY scenarios with superpartner masses around the electroweak scale.
No evidence for supersymmetric particles has been found thus
far; see, e.g., the recent SUSY searches by the LHC experiments
ATLAS~\cite{Aaboud:2017bac} and CMS~\cite{Sirunyan:2017cwe}.
Therefore, if supersymmetry exists in nature,
the \textit{superpartners of the SM particles} (sparticles)
must have masses of at least $\mathcal{O}\left(1\right)\,\textrm{TeV}$, if not much higher.
The idea of heavy sparticles is also corroborated by the measured value of the
Higgs boson mass~\cite{Aad:2012tfa,Chatrchyan:2012xdj}.
In the \textit{Minimal Supersymmetric Standard Model} (MSSM), a
$125\,\textrm{GeV}$ Higgs boson~\cite{Aad:2015zhl}
can only be explained by means of large radiative
corrections~\cite{Okada:1990vk,Okada:1990gg,Ellis:1990nz,Ellis:1991zd,Haber:1990aw}.
This points towards a large SUSY-breaking mass splitting between the
top quark and its scalar partners, the stop squarks.


If realized around the electroweak scale, supersymmetry provides a natural
solution to the large hierarchy problem in the standard model.
For decades, this observation has reinforced the paradigm
of low-scale supersymmetry that would be testable in collider experiments.
In the present experimental situation, the null results at the
LHC, however, bring about the little hierarchy problem.
Sparticles with masses of at least $\mathcal{O}\left(1\right)\,\textrm{TeV}$
can only be reconciled with the $\mathcal{O}\left(100\right)\,\textrm{GeV}$
value of the electroweak scale at the cost of fine-tuning.
One is therefore led to adopt one of two possible attitudes.
Either one gives up on supersymmetry as a well-motivated extension of the standard model, or
one challenges the concept of naturalness and accepts a certain degree of fine-tuning.
In this paper, we shall take the latter approach.
Our understanding of naturalness may be flawed and, for
some reason or another, not apply to the physics of the electroweak scale.
Moreover, dismissing supersymmetry altogether would do injustice to
supersymmetry's other merits.
One must not forget that,
irrespective of its relation to the electroweak scale, supersymmetry also
(i) provides a natural particle candidate for dark matter,
(ii) facilitates the unification of the SM gauge coupling constants
at a high energy scale, and
(iii) sets the stage for the \textit{ultraviolet} (UV) completion of the
standard model in string theory.


In fact, supersymmetry broken at a high
scale~\cite{Giudice:1998xp,Wells:2003tf,Wells:2004di}
has received significant interest in recent years.
Many authors have proposed models of high-scale SUSY breaking
and its mediation to the visible sector, including scenarios such as
universal high-scale supersymmetry~\cite{Hall:2009nd},
split supersymmetry~\cite{ArkaniHamed:2004fb,Giudice:2004tc,ArkaniHamed:2004yi},
mini-split supersymmetry~\cite{Arvanitaki:2012ps},
minimal split supersymmetry~\cite{ArkaniHamed:2012gw},
spread supersymmetry~\cite{Hall:2011jd,Hall:2012zp}, and
pure gravity mediation~\cite{Ibe:2006de,Ibe:2011aa,Ibe:2012hu}
(see also~\cite{Buchmuller:2015jna}).
Another intriguing feature of these models is that a high SUSY-breaking scale
implies a very heavy gravitino.
The gravitino, thus, decays very fast in the early Universe, which solves
the cosmological gravitino problem~\cite{Pagels:1981ke,Weinberg:1982zq,Khlopov:1984pf,Ellis:1984er}.
In addition, large sfermion masses help reduce the tension with constraints
on the SUSY parameter space from flavor-changing
neutral currents and $CP$ violation~\cite{Gabbiani:1996hi}.
For these reasons, we consider supersymmetry broken at a high energy scale 
to be a leading candidate for new physics beyond the standard model.


In this paper, we will take the idea of high-scale SUSY breaking to the extreme and
consider SUSY-breaking scales $\Lambda_{\rm SUSY}$ as large as the scale of gauge
coupling unification in typical \textit{grand unified theories} (GUTs),
$\Lambda_{\rm SUSY} \sim \Lambda_{\rm GUT} \sim 10^{16}\,\textrm{GeV}$.
Electroweak naturalness is then certainly lost.
But at the same time, another intriguing possibility emerges which is
out of reach in low-scale supersymmetry.
If $\Lambda_{\rm SUSY}$ is large enough, cosmic inflation in the early
Universe~\cite{Starobinsky:1980te,Guth:1980zm,Linde:1981mu,Albrecht:1982wi}
can be driven by the vacuum energy density associated
with the spontaneous breaking of supersymmetry,
$\left<V\right> \supset \Lambda_{\rm SUSY}^4$.
This represents a remarkable connection between particle physics and cosmology.
Inflation is a pillar of the cosmological standard model.
Not only does inflation explain the large degree of homogeneity and isotropy
of our Universe on cosmological scales, it is also the origin of the primordial
fluctuations that eventually seed structure formation on galactic scales
(see, e.g., \cite{Lyth:1998xn,Linde:2005ht} for reviews on inflation).
At present, there is, however, no consensus on how to make contact between
inflation and particle physics.
Against this background, the unification of inflation with the
dynamics of spontaneous SUSY breaking provides an elegant and economical
embedding of inflation into a microscopic theory.


The interplay between inflation and SUSY breaking has been studied
from different angles in the past
(see, e.g., \cite{Randall:1994fr,Riotto:1997iv,Buchmuller:2000zm}).
In the context of \textit{supergravity} (SUGRA), SUSY breaking in a hidden sector
can, in particular, result in severe gravitational corrections
to the inflationary dynamics~\cite{Hardeman:2010fh}.
In this case, one can no longer perform a naive slow-roll analysis
that only considers the properties of the inflaton sector
and disregards its gravitational coupling to other sectors.
Instead\,---\,and this is exactly what we will do in this paper\,---\,one has
to resort to a global and combined analysis that accounts for the presence and
interaction of all relevant sectors,
including the inflaton sector, SUSY-breaking sector, and visible sector.
The first unified model that illustrates how inflation and
soft SUSY breaking in the visible sector may originate from the same dynamics
has been presented in~\cite{Izawa:1997jc}.
In this model, supersymmetry is broken dynamically by the nonperturbative
dynamics in a strongly coupled supersymmetric gauge theory.
\textit{Dynamical SUSY breaking} (DSB) first occurs in the hidden sector
and is then mediated to the MSSM.
More recently, a number of related models have been constructed
in~\cite{Schmitz:2016kyr,Domcke:2017xvu,Harigaya:2017jny}.%
\footnote{Besides, there is a more general class of models, sometimes referred to as
dynamical inflation, where inflation is a consequence of dynamical
SUSY breaking in a hidden sector.
In these models, the breaking of supersymmetry during inflation is, however, not responsible
for the breaking of supersymmetry in the MSSM at low energies
(see, e.g.,~\cite{Dimopoulos:1997fv,Izawa:1997df,Hamaguchi:2008uy,
Harigaya:2012pg,Harigaya:2014sua,Harigaya:2014wta}).\smallskip}
All these models have in common that the energy scales of inflation
and SUSY breaking end up being related to the dynamical scale
$\Lambda_{\rm dyn}$ of the strong dynamics.
The scale $\Lambda_{\rm dyn}$ is generated via dimensional transmutation,
analogously to the scale of \textit{quantum chromodynamics}
(QCD) in the standard model.
That is, at energies around $\Lambda_{\rm dyn}$, the gauge coupling
constant in the hidden sector formally diverges.
The unified models in~\cite{Izawa:1997jc,Schmitz:2016kyr,Domcke:2017xvu,Harigaya:2017jny}
therefore do not require
any dimensionful input parameters to explain the origin of
the energy scales of inflation and SUSY breaking.
The generation of $\Lambda_{\rm dyn}$ is, in particular, a nonperturbative
effect in the \textit{infrared} (IR).
This explains the exponential hierarchy between $\Lambda_{\rm SUSY}$
and $\Lambda_{\rm inf}$ on the one hand and the Planck scale $M_{\rm Pl}$
on the other hand.
Meanwhile, several perturbative models of inflation and SUSY breaking
have recently been discussed in the
literature~\cite{Argurio:2017joe,Antoniadis:2017gjr,Dudas:2017kfz}.
These models also draw a unified picture of inflation and SUSY breaking.
But in contrast to strongly coupled models,
they depend on dimensionful input parameters which need to
be put in by hand.
They, thus, fail to provide a dynamical explanation
for the separation of scales.


In this paper, we shall revisit our model in~\cite{Domcke:2017xvu},
which gives rise to a viable scenario of 
\textit{D-term hybrid inflation} (DHI)~\cite{Binetruy:1996xj,Halyo:1996pp}
in the context of high-scale SUSY breaking.%
\footnote{This model has recently been employed in a supersymmetric
realization of the relaxion mechanism~\cite{Evans:2017bjs}.
For other models of hybrid inflation and high-scale
SUSY breaking, see~\cite{Higaki:2012iq,Pallis:2014xva}.
For more recent work on D-term inflation, see~\cite{Kadota:2017dbz}.}
Hybrid inflation~\cite{Linde:1991km,Linde:1993cn} is an interesting
scenario on general grounds, as it
establishes another connection between particle physics and cosmology.
In hybrid inflation, the inflationary era ends in a rapid second-order phase transition,
the so-called waterfall transition, which can be identified with the spontaneous
breaking of a local gauge symmetry in models of grand unification.
As shown in~\cite{Domcke:2017xvu}, this symmetry can
be chosen to correspond to $U(1)_{B-L}$, where $B$ and $L$ denote baryon
and lepton number, respectively.
Inflation therefore ends in what is referred to as the $B$$-$$L$ phase
transition~\cite{Buchmuller:2010yy,Buchmuller:2011mw,Buchmuller:2012wn,
Buchmuller:2012bt,Schmitz:2012kaa,Buchmuller:2013lra,Buchmuller:2013dja,Domcke:2013pma}.
That is, the end of inflation coincides with the spontaneous breaking
of $B$$-$$L$ in the visible sector.
This can be used to generate $L$-violating Majorana masses for
a number of sterile right-handed neutrinos.
These neutrinos then lead to
baryogenesis via leptogenesis~\cite{Fukugita:1986hr}
and generate the SM neutrino masses via the seesaw
mechanism~\cite{Minkowski:1977sc,Yanagida:1979as,Yanagida:1980xy,GellMann:1980vs,Mohapatra:1979ia}.
The model in~\cite{Domcke:2017xvu}, thus, provides a consistent picture of particle
physics and early Universe cosmology.
It unifies the dynamics of inflation,
high-scale SUSY breaking, and spontaneous $B$$-$$L$ breaking.
Remarkably enough, all of these phenomena occur at energies close to the GUT scale.


The discussion in~\cite{Domcke:2017xvu} mostly focused on model building aspects.
We gave a brief summary of our model construction and only touched upon phenomenology.
The purpose of the present paper is therefore threefold.
We will (i) review the construction of our model in more detail, including many aspects
that were left out in~\cite{Domcke:2017xvu} (see Sec.~\ref{sec:model}).
We will (ii) perform a more comprehensive scan of parameter space
(see Sec.~\ref{sec:inflation}).
In particular, we will identify new regimes of successful inflation that were overlooked
in~\cite{Domcke:2017xvu}.
And finally, we will (iii) provide a much broader phenomenological discussion,
including the implications of our model for the MSSM particle spectrum, the
$B$$-$$L$ phase transition, dark matter, and cosmic strings
(see Sec.~\ref{sec:discussion}).
In the last section of this paper, we will conclude and give an outlook on 
how our main observation\,---\,the fact that SUSY breaking close to the GUT scale
might be the key to a unified picture of particle physics and cosmology\,---\,could
lead to a new understanding of SUSY's role in nature (see Sec.~\ref{sec:conclusions}).
In the appendix, as a supplement to
Secs.~\ref{sec:model} and \ref{sec:inflation},
we collect various technical formulas that help to translate
between the Einstein-frame and Jordan-frame formulations of
supergravity (see Appendix~\ref{app:frames}).
This also includes a detailed comparison of our Jordan/Einstein-frame expressions
for the inflationary slow-roll parameters.




\section{Model: Strong dynamics and gauged
\texorpdfstring{\boldmath{$B$$-$$L$}}{B-L} in the Jordan frame}
\label{sec:model}


We begin by reviewing the model constructed in~\cite{Domcke:2017xvu}.
This will also allow us to introduce our notation and conventions.
The starting point of our analysis is the idea to build a viable SUGRA model of
hybrid inflation that ends in the spontaneous breaking of $B$$-$$L$, i.e.,
in the $B$$-$$L$ phase transition.


\subsection{Preliminary remarks on hybrid inflation in supergravity}
\label{subsec:HI}


In the absence of supersymmetry,
hybrid inflation is incompatible with the statistical properties of
the temperature fluctuations in the \textit{cosmic microwave background} (CMB).
Nonsupersymmetric hybrid inflation predicts the primordial scalar CMB power spectrum
to be blue-tilted, i.e., it predicts a scalar spectral index
$n_s$ greater than one~\cite{Linde:1991km,Linde:1993cn}.
This needs to be compared with the recent measurement by the PLANCK
collaboration, $n_s^{\rm obs} = 0.9677 \pm 0.0060$ \cite{Ade:2015lrj}.
Nonsupersymmetric hybrid inflation is, thus, ruled out
with a statistical significance of more than $5\,\sigma$.
This conclusion serves as an independent motivation
to introduce supersymmetry, in addition to supersymmetry's other advantages
(see Sec.~\ref{sec:introduction}).


In supersymmetry, hybrid inflation is understood to be a consequence
of (temporary) spontaneous SUSY breaking, which can be accomplished either
by a nonvanishing F term~\cite{Copeland:1994vg,Dvali:1994ms} or
D term~\cite{Binetruy:1996xj,Halyo:1996pp}.
In the following, we will outline  both scenarios
and explain why we will eventually focus on D-term inflation.
In both cases, the scalar inflaton $\sigma$ is contained in a chiral multiplet $S$
that transforms as a singlet under all gauge symmetries.
$S$ couples to charged chiral multiplets in the
superpotential $W_{\rm inf}$,
\begin{align}
W_{\rm inf} \supseteq \kappa\, S\, \Phi \bar{\Phi} \,,
\label{eq:SYukawa}
\end{align}
where $\kappa$ is a dimensionless Yukawa coupling.
$\Phi$ and $\bar{\Phi}$ denote the so-called waterfall fields,
which transform in conjugate representations of a gauge group $G$.
We shall identify $G$ with $U(1)_{B-L}$ and assign $B$$-$$L$ charges $+q$ and $-q$
to $\Phi$ and $\bar{\Phi}$.
This sets the stage for the $B$$-$$L$ phase transition at the end of inflation.
Any self-interaction of $S$ can be forbidden by invoking $R$ symmetry.
If we assign $R$ charges such that $\big[S\big]_R = 2$ and 
$\big[\Phi\bar{\Phi}\big]_R = 0$, terms such as $S^2$ and $S^3$ are not allowed.
The Yukawa coupling in Eq.~\eqref{eq:SYukawa} does not suffice
to break supersymmetry. 
This is, however, necessary to obtain a nonvanishing vacuum energy density
that can drive inflation. 
In \textit{F-term hybrid inflation} (FHI), one therefore equips $S$
with an F term, $\left|F_S\right| = \mu_S^2$.
Supersymmetry is then broken \`a la O'Raifeartaigh~\cite{ORaifeartaigh:1975nky}
during inflation.
In D-term inflation, one assumes instead a nonzero
\textit{Fayet-Iliopoulos} (FI) D term in the D-term scalar potential $V_D$.
This breaks supersymmetry during inflation via the FI mechanism~\cite{Fayet:1974jb}.
F-term and D-term inflation are then characterized by the following expressions
for $W_{\rm inf}$ and $V_D$,
\begin{align}
\label{eq:FHIDHI}
\textrm{FHI:} \qquad & W_{\rm inf} = \kappa\, S\, \Phi \bar{\Phi} + \mu_S^2\, S
\,, \quad V_D = \frac{g^2}{2}
\left[q\left(\left|\phi\right|^2 - \left|\bar{\phi}\right|^2 \right)\right]^2
\,, \\ \nonumber
\textrm{DHI:} \qquad & W_{\rm inf} = \kappa\, S\, \Phi \bar{\Phi} \:\phantom{+ M^2 S}
\,, \quad V_D = \frac{g^2}{2} \left[q_0\, \xi
- q\left(\left|\phi\right|^2 - \left|\bar{\phi}\right|^2 \right)\right]^2
\,.
\end{align}
The gauge coupling constant $g$ belongs to the gauge group $G$.
In our case, $g$ consequently denotes the $B$$-$$L$ gauge coupling.
The gauge charge $q_0$ is factored out of the FI parameter $\xi$ for later convenience.


In both F-term and D-term inflation, the field $S$ parametrizes
a completely flat direction in the scalar potential, at least
at tree level and in global supersymmetry.
This explains why it is natural to identify $S$ with the chiral inflaton field.
At large field values of $S$, the Yukawa coupling in Eq.~\eqref{eq:SYukawa}
induces large $S$-dependent masses for the waterfall fields.
$\Phi$ and $\bar{\Phi}$ can therefore be integrated out, which results
in a logarithmic effective potential for $S$.
This singles out the origin in field space, $S = 0$, as the unique vacuum
in the quantum theory.
Inflation is, thus, characterized by the slow-roll motion
of $S$ from large field values towards the origin in field space.
The SUSY-breaking parameters $\mu_S$ (FHI) and $\xi$ (DHI) induce a mass splitting 
between the scalar components of $\Phi$ and $\bar{\Phi}$.
Below a certain critical value of the inflaton field, this results in
one (linear combination of) scalar waterfall field(s) becoming unstable. 
This triggers the $B$$-$$L$ phase transition.
Inflation ends and the system approaches a ground state in which
$B$$-$$L$ is spontaneously broken and supersymmetry restored.
As we will discuss in Sec.~\ref{subsec:phasetransition},
this can be used to generate Majorana masses for the right-handed
neutrinos $N_i$ in the seesaw extension of the MSSM.
All we have to do is to choose appropriate gauge charges
and introduce Yukawa couplings between $\Phi$
and three generations of  sterile neutrinos,
$W \supset \frac{1}{2}\, h_{ij}\, \Phi\, N_i N_j$.


Hybrid inflation is sensitive to gravitational corrections
in supergravity~\cite{Panagiotakopoulos:1997qd,Linde:1997sj}.
In the case of F-term inflation, $R$ symmetry breaking leads, in particular,
to a SUGRA term in the F-term scalar potential $V_F$ that is linear
in the complex inflaton field
$s \subset S$~\cite{Buchmuller:2000zm,Rehman:2009nq,Rehman:2009yj,
Nakayama:2010xf,Buchmuller:2014epa}.
To see this, recall that $R$ symmetry breaking can be accounted for by introducing
a constant term in the superpotential, $w \subset W$.
This term ultimately sets the gravitino mass $m_{3/2}$ in the low-energy vacuum,
$w \propto m_{3/2}M_{\rm Pl}^2$ (where $M_{\rm Pl}$ denotes the reduced Planck
mass, $M_{\rm Pl} \simeq 2.44 \times 10^{18}\,\textrm{GeV}$).
In F-term inflation, the inflaton field itself possesses a nonzero F term,
$\left|F_S\right| = \mu_S^2$.
This F term couples to $w$ in the SUGRA scalar potential,
\begin{align}
V_F \supset \frac{\mu_S^2\,w^*}{M_{\rm Pl}^2}
\left[\left(\mathcal{K}^{-1}\right)_{ss^*} K_{s^*} - 3\,s\right]
+ \textrm{h.c.} \,. 
\label{eq:tadpole}
\end{align}
Here, $K_{s^*}$ stands for the derivative of the K\"ahler
potential $K$ w.r.t.\ $s^*$ and $\left(\mathcal{K}^{-1}\right)_{ss^*}$
denotes the $ss^*$ entry in the inverse of the K\"ahler metric $\mathcal{K}$.
For a canonical K\"ahler potential, $K \supset S^\dagger S$, one obtains
\begin{align}
V_F \supset -\frac{2\,\mu_S^2\,w^*}{M_{\rm Pl}^2}\,s
+ \textrm{h.c.} \,. 
\end{align}
This tadpole term is a vivid example for the potential impact of hidden-sector
SUSY breaking on the dynamics of inflation (see the discussion in Sec.~\ref{sec:introduction}).
Any analysis of F-term inflation that ignores Eq.~\eqref{eq:tadpole} is incomplete.
The tadpole term breaks the rotational invariance in the complex
inflaton plane and, hence, complicates the analysis of the
inflationary dynamics.
In fact, it renders F-term inflation a two-field
model~\cite{Buchmuller:2014epa}, which needs to be treated with special care.
Depending on the size of $m_{3/2}$, the tadpole term also potentially spoils
slow-roll inflation.
In addition, it generates a false vacuum at large field values, which limits
the set of viable initial conditions for inflation in phase space.


For these reasons, we will restrict ourselves to D-term inflation
in this paper.
A dynamical realization of F-term inflation that avoids the tadpole problem
can be found in~\cite{Schmitz:2016kyr}.
The model in~\cite{Schmitz:2016kyr} also establishes a connection
between inflation and dynamical SUSY breaking around the scale of grand unification.
In~\cite{Schmitz:2016kyr}, there is, however, no $U(1)$ symmetry that could
be identified with $U(1)_{B-L}$.
Moreover, inflation does not end in a phase transition in the waterfall
sector.
Together, these features of~\cite{Schmitz:2016kyr} eliminate the possibility to unify
inflation with the dynamics of the $B$$-$$L$ phase transition.
A dynamical model of F-term inflation that ends in the 
$B$$-$$L$ phase transition will be presented elsewhere.


\subsection{Ingredients for a unified model of D-term inflation}
\label{subsec:ingredients}


The absence of the linear tadpole term in D-term inflation tends to make
SUGRA corrections more manageable.
In particular, the issue of initial conditions appears more favorable
compared to F-term inflation
(however, see also Sec.~\ref{subsec:iniconds}).
Nonetheless, D-term inflation still faces a number of challenges.
In this section, we will discuss these challenges one by one and outline
how they are respectively met in our model.
This serves the purpose to explain the bigger physical picture
behind our specific setup.
In Sec.~\ref{subsec:iyit}, we will then become more explicit and
present the details of our construction.


\subsubsection*{Dynamical generation of the FI term in the strongly coupled SUSY-breaking sector}


First of all, the origin of the FI term and its embedding
into supergravity are subtle issues that have been the subject
of a long debate in the literature. 
At this point, it is important to distinguish between genuine FI terms
and effective FI terms.
The former refer to field-independent FI parameters $\xi_a$ that
parametrize constant shifts in the auxiliary $D_a$ components of Abelian 
vector multiplets $V_a$.
The latter denote field-dependent FI parameters $\xi_a$ that depend on
the \textit{vacuum expectation values} (VEVs) of dynamical scalar moduli $\psi_i$.
Genuine FI terms, $\xi_a = \textrm{const}$, are FI terms in the original sense.
That is, they preserve the underlying gauge symmetry and are
compatible with massless vector multiplets. 
By contrast, in the case of effective FI terms, $\xi_a = \xi_a\left(\left<\psi_i\right>\right)$, 
the underlying gauge symmetry is always spontaneously
broken by the modulus VEVs $\left<\psi_i\right>$.
Effective FI terms are therefore FI terms only in a slightly more general sense.
The embedding of genuine FI terms into supergravity
always requires the underlying gauge symmetry to be promoted to a gauged $U(1)_R$
symmetry as well as the presence of an exact
global continuous symmetry~\cite{Komargodski:2009pc,Dienes:2009td}.
While the requirement of a gauged $U(1)_R$ symmetry poses a challenge as
soon as one wants to make contact with low-energy phenomenology, the requirement
of a global continuous symmetry is problematic from the viewpoint of quantum gravity.
As can be shown on very general grounds, quantum gravity is likely
to violate any global symmetry~\cite{Banks:2010zn}.
Coupling genuinely constant FI terms to gravity, thus, appears to be almost impossible.%
\footnote{Shortly after completion of our work, two proposals appeared
in the literature that demonstrate how a novel type of genuine FI
terms, based on nonstandard supersymmetric invariants, can be
consistently coupled to supergravity~\cite{Cribiori:2017laj,Kuzenko:2018jlz}.
These FI terms do not require $R$ symmetry to be gauged and, hence, do not suffer
from the presence of a global symmetry.
On the other hand, they result in highly nonlinear terms in the fermionic action.
It would be interesting to employ these novel FI terms in phenomenological applications
in future work and investigate, e.g., their potential use for inflation.}
Effective FI terms promise to offer a possible way out of this problem.
Effective FI terms are frequently encountered in string theory~\cite{Dine:1987xk,Atick:1987gy},
where they arise in consequence of the Green-Schwarz mechanism of
anomaly cancellation~\cite{Green:1984sg}.
However, such constructions typically suffer from the presence of a shift-symmetric
modulus field~\cite{Komargodski:2010rb}.
This modulus field needs to be adequately stabilized~\cite{Binetruy:2004hh}.
Otherwise, it will absorb the effective FI term in its VEV
or cause a cosmological
modulus problem~\cite{Coughlan:1983ci,Ellis:1986zt}.


To avoid all of the problems listed above, we will
assume that the FI term responsible for inflation
is dynamically generated in the strongly coupled SUSY-breaking
sector.
That is, we will not resort to string theory, but work in the context
of field theory.
To be precise, we will employ the dynamical mechanism devised in~\cite{Domcke:2014zqa}.
A dynamical FI term generated via this mechanism is automatically an
effective field-dependent FI term that is controlled by
the VEVs of moduli in the hidden sector,
\begin{align}
\label{eq:xidyn}
\xi = \big<\bar{\psi}\bar{\psi}^*\big> - \big<\psi\psi^*\big> \sim \Lambda_{\rm dyn}^2 \,.
\end{align}
Here, $\psi$ and $\bar{\psi}$ belong to chiral multiplets $\Psi$ and $\bar{\Psi}$
that carry $B$$-$$L$ charges $+q_0$ and $-q_0$, respectively.
This ansatz complies with our philosophy outlined in the introduction
(see Sec.~\ref{sec:introduction}).
Being a dynamically generated quantity, the FI parameter $\xi$
does not need to be added by hand.
Instead, it is related to the dynamical scale $\Lambda_{\rm dyn}$ in the hidden sector
which is generated via dimensional transmutation.
Let us identify the UV embedding scale of our theory with the Planck scale,
$\Lambda_{\rm UV} = M_{\rm Pl}$.
The \textit{renormalization group} (RG) running of the hidden-sector
gauge coupling $g_{\rm hid}$ then results in the following relation,
\begin{align}
\label{eq:Lambdadyn}
\Lambda_{\rm dyn} = M_{\rm Pl}\, \exp\left[-\frac{1}{b_{\rm hid}^{\vphantom{2}}}
\frac{8\pi^2}{g_{\rm hid}^2\left(M_{\rm Pl}\right)}\right] \,,
\end{align}
where $b_{\rm hid}$ is the coefficient of the hidden-sector RG beta function.
This relation explains why $\Lambda_{\rm dyn}$ and, hence, $\xi$
end up being exponentially suppressed compared to the Planck scale
(which is the only available mass scale in our setup).
The advantage of our approach is that it comes with a built-in mechanism
for modulus stabilization.
As usual, the generation of $\xi$ results in a shift-symmetric modulus field.
In our case, this will be a linear combination of $\Psi$ and $\bar{\Psi}$
(see Sec.~\ref{subsec:iyit}).
However, as the FI parameter $\xi$ is generated in the SUSY-breaking sector,
the shift-symmetric modulus couples to \textit{degrees of freedom} (DOFs)
involved in the dynamical breaking of supersymmetry.
The F term of the SUSY-breaking Polonyi field therefore induces
a mass for the modulus field.
This stabilizes all dangerous directions in field space and
prevents us from running into any modulus problem.


\subsubsection*{Spontaneous \texorpdfstring{\boldmath{$B$$-$$L$}}{B-L}
breaking in the hidden sector before the end of inflation}


The $B$$-$$L$ phase transition at the end of D-term inflation
is accompanied by the production of topological defects in the form of cosmic
strings~\cite{Kibble:1976sj,Jeannerot:2003qv}
(for reviews on cosmic strings, see~\cite{Vilenkin:1984ib,Hindmarsh:1994re}).
Such cosmic strings can leave an imprint in the CMB temperature anisotropies,
the spectrum of \textit{gravitational waves} (GWs),
and in the \textit{diffuse gamma-ray background} (DGRB). 
Cosmic string decays can also affect outcome of
\textit{big bang nucleosynthesis} (BBN).
However, no signs of cosmic strings have been detected thus far.
Recent limits on the properties of cosmic strings
can be found in~\cite{Ade:2013xla,Ade:2015xua} (CMB),
\cite{Sanidas:2012ee,Blanco-Pillado:2017rnf,Ringeval:2017eww} (GWs),
and~\cite{Mota:2014uka} (DGRB and BBN).
These bounds allow to put severe constraints on the parameter space
of hybrid inflation~\cite{Charnock:2016nzm,Lizarraga:2016onn}.
In fact, the minimal scenario of D-term inflation is already ruled out by
the nonobservation of cosmic strings~\cite{Battye:2010hg}.
A possible way out of this problem is to consider scenarios
in which $B$$-$$L$ is spontaneously broken, in one way or another,
already during inflation.
Cosmic strings then form at early times and are sufficiently diluted
before the end of inflation.


Fortunately, the dynamical generation of $\xi$ in the hidden
sector (see Eq.~\eqref{eq:xidyn}) provides the ideal starting
point for implementing this solution to the cosmic string problem.
The modulus VEVs $\big<\psi\big>$ and $\big<\bar{\psi}\big>$ spontaneously
break $B$$-$$L$ in the hidden sector already during inflation.
Therefore, to prevent the formation of cosmic strings in the waterfall sector
at the end of inflation, all we have to do is to communicate
the breaking of $B$$-$$L$ in the hidden sector to the waterfall fields.
This is readily done by adding marginal couplings between the two sectors
in the superpotential or K\"ahler potential
that are otherwise irrelevant for the dynamics of inflation~\cite{Evans:2017bjs}.
We will come back to this issue in Sec.~\ref{subsec:strings}.


\subsubsection*{Sequestered sectors in Jordan-frame supergravity}


A notorious problem of any model of D-term inflation is that additional
charged scalar fields, other than the waterfall fields $\Phi$ and $\bar{\Phi}$,
threaten to destabilize the FI term during inflation.
Without any additional physical assumption, there is no reason why 
$\Phi$ and $\bar{\Phi}$ should be the only fields charged under the $U(1)$
symmetry whose FI term drives inflation.
In general, one should rather expect a whole set of $N$ charged pairs,
$\left\{\Phi_i,\bar{\Phi}_i\right\}$,
with the inflaton only coupling to a subset of $M$ such pairs,
\begin{align}
W \supset \sum_{i=1}^N \kappa_i\,S\,\Phi_i\bar{\Phi}_i \,, \qquad
\kappa_i
\begin{cases}
\neq 0 & ; \quad \forall i \leq M \\
= 0    & ; \quad \forall i > M     
\end{cases} \,.
\end{align}
In this case, there are $N-M$ pairs that are not sufficiently stabilized by an
inflaton-induced mass during inflation.
At the same time, these fields also enter into the D-term scalar potential, where
they threaten to absorb the FI parameter $\xi$ in their VEVs.
In our scenario based on $U(1)_{B-L}$, the role of these dangerous scalar
directions is played by the scalar partners of the MSSM quarks and leptons.
The squarks $\tilde{q}_i$ and sleptons $\tilde{\ell}_i$
are also charged under $B$$-$$L$, but do not couple to
the inflation field.
Accounting for the presence of these fields, the D-term scalar potential
needs to be rewritten as follows,
\begin{align}
\label{eq:VDMSSM}
V_D = \frac{g^2}{2} \left[q_0\, \xi
- q\left(\big|\phi\big|^2 - \big|\bar{\phi}\big|^2 \right)
- \sum_i q_{q_i}\left(\big|\tilde{q}_i\big|^2 - \big|\tilde{\bar{q}}_i\big|^2 \right)
- \sum_i q_{\ell_i}\left(\big|\tilde{\ell}_i\big|^2 - \big|\tilde{\bar{\ell}}_i\big|^2 \right)
\right]^2 \,.
\end{align}
The squarks and sleptons therefore acquire
D-term-induced masses proportional to $m_D = g \left|q_0\xi\right|^{1/2}$.
As evident from the sign relations in Eq.~\eqref{eq:VDMSSM}, half of these masses
end up being tachyonic (see also~\cite{Babu:2015xba}).
This renders the corresponding directions in field space tachyonically unstable.
The inflationary trajectory, thus, decays into a vacuum
in which $B$$-$$L$ is broken by nonvanishing sfermion VEVs.


To avoid this problem, one needs to stabilize the MSSM sfermions
by means of additional mass contributions during inflation.
Here, the simplest solution is to make use of the soft scalar
masses induced by the spontaneous breaking of supersymmetry in the hidden sector.
Let us assume that these soft masses are all more or less close to a common
value $m_0$.
Then, to stabilize the MSSM sfermions during inflation, we must require
that $m_0 \gg m_D$.
This is, however, too strong a condition if SUSY breaking is communicated
to the visible sector only via ordinary gravity mediation in the Einstein frame
(for a review on gravity mediation, see~\cite{Nilles:1983ge}).
In gravity mediation, we expect the soft scalar masses to be of the
order of the gravitino mass, $m_0 \sim m_{3/2}$.
This soft mass is universal such that also the waterfall fields
obtain soft masses of $\mathcal{O}\left(m_{3/2}\right)$.
As a consequence, the stabilization of the MSSM sfermions
also stabilizes the waterfall fields.
This is an unwanted but unavoidable side effect.
In such a scenario, the waterfall fields would never become
unstable and inflation would never end.


A possible solution to this problem is to presume a separation
of scales of the following form,
\begin{align}
\label{eq:mmm}
m_{3/2} \ll m_D \ll m_0 \,.
\end{align}
In this case, the MSSM sfermions remain stabilized at all times, while
the waterfall fields can become unstable at the end of inflation.
Parametrically large soft sfermion masses can, e.g., be achieved
by adding a direct coupling between the visible and the hidden sector in the 
K\"ahler potential,
\begin{align}
\label{eq:KQQXX}
K \supset \frac{a_{ij}}{M_*^2}\,Q_i^\dagger Q_j^{\vphantom{\dagger}} X^\dagger X \,.
\end{align}
Here, $Q_i$ and $X$ stand for a generic MSSM matter field and the SUSY-breaking Polonyi 
field, respectively.
$M_*$ denotes the mass scale at which the effective operator in Eq.~\eqref{eq:KQQXX}
is generated.
The constants $a_{ij}$ are dimensionless Wilson coefficients that are expected
to be of $\mathcal{O}\left(1\right)$.
Any speculations regarding the underlying UV physics
are left for future work.
In this paper, we will content ourselves with the observation
that Eq.~\eqref{eq:KQQXX} results in soft masses
that can be parametrically large compared to $m_{3/2}$.
Provided that the cutoff scale is sufficiently below the Planck scale,
$M_* \ll M_{\rm Pl}$ , one finds
\begin{align}
m_0 \sim \frac{M_{\rm Pl}}{M_*}\,m_{3/2} \gg m_{3/2} \,.
\end{align}
Now, however, we need a conspiracy among certain parameters.
Successful inflation is only possible as long as 
the parameters $g$, $q_0$, $\xi$, $m_{3/2}$, and $M_*$ conspire in order
to satisfy the following relation,
\begin{align}
\label{eq:mDm32}
\textrm{Einstein frame:} \qquad
1 \ll \frac{m_D}{m_{3/2}} \ll \frac{M_{\rm Pl}}{M_*} \,.
\end{align}
We do not see any compelling argument why this relation should be automatically
fulfilled.
For this reason, we will go one step further and solve the MSSM sfermion
problem in a more elegant way.


Let us assume that the canonical description of hybrid inflation in supergravity
corresponds to an embedding into a (specific) Jordan frame rather than an embedding
into the Einstein frame~\cite{Ferrara:2010yw,Ferrara:2010in}.
At this point, recall that every Jordan-frame formulation of supergravity is characterized
by a specific choice for the so-called frame function $\Omega$.
The frame function is an arbitrary function of the complex scalar fields
in the theory, $\Omega = \Omega\left(\phi_i,\phi_{\bar{\imath}}^*\right)$.
For a given $\Omega$, the metric tensor in the Jordan
frame, $g_{\mu\nu}^J$, is related to the metric tensor
in the Einstein tensor, $g_{\mu\nu}$, via the following Weyl rescaling,
\begin{align}
\label{eq:Weyl}
g_{\mu\nu}^J = \mathcal{C}^2\,g_{\mu\nu} \,, \quad
\mathcal{C} = \left(-\frac{3 M_{\rm Pl}^2}{\Omega}\right)^{1/2} \,,
\end{align}
Here, $\mathcal{C}$ denotes what we will refer to as the conformal factor.
We emphasize that the Weyl transformation in Eq.~\eqref{eq:Weyl}
does not change the physical predictions of the theory.
The physical content of the Jordan frame is equivalent to
the physical content of the Einstein frame,
even at the quantum level~\cite{Postma:2014vaa,Kamenshchik:2014waa}.
In what follows, we will simply assume that the SUGRA embedding
of hybrid inflation is most conveniently described in the Jordan frame.
More details on the conversion between the Einstein-frame 
and Jordan-frame formulations of supergravity can be
found in Appendix~\ref{app:frames}.


Given the freedom in defining the frame function $\Omega$,
there is, in principle, an infinite number of possible Jordan frames.
In the following, we will, however, focus on one particular choice
which stands out for several reasons.
In this frame, the frame function $\Omega$ is
determined by the K\"ahler potential $K$,
\begin{align}
\label{eq:OmegaK}
\Omega = -3 M_{\rm Pl}^2\, \exp\left[-\frac{K}{3M_{\rm Pl}^2}\right] \,.
\end{align}
This relation is understood to hold in superspace,
such that the frame function $\Omega$ becomes a function of chiral multiplets,
$\Omega = \Omega\big(\Phi_i,\Phi_{\bar{\imath}}^\dagger\big)$.
This choice for $\Omega$ is motivated by the curved superspace approach
to old minimal supergravity in the Einstein frame~\cite{Stelle:1978ye,Ferrara:1978em}.
In this derivation of the SUGRA action, the function $\Omega$ as defined
in Eq.~\eqref{eq:OmegaK} is identified as the generalized kinetic energy
on curved superspace.
Meanwhile, $\Omega$ is also a meaningful quantity in the derivation
of the Einstein-frame action based on local superconformal
symmetry~\cite{Cremmer:1982en,Kugo:1982mr}.
In this approach to old minimal supergravity,
the function $\Omega$ is identified as the prefactor
of the kinetic term of the chiral compensator superfield.
For our purposes, the advantage of the choice in Eq.~\eqref{eq:OmegaK}
is that it sets the stage for canonically normalized kinetic
terms for the complex scalar fields in the Jordan frame.
Indeed, to obtain canonically normalized kinetic terms, the defining
relation in Eq.~\eqref{eq:OmegaK} needs to be combined with
the following ansatz for $\Omega$~\cite{Ferrara:2010yw},
\begin{align}
\label{eq:OmegaPhi}
\Omega = -3 M_{\rm Pl}^2 + F \,, \quad
F = \delta_{\bar{\imath}j}\, \Phi_{\bar{\imath}}^\dagger \Phi_j +
\left[J\left(\Phi_i\right) + \textrm{h.c.}\right] \,,
\end{align}
Here, we introduced $F$ as the \textit{kinetic function}
of the chiral matter fields.
The additional $-3M_{\rm Pl}^2$ term in $\Omega$ accounts for the kinetic term
of the gravitational DOFs.
$J$ is an arbitrary holomorphic function.


We mention in passing that the relations in
Eqs.~\eqref{eq:OmegaK} and \eqref{eq:OmegaPhi} also
provide the basis for a class of SUGRA models known as
\textit{canonical superconformal supergravity} (CSS) models~\cite{Ferrara:2010in}.
In these models, the pure supergravity part of the total action is invariant
under local Poincar\'e transformations as usual.
At the same time, the matter and gauge sectors of the theory can be
made invariant under a local superconformal symmetry
by setting the holomorphic function $J$ to zero.
This larger set of symmetry transformations renders CSS models particularly simple.
In the Jordan frame, one obtains canonical kinetic terms for all fields.
Moreover, one finds that the Jordan-frame scalar potential
coincides with the scalar potential in global supersymmetry.
In our scenario, we will, however, break the superconformal symmetry
via the holomorphic function $J$ to a large degree (see Eq.~\eqref{eq:FJ} below).
For this reason, one should not regard our model to be of the CSS type.
A model of D-term inflation based on the idea of superconformal symmetry has
been constructed in~\cite{Buchmuller:2012ex}.
This model employs a constant FI term that does not
depend on the inflaton field value in the Einstein frame.
Our model will by contrast involve an effective FI term that does not
depend on the inflaton field value in the Jordan frame (see Sec.~\ref{subsec:potential}). 
In~\cite{Buchmuller:2012ex}, the dynamics of inflation are moreover 
described by a two-field model, whereas we will
only deal with a single inflaton field.
Interestingly enough, the model in~\cite{Buchmuller:2012ex}
reproduces the predictions of Starobinsky inflation~\cite{Starobinsky:1980te}
in the limit of large inflaton field values~\cite{Buchmuller:2013zfa}.


Together with Eq.~\eqref{eq:OmegaK},
the ansatz in Eq.~\eqref{eq:OmegaPhi} results in the following K\"ahler potential,
\begin{align}
\label{eq:KF}
K = -3 M_{\rm Pl}^2\,\ln\left[1-\frac{F}{3M_{\rm Pl}^2}\right] \,.
\end{align}
This is an important result.
If we choose the kinetic function $F$ or, equivalently,
the holomorphic function $J$ appropriately, this K\"ahler
potential can be readily used to sequester the different
sectors of our model.
In fact, Eq.~\eqref{eq:KF} turns into a K\"ahler potential
of the sequestering type~\cite{Randall:1998uk} if the
function $F$ can be split into separate (canonical) contributions
from the hidden, visible, and inflation sectors,
\begin{align}
\label{eq:sequestering}
K = -3 M_{\rm Pl}^2\,\ln\left[1-\frac{F_{\rm hid} + F_{\rm vis} + F_{\rm inf}
}{3M_{\rm Pl}^2}\right] \,.
\end{align}
K\"ahler potentials of this form have been derived in the context of
extra dimensions~\cite{Randall:1998uk} as well as in strongly coupled
\textit{conformal field theories} (CFTs)~\cite{Luty:2001jh,Luty:2001zv,Ibe:2005pj,Ibe:2005qv}.
They are also similar to the K\"ahler potential in models of no-scale
supergravity~\cite{Cremmer:1983bf,Ellis:1983sf,Lahanas:1986uc} that
can be derived from string theory~\cite{Witten:1985xb}.%
\footnote{In no-scale supergravity, the kinetic term of the gravitational DOFs
depends on a dynamical modulus field $T$.
This is accounted for by replacing the $1$
inside the logarithm in Eq.~\eqref{eq:sequestering} by a field-dependent quantity:
$1\rightarrow \left(T+T^\dagger\right)/M_{\rm Pl}$.}
If the various sectors do not couple to each other in the superpotential,
the K\"ahler potential in Eq.~\eqref{eq:sequestering} leads to
vanishing soft scalar masses at tree level in all sectors except for the hidden sector.


The possibility to sequester different sectors is a crucial
property of Eq.~\eqref{eq:KF} which we will use to solve the MSSM sfermion problem.
Altogether, we will choose the function $F$ in our model as follows,%
\begin{align}
\label{eq:Ftot}
F \quad\rightarrow\quad F_{\rm tot} = F_{\rm hid} + F_{\rm vis} + F_{\rm inf} +
\frac{a_{ij}}{M_*^2}\,Q_i^\dagger Q_j^{\vphantom{\dagger}} X^\dagger X \,.
\end{align}
From now on, we will refer to the total kinetic function as $F_{\rm tot}$
and reserve the symbol $F$ for the kinetic function of the inflaton field $S$ (see further below).
We will also assume that the holomorphic function $J$ in Eq.~\eqref{eq:OmegaPhi}
is a function of $S$ only.
That is, the kinetic functions $F_{\rm hid}$, $F_{\rm vis}$, and $F_{\rm inf}$ are supposed
to consist of standard canonical terms for all fields except for $S$.
The kinetic function $F_{\rm tot}$ as defined in Eq.~\eqref{eq:Ftot} combines 
two important features.
(i) It leads to
a sequestering between the hidden sector and the inflaton sector.
The waterfall fields consequently obtain no soft masses at tree level.
This is necessary to be able to trigger the $B$$-$$L$ phase transition
at the end of inflation, irrespective of the size of the gravitino mass.
(ii) The MSSM sfermions are stabilized thanks to higher-dimensional
operators in $F_{\rm tot}$ that couple the visible sector to the SUSY-breaking sector.
At this point, we stick to the mechanism that we already discussed
in the case of the Einstein frame (see Eq.~\eqref{eq:KQQXX}).
Together, these two features allow us to realize successful inflation
and solve the MSSM sfermion problem.


Our solution of the MSSM sfermion problem
in the Jordan frame is conceptually different from the solution in the Einstein
frame discussed around Eq.~\eqref{eq:mDm32}.
Now, as the waterfall fields do not acquire a soft mass at tree level, 
the requirement in Eq.~\eqref{eq:mDm32} turns into the following two conditions,
\begin{align}
\label{eq:scalesJF}
\textrm{Jordan frame:} \qquad
0 \ll \frac{m_D}{m_{3/2}} \ll \frac{M_{\rm Pl}}{M_*} \,.
\end{align}
The first inequality is a consequence
of our decision to work in a Jordan frame with canonical kinetic terms.
It is trivially fulfilled.
We are, thus, left with only one sensible physical condition,
$M_* \ll m_{3/2} M_{\rm Pl} / m_D$.
To satisfy this condition, we no longer have to rely on a conspiracy
among different parameter values.
Instead, we simply have to deal with an upper bound on the scale
$M_*$ which derives from the requirement that
all dangerous scalar directions in field space must be sufficiently
stabilized during inflation.
We therefore manage to solve the MSSM sfermion problem in $B$$-$$L$
D-term inflation by means of
model-building decisions rather than by resorting to a specific
part of parameter space.


In the following, we will remain agnostic as to the UV origin of $F_{\rm tot}$
in Eq.~\eqref{eq:Ftot}.
We settle for the observation that, apart from additional Planck-suppressed
interactions, Eq.~\eqref{eq:Ftot} can be motivated by demanding canonically normalized
kinetic terms in the (standard) Jordan frame.
This is the reason why we will formulate parts of our analysis in the language
of Jordan-frame supergravity.
Beyond that, it might be possible to embed our model into
extra dimensions, strongly coupled CFTs, no-scale supergravity
and/or string theory.
But such a task is beyond the scope of this paper and left for future work.
For our purposes, the formalism of Jordan-frame supergravity simply
provides a convenient technical framework.
We shall not speculate about the underlying physics at higher energies.


\subsubsection*{Shift symmetry in the direction of the inflaton field}


Working in the Jordan frame not only helps to protect the waterfall fields
against large soft masses.
In ordinary gravity mediation in the Einstein frame, also the inflaton
acquires a soft mass of the order of the gravitino mass.
This results in the notorious eta problem in
supergravity~\cite{Copeland:1994vg,Stewart:1994ts}.
To see this, recall that
the gravitino mass is related to the F term of the SUSY-breaking Polonyi field as follows,
\begin{align}
m_{3/2} = \frac{\left<\left|F_X\right|\right>}{\sqrt{3}M_{\rm Pl}} \,.
\end{align}
At the same time, the Hubble rate during D-term inflation is controlled
by the size of the FI term,
\begin{align}
H_{\rm inf} = \frac{\left<V_D\right>^{1/2}}{\sqrt{3}M_{\rm Pl}} =
\frac{\left<D\right>}{\sqrt{6}M_{\rm Pl}} = 
\frac{gq_0\xi}{\sqrt{6}M_{\rm Pl}} \,.
\end{align}
General arguments in supergravity indicate that
D terms are always accompanied by an F term which is at least
as large or even larger~\cite{Dumitrescu:2010ca,Kawamura:2010mb}.
In our case, we intend to dynamically generate the FI term
in conjunction with the Polonyi F term in the SUSY-breaking sector.
On general grounds, we, thus, expect that
$\left<\left|F_X\right|\right> \gtrsim \left<D\right>$.
Therefore, if the inflaton indeed obtained a soft mass of
$\mathcal{O}\left(m_{3/2}\right)$, we would immediately encounter
an eta problem, i.e., a slow-roll parameter $\eta$ much larger than one,
\begin{align}
\eta = M_{\rm Pl}^2 \, \frac{V''}{V} \sim \left(\frac{m_{3/2}}{H_{\rm inf}}\right)^2 \gg 1 \,,
\end{align}
where $V''$ denotes the second derivative of the scalar potential w.r.t.\ the inflaton field.
This serves as an additional motivation for our specific Jordan frame.
There, the soft mass of the inflaton vanishes (at least as long as
$F = S^\dagger S$ and $J = 0$), which renders the most dangerous contribution to $\eta$ zero.


This is, however, not the end of the story. 
To fully solve the eta problem, we need to work a bit harder.
In the Jordan frame, the complex scalars are nonminimally coupled
to the Ricci scalar $R_J$ via the frame function $\Omega$.
This follows from
the Jordan-frame equivalent of the Einstein-Hilbert action,
\begin{align}
S \supset \frac{1}{2}\int d^4 x\, \sqrt{-g_J} \left(-\frac{\Omega_{\rm tot}}{3}\right) R_J = 
\frac{1}{2} \int d^4 x\, \sqrt{-g_J} \left(M_{\rm Pl}^2 -\frac{F_{\rm tot}}{3}\right) R_J \,,
\end{align}
which contains the nonminimal term $-F_{\rm tot}/3\,R_J$.
This coupling yields additional mass contributions for the scalar fields.
Consider, e.g., the canonical terms in the total kinetic function,
$F_{\rm tot} \supset \delta_{\bar{\imath}j}\, \Phi_{\bar{\imath}}^\dagger \Phi_j$,
\begin{align}
\label{eq:Snonmin}
S \supset - \int d^4 x\, \sqrt{-g_J}\, \sum_i \zeta_i\, R_J \left|\phi_i\right|^2 \,, \quad
\zeta_i = \zeta = \frac{1}{6} \,.
\end{align}
which describes the special case of a set of conformally coupled scalars.
Each complex scalar with a canonical term in the kinetic function $F_{\rm tot}$
therefore acquires a universal gravity-induced mass $m_R$, 
\begin{align}
\label{eq:mR}
m_R^2 = \zeta R_J = 2 H_J^2 \,,
\end{align}
where $H_J$ denotes the Hubble parameter in the Jordan frame
and where we used the relation between Ricci scalar and Hubble parameter
in exact de Sitter space, $R_J = 12 H_J^2$.
This gravitational mass correction spoils slow-roll inflation
as long as it is not sufficiently suppressed.
That is, an inflaton kinetic function that only consists of a
canonical term, $F = S^\dagger S$, results in too large an $\eta$ parameter,
\begin{align}
\label{eq:eta23}
\eta \sim \frac{m_R^2}{3H_J^2} = \frac{2}{3} \,.
\end{align}


Thus, to fully solve the eta problem, we have to make use of
the holomorphic function $J$ in Eq.~\eqref{eq:OmegaPhi}.
The freedom in defining $J$ allows us to realize an approximate shift symmetry
in the direction of the inflaton field.
Such a shift symmetry is a common tool in SUGRA models
of inflation, as it allows to suppress the most dangerous
SUGRA contributions to the inflaton potential~\cite{Kawasaki:2000yn}.
In our case, an approximate shift symmetry is realized for
the following kinetic function of the inflaton field,
\begin{align}
\label{eq:FJ}
F = S^\dagger S + \left[ J\left(S\right) + \textrm{h.c.} \right] \,, \quad 
J\left(S\right) = - \frac{1}{2}\left(1-2\chi\right) S^2 \,.
\end{align}
Here, $\chi$ is a positive shift-symmetry-breaking parameter which we will
assume to be small, $0 < \chi \ll 1$.
To see that Eq.~\eqref{eq:FJ} indeed features a shift symmetry,
it is convenient to rewrite $F$ as follows,
\begin{align}
\label{eq:F}
F = \frac{1}{2}\,\chi\left(S^\dagger + S\right)^2
- \frac{1}{2}\left(1-\chi\right) \left(S^\dagger - S\right)^2 =
\chi\,\sigma^2 + \left(1-\chi\right) \tau^2 \,, \quad
S = \frac{1}{\sqrt{2}}\left(\sigma + i \tau\right)\,.
\end{align}
This form of $F$ illustrates that, for $\chi \ll 1$, the kinetic
function is approximately invariant under shifts in $\sigma$, i.e.,
the real scalar part of the inflaton field $S$.
Conversely, $\chi$ values close to one, $1-\chi \ll 1$, lead to an approximate
shift symmetry in $\tau$, i.e., the imaginary scalar component of $S$.
In the following, we will focus w.l.o.g.\ on the first of these two cases.
In passing, we also mention that the trivial case $\chi = \chi_{\rm CSS} = 1/2$ (which renders the
holomorphic function $J$ vanishing) corresponds to an inflaton field
that is conformally coupled to the Ricci scalar.
This choice for the parameter $\chi$ would allow to construct a SUGRA model
that is invariant under local superconformal symmetry (see the discussion
below Eq.~\eqref{eq:OmegaPhi}).
However, as argued above, we would then fail to solve the eta problem
(see Eq.~\eqref{eq:eta23}).
For this reason, we need to break the superconformal symmetry.
In fact, by choosing $\chi \ll \chi_{\rm CSS}$, we break the superconformal symmetry
in a maximal sense in favor of an approximate shift symmetry.


Given the kinetic function in Eq.~\eqref{eq:F}, it is straightforward
to solve the eta problem.
In consequence of the approximate shift symmetry, all contributions
to the inflaton mass $m_\sigma$ end up being suppressed by $\chi$.
This follows from an explicit computation of $m_\sigma$ in the Einstein frame
(see Eq.~\eqref{eq:msigmaDF} in Sec.~\ref{subsec:potential}),
\begin{align}
\label{eq:msigma}
m_\sigma^2 \approx 2\chi\left[m_R^2 - \left(1-2\chi\right) m_{3/2}^2\right] \,.
\end{align}
As expected, $m_\sigma$ reduces to $m_R$ in the limit $\chi\rightarrow 1/2$.
On the other hand, if $\chi$ is chosen small enough, $m_\sigma$ becomes suppressed,
so that the slow-roll parameter $\eta$ remains sufficiently small during inflation.%
\footnote{Imposing an approximate shift symmetry in the direction of the
inflaton field would also allow to solve the eta problem in the Einstein frame.
There, the inflaton mass also vanishes in the limit of an exact shift symmetry.
From this perspective, our solution to the eta problem actually does not
serve as an additional motivation to work in the Jordan frame.
However, our arguments regarding the MSSM sfermion problem remain unchanged.
This problem is best solved in the Jordan frame
(see the discussion around Eq.~\eqref{eq:scalesJF}).
We will therefore continue to work in the Jordan frame.}
 

\subsubsection*{Explicit breaking of the shift symmetry}


An exact shift symmetry is out of reach in our model, as the Yukawa coupling
in the superpotential,
$W_{\rm inf} = \kappa\,S\,\Phi\bar{\Phi}$, breaks any inflaton (or waterfall
field) shift symmetry explicitly.
Therefore, while $\chi$ may be zero at tree level, a nonvanishing value
of the shift-symmetry-breaking parameter $\chi$ is
always generated via radiative corrections.
To see this, let us consider the \textit{one-particle-irreducible} (1PI)
effective action in global supersymmetry.
The superpotential does not receive any quantum corrections
in consequence of the SUSY nonrenormalization theorem~\cite{Grisaru:1979wc}.
The renormalization of the K\"ahler potential is, however, nontrivial
and results in a one-loop effective K\"ahler
potential $K_{1\ell}$~\cite{Buchbinder:1994iw,Grisaru:1996ve,Brignole:2000kg,Nibbelink:2006si}.
Along the inflationary trajectory, $\Phi = \bar{\Phi} = 0$, 
a calculation in the {\footnotesize$\overline{\textrm{MS}}$}
renormalization scheme yields
\begin{align}
\label{eq:K1l}
K_{1\ell} = 2\, \chi_{1\ell} \left[1-\frac{1}{2}
\ln\left(\frac{\kappa^2 S^\dagger S}{\bar{\mu}^2}\right)\right]
S^\dagger S \,, \quad \chi_{1\ell} = \frac{\kappa^2}{16\pi^2} \,,
\end{align}
where $\bar{\mu}$ denotes the {\footnotesize$\overline{\textrm{MS}}$}
renormalization scale.
Note that this result for $K_{1\ell}$ corresponds to a wave-function
renormalization of the inflaton field $S$.
Next, let us embed the effective K\"ahler
potential in Eq.~\eqref{eq:K1l} into supergravity.
In the Einstein frame, the relevant quantity is 
the total K\"ahler potential $K_{\rm tot}$, which simply follows from 
adding $K_{1\ell}$ to the tree-level K\"ahler potential, $K_{\rm tot} = K_{\rm tree} + K_{1\ell}$.
In the Jordan frame, we are by contrast interested in the total frame
function, $\Omega_{\rm tot} = \Omega_{\rm tree} + \Omega_{1\ell}$.
One can show that the one-loop correction to the tree-level frame function
is related to $K_{1\ell}$ as follows,
\begin{align}
\label{eq:O1l}
\Omega_{1\ell} = \Omega_{\rm tree} \left(
\exp\left[-\frac{K_{1\ell}}{3M_{\rm Pl}^2}\right]-1\right) =
K_{1\ell} + \mathcal{O}\left(M_{\rm Pl}^{-2}\right) \,.
\end{align}
Here, the higher-order terms correspond to Planck-suppressed radiative corrections
which are negligibly small.
Together, Eqs.~\eqref{eq:F}, \eqref{eq:K1l}, and \eqref{eq:O1l}
allow us to determine the effective $\chi$ parameter that is induced by
the breaking of shift symmetry in the superpotential.
Along the direction of the real inflaton component, $\tau = 0$, 
we obtain the following one-loop kinetic function for the inflaton field,
\begin{align}
F_{1\ell} \simeq K_{1\ell} \:\:\overset{\tau = 0}{\longrightarrow}\:\:
\frac{1}{2}\,\chi_{\rm eff}\left(S^\dagger + S\right)^2 \,, \quad 
\chi_{\rm eff} = \chi_{1\ell}
\left[1-\frac{1}{2}\ln\left(\frac{\kappa^2 S^\dagger S}{\bar{\mu}^2}\right)\right]
\sim \chi_{1\ell} \,.
\end{align}


In the absence of any tree-level contribution, the shift-symmetry-breaking
parameter $\chi$ is therefore expected to be of the
order of $\kappa^2/ \left(16\pi^2\right)$.
This is an important result which was overlooked in~\cite{Domcke:2017xvu}.
There, we simply varied $\chi$ as a free parameter for fixed $\kappa$.
Of course, this is a valid procedure, given the fact that $\chi$ can very well
receive further tree-level contributions (or further radiative corrections
from inflaton couplings to extra heavy states).
In this case, $\chi$ is simply the sum of various contributions,
$\chi = \chi_{\rm tree} + \chi_{1\ell}$, which can take any arbitrary value.
But the case $\chi = \chi_{1\ell}$\,---\,which we had neglected thus far\,---\,is
special, as it corresponds to a scenario with minimal field content and number
of free parameters.
We will study this scenario in more detail in Sec.~\ref{subsec:scan}.
This will represent one of the main results of this
paper and a significant step forward beyond our analysis in~\cite{Domcke:2017xvu}.
In particular, we will find that
$\chi = \chi_{1\ell}$ leads to inflation in new parts of parameter space
that we had dismissed before.


Finally, we point out that the fact that we are unable to realize an
exact shift symmetry is a virtue rather than a shortcoming of our model.
A slightly broken shift symmetry allows us to get a handle on
the scalar spectral index $n_s$ which we would otherwise lack
in the case of an exact shift symmetry.
The prediction for $n_s$ in standard D-term inflation in global supersymmetry
roughly corresponds to
\begin{align}
\label{eq:ns098}
n_s = 1 + 2\,\eta - 6\,\varepsilon \simeq 1 + 2\,\eta \gtrsim 1 - \frac{1}{N_e} \simeq 0.98 \,,
\end{align}
where $N_e$ denotes the number of e-folds between the end of inflation
and the time $t_{\rm CMB}$ when the CMB pivot scale, $k_{\rm CMB} = 0.05\,\textrm{Mpc}^{-1}$,
exits the Hubble horizon during inflation.
The prediction in Eq.~\eqref{eq:ns098} exceeds the current best-fit value,
$n_s^{\rm obs} = 0.9677 \pm 0.0060$ \cite{Ade:2015lrj}, by at least $2\,\sigma$.
This puts some phenomenological pressure on the simplest version of D-term inflation.
To improve on the predicted value of $n_s$, various SUGRA models have been proposed
in the literature~\cite{Buchmuller:2012ex,Buchmuller:2013zfa,Seto:2005qg,
Lin:2006xta,Rocher:2006nh,Buchmuller:2013uta}.
However, in our scenario, no extra effort is needed to enhance
the absolute value of the slow-roll parameter $\eta$ and, thus,
reproduce the best-fit value for $n_s$. 
The inflaton mass in Eq.~\eqref{eq:msigma} approximately results in
\begin{align}
\eta \sim - \frac{2\chi}{3}\left(\frac{m_{3/2}}{H_J}\right)^2 \,.
\end{align}
Therefore, to realize $n_s$ values around $n_s \simeq 0.96$, all we have to do is
to choose $\chi$ small enough,
\begin{align}
n_s \simeq 0.96 \qquad\Rightarrow\qquad
\eta \sim -0.02 \qquad\Rightarrow\qquad
\chi \sim 0.03 \left(\frac{H_J}{m_{3/2}}\right)^2 \,.
\end{align}
In this sense, the approximate shift symmetry in the kinetic
function of the inflaton field automatically provides a possibility
to achieve a scalar spectral index consistent with the observational data.


\subsubsection*{Three physical assumptions to solve five problems of D-term inflation}


So far, we have mainly outlined the ingredients
of our construction in physical and less technical terms.
We hope that this part of our discussion will be accessible also to readers
without a strong background in SUGRA model building.
A more technical description of our model will be given in the next three
sections (see
Secs.~\ref{subsec:iyit}, \ref{subsec:polonyi}, and \ref{subsec:potential}).
Readers less interested in the technical aspects of our model and more
interested in its phenomenological implications may skip directly to Sec.~\ref{sec:inflation}.


Before entering the technical part of our discussion, let us summarize our insights
up to this point.
On the one hand, we showed that D-term inflation faces a number of challenges.
We discussed the following five problems:
(i) The generation of the FI term in the D-term scalar potential and its embedding
into supergravity,
(ii) the production of dangerous cosmic strings at the end of inflation,
(iii) the stabilization of dangerous MSSM sfermion directions in the scalar potential
during and after inflation,
(iv) the eta problem in supergravity, and 
(v) the tension between the lower bound on $n_s$ in D-term inflation
and the current best-fit value.
On the other hand, we argued that all five of these problems can be solved
if one makes the following three assumptions:
(i) The FI term is dynamically generated in the hidden SUSY-breaking sector.
(ii) The canonical description of D-term inflation in supergravity corresponds
to the embedding into the (standard) Jordan frame with canonically normalized kinetic
terms for all scalar fields.
(iii) The kinetic function of the inflaton field exhibits a slightly broken
shift symmetry.
Our model therefore turns out to be a viable SUGRA realization of $B$$-$$L$ D-term
inflation that is consistent with all theoretical and phenomenological constraints.


\subsection{SUSY-breaking dynamics in the hidden sector}
\label{subsec:iyit}


In the previous section, we summarized our physical ideas about how to realize
a viable SUGRA model of D-term inflation that (i) unifies the dynamics of
supersymmetry breaking and inflation and that (ii) ends in the $B$$-$$L$ phase
transition at energies around the GUT scale.
In the following, we will show how these ideas can be implemented into
a specific model of dynamical SUSY breaking: the
\textit{Izawa-Yanagida-Intriligator-Thomas} (IYIT)
model~\cite{Izawa:1996pk,Intriligator:1996pu}, which represents the
simplest vector-like model of dynamical SUSY breaking.
Despite this choice, we believe that our general ideas
extend beyond our specific model. 
In future work, it would be interesting to study alternative DSB models
and assess which other models might give rise to unified dynamics of
supersymmetry breaking and inflation.


\subsubsection*{IYIT sector at high and low energies}


We begin by reviewing the IYIT model.
In its most general form, the IYIT model corresponds to a strongly coupled
supersymmetric $Sp(N)$ gauge theory.%
\footnote{In our notation, the compact symplectic group $Sp(N)$
is identical to the unitary group over the quaternions, $U(N,\mathbb{H})$.
Here, $N$ denotes the dimension of the quaternionic vector space $\mathbb{H}^N$
that $Sp(N)$ acts on in its fundamental representation.}
At high energies, its charged matter content consists
of $N_f = N+1$ vector-like pairs of quark flavors, where
each quark field $\Psi_i$ transforms in the fundamental representation of $Sp(N)$.
The theory becomes confining at energies around the dynamical scale $\Lambda_{\rm dyn}$
which is generated via dimensional transmutation.
Below $\Lambda_{\rm dyn}$, the dynamical DOFs in the IYIT sector
correspond to a set of $N_f\left(2N_f-1\right)$ gauge-invariant
composite meson fields $M_{ij}$,
\begin{align}
M_{ij} \simeq \frac{1}{\eta\, \Lambda_{\rm dyn}} \left<\Psi_i \Psi_j\right> \,, \quad
\eta \sim 4\pi \,, \quad i,j = 1,2,\cdots,2N_f \,,
\end{align}
where $M_{ji} = -M_{ij}$ and where $\eta$ denotes a numerical factor of
$\mathcal{O}\left(4\pi\right)$ that follows from \textit{naive dimensional analysis}
(NDA)~\cite{Manohar:1983md,Georgi:1986kr,Luty:1997fk,Cohen:1997rt}.
It turns out to be useful to absorb the NDA factor
$\eta$ in the dynamical scale $\Lambda_{\rm dyn}$.
In the following, we will therefore work with the reduced dynamical scale $\Lambda$,
\begin{align}
\Lambda = \frac{\Lambda_{\rm dyn}}{\eta} \sim \frac{\Lambda_{\rm dyn}}{4\pi} \,. 
\end{align}
At low energies, the scalar meson VEVs parametrize a moduli space
of degenerate supersymmetric vacua.
This moduli space is subject to a constraint equation, which,
in the classical limit, corresponds to the requirement
that the Pfaffian of the antisymmetric meson matrix $M_{ij}$ must vanish,
$\textrm{Pf}\left(M_{ij}\right) = 0$.
This constraint, however, becomes deformed
in the quantum theory.
There, it reads~\cite{Seiberg:1994bz}
\begin{align}
\label{eq:Pf}
\textrm{Pf}\left(M_{ij}\right) \simeq \Lambda^{N_f} \,.
\end{align}


To break supersymmetry in the IYIT sector, one needs to lift the
flat directions in moduli space.
This is readily achieved by coupling the IYIT quarks $\Psi^i$ to a 
set of $N_f\left(2N_f-1\right)$ singlet fields $Z_{ij}$,
\begin{align}
\label{eq:WhidHE}
\textrm{At high energies:} \qquad
W_{\rm hid} = \frac{1}{2}\, \lambda_{ij}\, Z_{ij}\, \Psi_i \Psi_j \,,
\end{align}
where $\lambda_{ij} = - \lambda_{ji}$ are dimensionless coupling constants.
At high energies, these couplings are nothing but ordinary Yukawa
couplings which do not affect the vacuum structure of the theory.
At low energies, the terms in Eq.~\eqref{eq:WhidHE}, however,
turn into mass terms for the meson and singlet fields $M_{ij}$ and $Z_{ij}$,
\begin{align}
\label{eq:WhidLE}
\textrm{At low energies:} \qquad
W_{\rm hid} \simeq \frac{1}{2}\,\lambda_{ij}\Lambda\, Z_{ij}\, M_{ij} \,.
\end{align}
These mass terms single out the origin in field space as the true
supersymmetric ground state.
The quantum-mechanically deformed moduli constraint in Eq.~\eqref{eq:Pf},
however, prevents the system from reaching the origin in field space.
This breaks supersymmetry.
The theory is forced to settle into a vacuum away
from the origin, $\left<M_{ij}\right> \neq 0$, where some of
the singlet F-term conditions, $F_{Z_{ij}} = 0$, cannot be satisfied.
Supersymmetry is, hence, broken \`a la O'Raifeartaigh
by nonvanishing F terms~\cite{ORaifeartaigh:1975nky}.


In the following, we shall focus on the minimal $N = 1$ realization
of the IYIT model, for simplicity.
In this case, the $Sp(1)$ gauge dynamics are equivalent to those of an $SU(2)$
theory, $Sp(1) \cong SU(2)$, and we have to deal with four quark
fields $\Psi_i$ and six singlet fields $Z_{ij}$ at high energies.
This translates into six meson fields $M_{ij}$ (and six singlet fields
$Z_{ij}$) at low energies.
As we will see shortly, it turns out to be convenient to label the fields in
the low-energy theory in a suggestive manner.
To do so, we first note that Eq.~\eqref{eq:WhidHE} exhibits a global $U(1)_A$ flavor
symmetry that corresponds to an axial quark phase rotation.
The $U(1)_A$ charges of the two quark flavors at high energies can be chosen as follows,
\begin{align}
\left[\Psi_1\right]_A = \left[\Psi_2\right]_A = +\frac{q_0}{2} \,, \quad
\left[\Psi_3\right]_A = \left[\Psi_4\right]_A = -\frac{q_0}{2} \,.
\end{align}
This normalization ensures that the charged meson fields at low energies
carry $U(1)_A$ charges $\pm q_0$,
\begin{align}
\left[M_{12}\right]_A = + q_0 \,, \quad 
\left[M_{34}\right]_A = - q_0 \,, \quad 
\left[M_{13}\right]_A = \left[M_{14}\right]_A = \left[M_{23}\right]_A = \left[M_{24}\right]_A =  0 \,,
\end{align}
and similarly for the $Z_{ij}$.
In the second step, we relabel all fields according to their $U(1)_A$ charges,
\begin{align}
M_{12} & \rightarrow M_+   \,, &
M_{34} & \rightarrow M_-   \,, &
M_{13} & \rightarrow M_0^1 \,, &
M_{14} & \rightarrow M_0^2 \,, &
M_{23} & \rightarrow M_0^3 \,, &
M_{24} & \rightarrow M_0^4 \,,
\\ \nonumber
Z_{12} & \rightarrow Z_-   \,, &
Z_{34} & \rightarrow Z_+   \,, &
Z_{13} & \rightarrow Z_0^1 \,, &
Z_{14} & \rightarrow Z_0^2 \,, &
Z_{23} & \rightarrow Z_0^3 \,, &
Z_{24} & \rightarrow Z_0^4 \,.
\end{align}
In this notation, the low-energy superpotential in Eq.~\eqref{eq:WhidLE}
takes the following form,
\begin{align}
\label{eq:Whidpm}
W_{\rm hid} \simeq \Lambda\left(\lambda_+ M_+ Z_- + \lambda_- M_- Z_+
+ \lambda_0^1 M_0^1 Z_0^1
+ \lambda_0^2 M_0^2 Z_0^2
+ \lambda_0^3 M_0^3 Z_0^3
+ \lambda_0^4 M_0^4 Z_0^4 \right) \,,
\end{align}
where we also relabeled the $\lambda_{ij}$.
Meanwhile, the constraint in Eq.~\eqref{eq:Pf} can now be written as follows,
\begin{align}
\label{eq:Pfpm}
\textrm{Pf}\left(M_{ij}\right) = M_+ M_- - M_0^1 M_0^4 + M_0^2 M_0^3
\simeq \Lambda^2 \,.
\end{align}


Together, Eqs.~\eqref{eq:Whidpm} and \eqref{eq:Pfpm} allow to explicitly calculate
the VEVs in the SUSY-breaking vacuum.
As it turns out, the location of the true ground state in meson field space depends on 
the hierarchy among three geometric means of Yukawa couplings,
$\lambda = \left(\lambda_+\lambda_-\right)^{1/2}$,
$\lambda_0^{14} = \left(\lambda_0^1\lambda_0^4\right)^{1/2}$,
$\lambda_0^{23} = \left(\lambda_0^2\lambda_0^3\right)^{1/2}$,
\begin{align}
\lambda < \min\left\{\lambda_0^{14},\lambda_0^{23}\right\} \qquad\Rightarrow\qquad
\left<M_+M_-\right>     & \simeq \Lambda^2 \,, &
\left<M_0^1M_0^4\right> & =      0         \,, &
\left<M_0^2M_0^3\right> & =      0         \,,
\\ \nonumber
\lambda_0^{14} < \min\left\{\lambda,\lambda_0^{23}\right\} \qquad\Rightarrow\qquad
\left<M_+M_-\right>     & =      0         \,, &
\left<M_0^1M_0^4\right> & \simeq \Lambda^2 \,, &
\left<M_0^2M_0^3\right> & =      0 \,,
\\ \nonumber
\lambda_0^{23} < \min\left\{\lambda,\lambda_0^{14}\right\} \qquad\Rightarrow\qquad
\left<M_+M_-\right>     & =      0         \,, &
\left<M_0^1M_0^4\right> & =      0         \,, &
\left<M_0^2M_0^3\right> & \simeq \Lambda^2 \,.
\end{align}
We will assume that the first of these three cases is realized, 
$\lambda < \min\left\{\lambda_0^{14},\lambda_0^{23}\right\}$.
In this case, it is the charged meson fields that obtain a nonzero VEV,
$\left<M_+M_-\right> \simeq \Lambda^2$. 
This case is special in the sense that the global $U(1)_A$ flavor symmetry
becomes spontaneously broken at low energies.
In the other two cases, the flavor symmetry remains unbroken
even in the SUSY-breaking vacuum.


\subsubsection*{Properties of the low-energy vacuum}


Let us now discuss the properties of the $U(1)_A$-breaking vacuum
in a bit more detail.
In this vacuum, all neutral fields are stabilized
by their supersymmetric masses in Eq.~\eqref{eq:Whidpm}.
The relevant terms in the superpotential and Pfaffian constraint
are therefore only those involving
charged fields,
\begin{align}
\label{eq:WhidPf}
W_{\rm hid} \simeq \Lambda\left(\lambda_+ M_+ Z_- + \lambda_- M_- Z_+ \right) \,, \quad 
\textrm{Pf}\left(M_{ij}\right) = M_+ M_- \simeq \Lambda^2 \,.
\end{align}
The constraint is most easily accounted for by adding
a Lagrange multiplier term to the superpotential,
\begin{align}
\label{eq:WhidT}
W_{\rm hid} \simeq \Lambda\left(\lambda_+ M_+ Z_- + \lambda_- M_- Z_+ \right)
+ \lambda_T \, T \left(M_+ M_- - \Lambda^2\right) \,,
\end{align}
where the field $T$ represents the actual Lagrange multiplier.
The physical nature of the field $T$ depends on strong-coupling effects
in the K\"ahler potential.
If it acquires a nonperturbative kinetic term from the strong gauge dynamics,
$T$ becomes physical.
On the other hand, if no kinetic term is generated, $T$ is merely an auxiliary
field that remains unphysical.
Unfortunately, it is unknown which of these cases is realized, as the
K\"ahler potential for $T$ in the strong-coupling regime is incalculable.
At any rate, the difference between the two cases is mostly irrelevant
for our purposes.
All effects in the case of a physical Lagrange multiplier field $T$
are suppressed by powers of $\lambda/\left(4\pi\right)$~\cite{Chacko:1998si}.
Thus, as long as we stay in the perturbative regime, $\lambda \ll 4\pi$,
our results will not be affected by the physical status of the field $T$.
In the following, we will therefore assume that $T$ remains unphysical, for simplicity.
In practice, this means that we will take the limit $\lambda_T \rightarrow \infty$
wherever possible.
For discussions of the IYIT model based on the assumption of a physical
Lagrange multiplier field $T$, see, e.g.,~\cite{Schmitz:2016kyr,Harigaya:2015soa}.


Given the superpotential in Eq.~\eqref{eq:WhidT} (and taking the limit
$\lambda_T \rightarrow \infty$ at the end of the calculation), one can easily show that
the vacuum energy density is minimized for the following meson VEVs,
\begin{align}
\label{eq:MVEVs}
\left<M_\pm\right> = \frac{\lambda}{\lambda_\pm} \Lambda \,.
\end{align}
These meson VEVs induce SUSY-breaking F terms for the singlet fields $Z_\pm$.
To determine the total F-term SUSY breaking scale $\mu$, it is useful to
transform the fields $M_\pm$ and $Z_\pm$ to a new basis,
\begin{align}
\label{eq:AMXY}
\begin{pmatrix} A \\ M \end{pmatrix} = \frac{1}{f_A}
\begin{pmatrix} \left<M_+\right> & -\left<M_-\right> \\
                \left<M_-\right> & \phantom{-}\left<M_+\right> \end{pmatrix} 
\begin{pmatrix} M_+ \\ M_- \end{pmatrix} \,, \quad
\begin{pmatrix} X \\ Y \end{pmatrix} = \frac{1}{\sqrt{2}}
\begin{pmatrix} 1 & \phantom{-}1 \\ 1 & -1 \end{pmatrix} 
\begin{pmatrix} Z_+ \\ Z_- \end{pmatrix} \,,
\end{align}
where we introduced $f_A$ to denote the total energy scale of spontaneous $U(1)_A$ breaking,
\begin{align}
f_A = \left(\left<M_+^2\right> + \left<M_-^2\right>\right)^{1/2} =
\left(\frac{\lambda_+}{\lambda_-} + \frac{\lambda_-}{\lambda_+}\right)^{1/2}\Lambda\,.
\end{align}
In the new field basis, the fields $M$ and $T$ share a supersymmetric
Dirac mass term, $m_{MT} = \lambda_T f_A$, that formally
diverges in the limit $\lambda_T \rightarrow \infty$.
This allows us to identify $M$ as the meson field that becomes eliminated
by the Pfaffian constraint.
The remaining meson DOFs are then described by the orthogonal linear 
combination, i.e., by the field $A$.
Note that this automatically implies that the field $A$ plays the role of
the chiral Goldstone multiplet of spontaneous $U(1)_A$ breaking.
From this perspective, the energy scale $f_A$ may also be regarded
as the Goldstone decay constant.
To obtain the superpotential describing the low-energy dynamics of $A$, $X$, and $Y$,
we proceed as follows:
(i) We  perform the field rotation in Eq.~\eqref{eq:AMXY},
(ii) shift the meson fields $A$ and $M$ by their nonvanishing VEVs,
\begin{align}
A \rightarrow \frac{1}{f_A}\left(\left<M_+^2\right> - \left<M_-^2\right>\right) + A \,, \quad
M \rightarrow \frac{2}{f_A}\left<M_+\right>\left<M_-\right> + M \,,
\end{align}
and (iii) integrate out the heavy fields $M$ and $T$.
This results in the following superpotential,
\begin{align}
\label{eq:WhidAXY}
W_{\rm hid} \simeq \mu^2 X - m_F\, Y A + \frac{1}{2}\,\lambda_X X A^2 \,.
\end{align}
Here, $\mu$ denotes the total F-term SUSY breaking scale,
\begin{align}
\label{eq:mu}
\mu = \left[\left(\lambda_+^2\left<M_+^2\right> +
\lambda_-^2\left<M_-^2\right>\right)\Lambda^2\right]^{1/4} =
2^{1/4}\lambda^{1/2} \Lambda \,.
\end{align}


By construction, the singlet field $X$ is the only field with a nonzero F term,
$\left<\left|F_X\right|\right> = \mu^2$.
It can therefore be identified with the SUSY-breaking Polonyi field.
The orthogonal field $Y$ shares a Dirac mass term with the $U(1)_A$
Goldstone field $A$ which is given in terms of the mass scales $\mu$ and $f_A$,
\begin{align}
\label{eq:mF}
m_F = \frac{\mu^2}{f_A} = \rho\, \lambda\, \Lambda \,, \quad
\rho = \left[\frac{1}{2}\left(r_\lambda + \frac{1}{r_\lambda}
\right)\right]^{-1/2} \,, \quad r_\lambda = \frac{\lambda_+}{\lambda_-} \,.
\end{align}
Here, $\rho$ measures the degeneracy 
between $\lambda_+$ and $\lambda_-$.
For $\lambda_+ \rightarrow \lambda_-$, the parameter $\rho$ approaches one,
$\rho \rightarrow 1$.
For $\lambda_+ \ll \lambda_-$ or $\lambda_+ \ll \lambda_-$,
it goes to zero, $\rho \rightarrow 0$.
In the following, we will assume that both $\lambda_+$ and $\lambda_-$
are sufficiently small, so that we always stay in the perturbative regime.
For definiteness, let us require that both couplings
are always at least half an order of magnitude smaller than $4\pi$,
\begin{align}
\lambda_\pm < \lambda_{\rm pert} = 4 \simeq 10^{-1/2} 4\pi \,.
\end{align}
This translates into a lower bound on the hierarchy parameter $\rho$
in dependence of $\lambda = \left(\lambda_+\lambda_-\right)^{1/2}$,
\begin{align}
\label{eq:rhopert}
\lambda_\pm < \lambda_{\rm pert} \qquad\Rightarrow\qquad
\rho > \rho_{\rm pert} = \left[\frac{1}{2}
\left(r_{\rm pert} + \frac{1}{r_{\rm pert}} \right)\right]^{-1/2} \,, \quad
r_{\rm pert} = \frac{\lambda_+\lambda_-}{\lambda_{\rm pert}^2} \,.
\end{align}
Moreover, to simplify our analysis, we will
replace $\rho$ by its expectation value $\bar{\rho}$ in the following.
We compute $\bar{\rho}$ by averaging $\rho$ over
all possible values of $\lambda_\pm$, varying both couplings on a linear scale,
\begin{align}
\label{eq:rho0}
\bar{\rho} = \frac{1}{\lambda_{\rm pert}^2}
\int_0^{\lambda_{\rm pert}} \int_0^{\lambda_{\rm pert}}
\rho\: d\lambda_+ d\lambda_- \simeq 0.80 \,.
\end{align}
Note that this result is independent of the concrete value of $\lambda_{\rm pert}$.
With $\rho$ fixed at this value, the perturbativity constraint in Eq.~\eqref{eq:rhopert}
turns into an upper bound on the Yukawa coupling $\lambda$,
\begin{align}
\label{eq:lmax}
\lambda_\pm < \lambda_{\rm pert} \,, \quad \rho = \bar{\rho}
\qquad\Rightarrow\qquad \lambda <
\left[1-\left(1-\bar{\rho}^4\right)^{1/2}\right]^{1/2}
\frac{\lambda_{\rm pert}}{\bar{\rho}} \simeq 0.60\,\lambda_{\rm pert} \simeq 2.41 \,.
\end{align}
This yields in turn an upper bound on the F-term-induced mass in the superpotential,
$m_F \lesssim 1.94\,\Lambda$.
The lesson from this analysis is the following:
From now on, we will work with $\rho \simeq 0.80$ and $\lambda \lesssim 2.41$.
The Yukawa couplings $\lambda_\pm$ are then guaranteed to assume
``typical values'' (in the sense of the average in Eq.~\eqref{eq:rho0})
which are, at the same time, consistent with the requirement
of perturbativity, $\lambda < \lambda_{\rm pert}$.


Similar to the mass $m_F$, also the Yukawa coupling $\lambda_X$
in Eq.~\eqref{eq:WhidAXY} is given in terms of $\mu$ and $f_A$,
\begin{align}
\lambda_X = \left(\frac{\mu}{f_A}\right)^2 = \left(\frac{m_F}{\mu}\right)^2 =
\frac{\rho^2\lambda}{\sqrt{2}} \,.
\end{align}
This Yukawa coupling between the Polonyi field $X$ and the Goldstone field $A$
is is a direct consequence of the $T\,M_+M_-$ term in Eq.~\eqref{eq:WhidT}.
Similar couplings also exist between $X$ and the neutral meson fields.
This is shown explicitly in Appendix A of~\cite{Harigaya:2015soa}
(see also~\cite{Chacko:1998si}).
Together, these Yukawa couplings result in an effective Polonyi mass $m_{1\ell}$
at the one-loop level.
An explicit calculation yields~\cite{Harigaya:2015soa}
\begin{align}
\label{eq:m1l}
m_{1\ell} = c_{1\ell}\,\lambda^2 \Lambda \,,\quad
c_{1\ell} = \left(\frac{2\ln2-1}{32\pi^2}\,n_M^{\rm eff}\right)^{1/2} \,.
\end{align}
Here, $n_M^{\rm eff}$ denotes the effective number of meson loops
contributing to the Polonyi mass.
Let us assume that all neutral mesons share the same Yukawa coupling,
$\lambda_0^{1,2,3,4} \equiv \lambda_0$.
In this case, $n_M^{\rm eff}$ can be brought into the following compact form
(the full expression is complicated and can be found in~\cite{Harigaya:2015soa}),
\begin{align}
\label{eq:nMeff}
n_M^{\rm eff} = \rho^6 + 4\,\ell_0 \,,
\end{align}
where $\ell_0 = \ell\left(\lambda/\lambda_0\right)$ is a loop function
that can be approximated by a simple quadratic power law,
\begin{align}
\ell\left(x\right) = \frac{1}{2\ln2-1} \left[\frac{1}{2}
\left(1+\frac{1}{x^2}\right)^2\ln\left(1+x^2\right) - \frac{1}{2}
\left(1-\frac{1}{x^2}\right)^2\ln\left(1-x^2\right) - \frac{1}{x^2}\right] \approx x^2 \,.
\end{align}
In the following, we will set $\lambda_0 = 4\pi$ to account for the presence
of heavy composite states with masses around the dynamical scale,
$m_{\rm heavy} = \lambda_0\,\Lambda \sim \Lambda_{\rm dyn}$.
Just like in QCD, such heavy resonances are expected to appear in the spectrum.
Our perturbative language, however, does
not suffice to capture their dynamics at low energies.
For this reason, we will instead follow an effective approach and mimic the effect 
of additional heavy states by means of a particular choice for $\lambda_0$.
For a Yukawa coupling $\lambda$ of $\mathcal{O}(1)$,
the one-loop coefficient $c_{1\ell}$ in Eq.~\eqref{eq:m1l} is then roughly given
by $c_{1\ell} \simeq 0.02$.
In global supersymmetry, the Polonyi field $X$
corresponds to a tree-level flat direction (see Eq.~\eqref{eq:WhidAXY}).
The loop-induced Polonyi mass $m_{1\ell}$ is therefore crucial to stabilize the
SUSY-breaking vacuum against gravitational corrections in 
supergravity~\cite{Chacko:1998si}.
We will discuss this issue in more detail in Sec.~\ref{subsec:polonyi}.


\subsubsection*{Dynamical generation of an effective FI term}


The IYIT model can also be used to dynamically generate an effective FI term.
This was pointed out for the first time in~\cite{Domcke:2014zqa}.
In this paper, we will make use of this mechanism to generate
the effective FI term required for $B$$-$$L$ D-term inflation.
All we have to do now is to promote the global $U(1)_A$ flavor symmetry
in the IYIT superpotential (see Eq.~\eqref{eq:WhidT})
to a local $U(1)_{B-L}$ gauge symmetry.
The $B$$-$$L$ gauge interactions then result in the following D-term 
scalar potential in the IYIT sector,
\begin{align}
\label{eq:VDIYIT}
V_D = \frac{g^2 q_0^2}{2} \left(\left|m_+\right|^2 - \left|m_-\right|^2
+ \left|z_+\right|^2 - \left|z_-\right|^2\right)^2 \,.
\end{align}
Here and in the following, lowercase symbols ($m_\pm$, $z_\pm$, etc.)
denote the complex scalar components of the corresponding
chiral multiplets ($M_\pm$, $Z_\pm$, etc.).
The charged mesons $M_\pm$ acquire nonzero VEVs as a result
of the dynamical breaking of supersymmetry (see Eq.~\eqref{eq:MVEVs}).
These VEVs spontaneously break $B$$-$$L$ which leads to the following effective FI term,
\begin{align}
\label{eq:xiIYIT}
\xi = \left<M_-^2\right> - \left<M_+^2\right> \,.
\end{align}
This FI parameter is exactly of the form that we anticipated in Eq.~\eqref{eq:xidyn}
in Sec.~\ref{subsec:ingredients}.
In particular, we now see that, in our dynamical model, the roles of
the moduli $\Psi$ and $\bar{\Psi}$ are played by the mesons $M_\pm$.
To evaluate the expression in Eq.~\eqref{eq:xiIYIT}, it is, in principle, necessary
to account for the backreaction of the D-term scalar potential in Eq.~\eqref{eq:VDIYIT}
on the meson VEVs in Eq.~\eqref{eq:MVEVs}.
In the following, we will, however, restrict ourselves to the weakly gauged limit,
$\left|gq_0\right| \lesssim \lambda$, where this backreaction is negligible.
In this case, we can simply continue to work with our results in Eq.~\eqref{eq:MVEVs},
such that Eq.~\eqref{eq:xiIYIT} turns into
\begin{align}
\xi = \left(\frac{\lambda_+}{\lambda_-} - \frac{\lambda_-}{\lambda_+}\right) \Lambda^2 
= \frac{2\left(1-\rho^4\right)^{1/2}}{\rho^2}\,\Lambda^2 \,.
\end{align}
Here, we assumed w.l.o.g.\ that $\lambda_+ > \lambda_-$ such that the FI parameter
is always positive, $\xi > 0$.
Once we set the parameter $\rho$ to its expectation value, $\bar{\rho} \simeq 0.80$,
we obtain the following simple relation,
\begin{align}
\label{eq:xifaL}
\xi^{1/2} \simeq 0.88\,f_A \simeq 1.54\,\Lambda \,.
\end{align}

\begin{figure}
\centering
\includegraphics[width=0.48\textwidth]{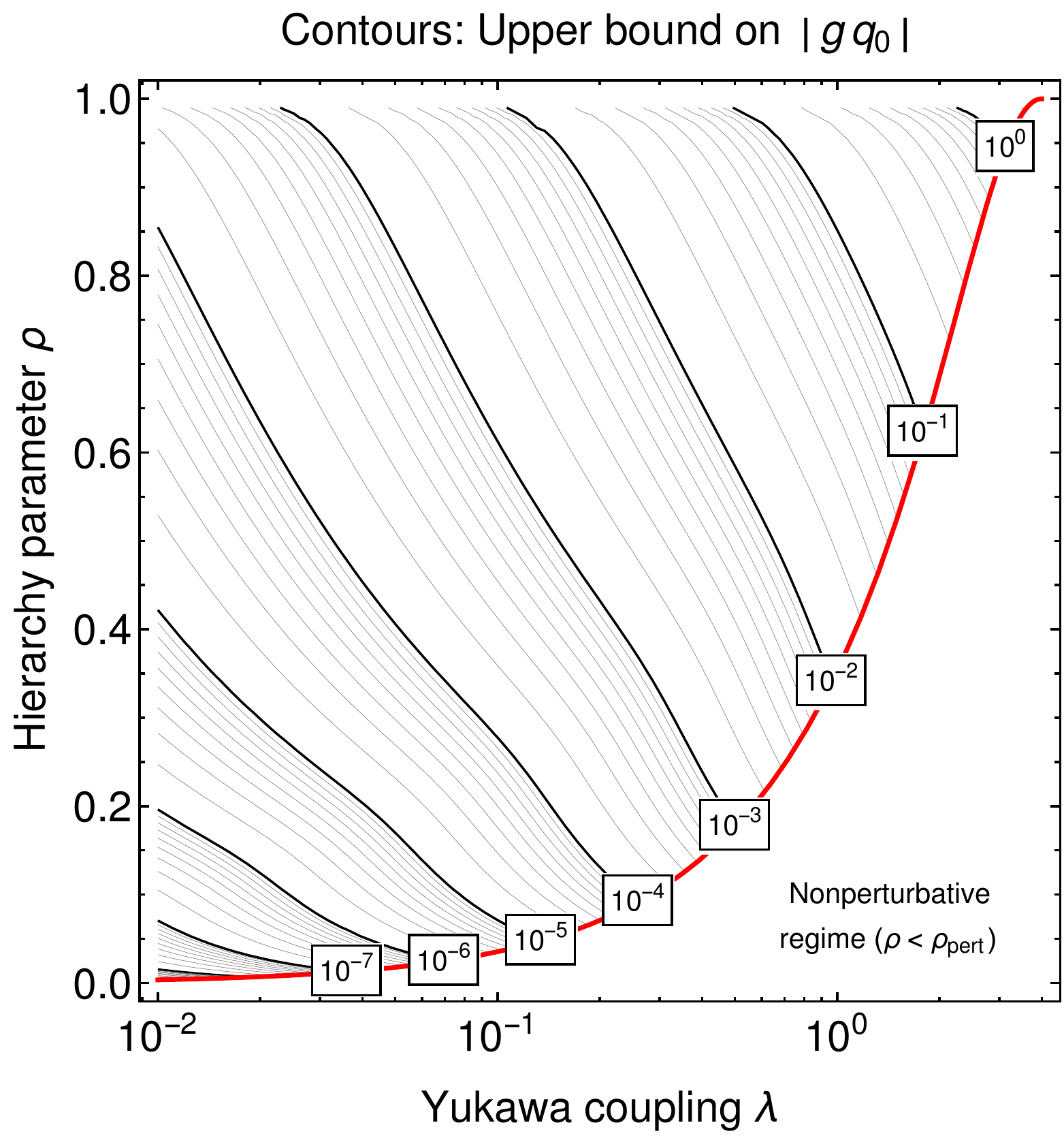}\hfill
\includegraphics[width=0.48\textwidth]{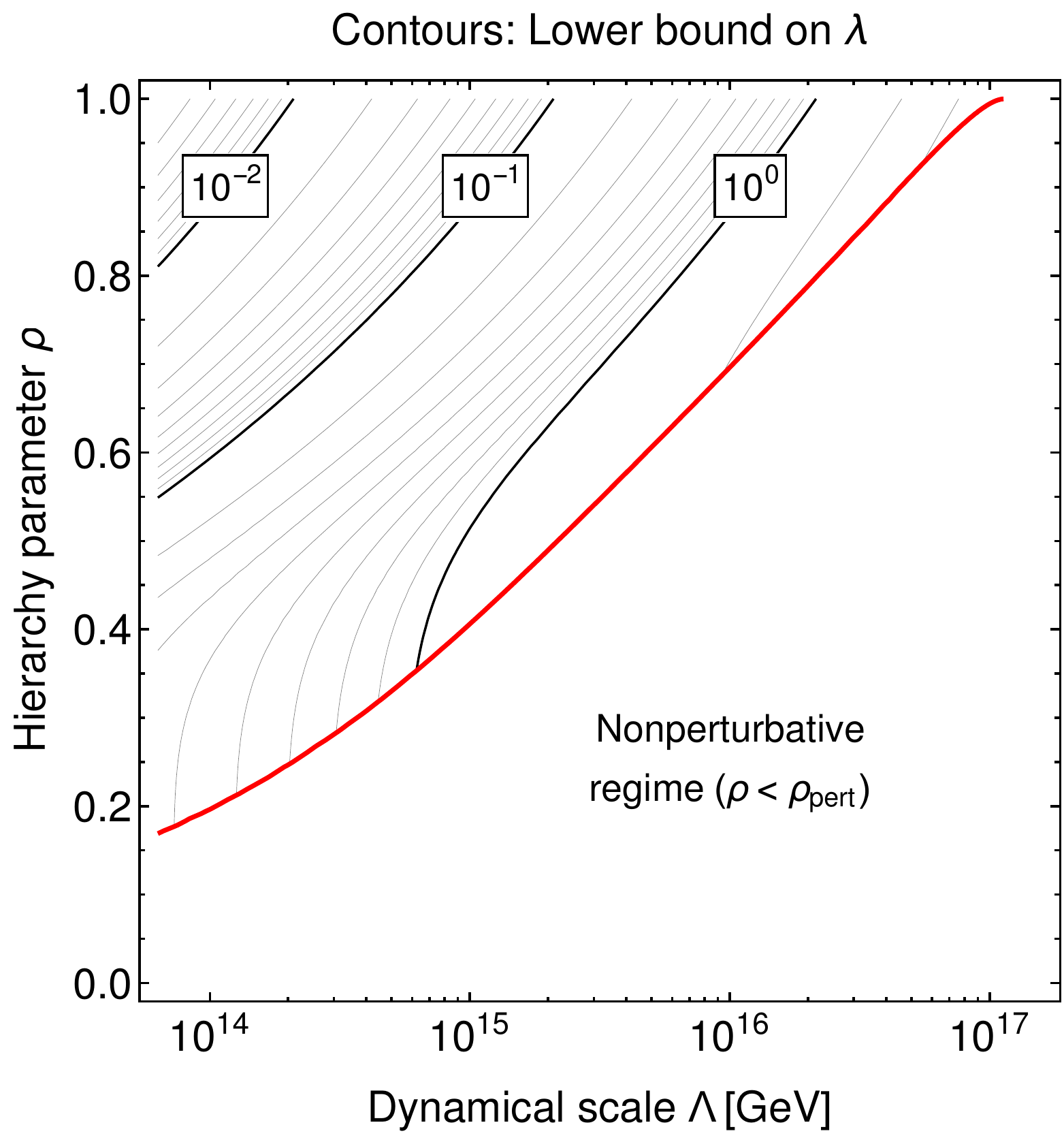}
\caption{Bounds on the gauge coupling $g$ \textbf{(left panel)}
and the Yukawa coupling $\lambda$ \textbf{(right panel)}.
The upper bound on $g$ ensures that the IYIT singlets $X$ and $Y$
do not absorb $\xi$ in their VEVs
(see Eq.~\eqref{eq:gmax}).
Similarly, the lower bound on $\lambda$ guarantees that the VEV of the Polonyi field
is located in the quadratic part of the effective potential (see Eq.~\eqref{eq:lmin}).
Both bounds need to be satisfied to sufficiently stabilize
the SUSY-breaking vacuum in the IYIT sector.
In both plots, the solid red line indicates where in parameter space
the perturbativity constraint in Eq.~\eqref{eq:rhopert} becomes violated.}

\label{fig:bounds}
\end{figure}

In Sec.~\ref{subsec:potential}, we will use this result
as a key ingredient in our construction of the inflaton potential.
However, before we are able to do so, we need to make sure that the FI parameter $\xi$
is not ``eaten up'' by the charged singlet fields $Z_\pm$ in the IYIT sector.
To this end, let us rewrite Eq.~\eqref{eq:VDIYIT} as follows,
\begin{align}
\label{eq:VDXY}
V_D = \frac{g^2 q_0^2}{2} \left(\xi - xy^* - x^*y\right)^2 \,,
\end{align}
where we integrated out the meson fields and replaced the fields
$Z_\pm$ by the Polonyi field $X$ and the stabilizer field $Y$.
From Eq.~\eqref{eq:VDXY}, it is evident that $\xi$ results
in a mass mixing between $X$ and $Y$,
\begin{align}
\label{eq:mXY}
\Delta m_{xy}^2 = -g^2q_0^2\,\xi \,.
\end{align}
At the same time, $X$ obtains an effective mass $m_{1\ell}$ at the one-loop level,
while $Y$ acquires a tree-level mass $m_F$ from the superpotential.
Taken as a whole, this results in the following two mass eigenvalues,
\begin{align}
\left(m_{xy}^\pm\right)^2 = \frac{m_F^2 + m_{1\ell}^2}{2} \pm
\left[\left(\frac{m_F^2 - m_{1\ell}^2}{2}\right)^2
+ \Delta m_{xy}^4\right]^{1/2} \,.
\end{align}
In the limit of vanishing mass mixing, $\Delta m_{xy} \rightarrow 0$,
these mass eigenvalues reduce to $m_{1\ell}$ and $m_F$, respectively.
For too large mass mixing, the eigenvalue $m_{xy}^-$ can, however, become
tachyonic.
In this case, one singlet mass eigenstate becomes unstable and absorbs
$\xi$ in its VEV.
To prevent this from happening, the absolute value of $\Delta m_{xy}$
needs to be smaller than the  geometric mean of $m_{1\ell}$ and $m_F$,
\begin{align}
m_{xy}^- > 0 \qquad\Rightarrow\qquad 
\left|\Delta m_{xy}\right| < \left(m_{1\ell}\,m_F\right)^{1/2} \,.
\end{align}
This requirement translates into an upper
bound on the gauge coupling constant $g$,
\begin{align}
\label{eq:gmax}
g < g_{\rm max} = \frac{1}{\left|q_0\right|}\left(\frac{c_{1\ell}}{2}\right)^{1/2}
\frac{\left(\rho\lambda\right)^{3/2}}{\left(1-\rho^4\right)^{1/4}} \,.
\end{align}
For $\rho = \bar{\rho}$ and $\lambda \ll \lambda_{\rm pert}$, we hence obtain
$\left|g_{\rm max}\,q_0\right| \simeq 7.8\times 10^{-2}\,\lambda^{3/2}$.
In Fig.~\ref{fig:bounds}, we plot the upper bound on $\left|gq_0\right|$
as a function of $\lambda$ and $\rho$.
This figure illustrates that the stability condition in Eq.~\eqref{eq:gmax}
is always stronger than the mere requirement of a weak gauge coupling,
$\left|gq_0\right| \lesssim \lambda$.
The constraint in Eq.~\eqref{eq:gmax} is therefore a sufficient
condition to justify our analysis in the weakly gauged limit.


\subsubsection*{Modulus stabilization and mass spectrum at low energies}


The dynamically generated FI term in Eq.~\eqref{eq:xiIYIT} is an effective
FI term that depends on the VEVs of the meson fields $M_\pm$.
As shown in~\cite{Komargodski:2010rb}, such FI terms are typically
accompanied by a shift-symmetric modulus field (see Sec.~\ref{subsec:ingredients}).
Our dynamical model is no exception to this statement.
In our case, the role of the modulus field is played by
the $B$$-$$L$ Goldstone multiplet $A$ which contains all meson DOFs
after imposing the Pfaffian constraint in Eq.~\eqref{eq:WhidPf}.
The field $A$ is a chiral multiplet.
It, hence, consists of a real scalar $c$, a real
pseudoscalar $\varphi$, and a Weyl fermion $\tilde{a}$.
In analogy to supersymmetric models of the QCD axion, these particles may
also be referred to as the saxion $c$, axion $\varphi$,
and axino $\tilde{a}$ (see, e.g.,~\cite{Harigaya:2015soa}).
In our model, the pseudoscalar $\varphi$ corresponds to the Goldstone
boson of spontaneous $B$$-$$L$ breaking.
It remains massless and exhibits a derivative coupling to
the $B$$-$$L$ vector boson $A_\mu$.
To see this, we have to apply the field transformation in Eq.~\eqref{eq:AMXY}
to the kinetic terms of the scalar meson fields.
The kinetic part of the Lagrangian then ends up containing the following terms,
\begin{align}
\label{eq:Stuckel}
- \mathcal{L}_{\rm kin} \supset
  \frac{1}{4} \left(\partial_\mu A_\nu - \partial_\nu A_\mu\right)
              \left(\partial^\mu A^\nu - \partial^\nu A^\mu\right)
+ \frac{1}{2} \left(\partial_\mu \varphi + m_V A_\mu\right)
              \left(\partial^\mu \varphi + m_V A^\mu\right) \,.
\end{align}
Note that this is nothing but the St\"uckelberg Lagrangian of an Abelian gauge field
with mass $m_V$,
\begin{align}
m_V = \sqrt{2}\,gq_0\,f_A \,.
\end{align}
In view of Eq.~\eqref{eq:Stuckel}, the Goldstone boson $\varphi$
can also be regarded as the St\"uckelberg
field of spontaneous $B$$-$$L$ breaking.
This is consistent with the fact that the $B$$-$$L$ Higgs multiplet\,---\,represented
by the meson field $M$\,---\,decouples once we insist on the
Pfaffian constraint in Eq.~\eqref{eq:WhidPf} being satisfied exactly.


After spontaneous $B$$-$$L$ breaking, the pseudoscalar $\varphi$ parametrizes
the longitudinal polarization state of the massive $B$$-$$L$ vector boson $A_\mu$.
It is ``eaten up'' by $A_\mu$ and does not cause any further problems.
At the same time, the real scalar $c$ may become problematic, as it threatens
to destabilize the FI parameter $\xi$.
This is the notorious modulus problem in the presence of an
effective FI term~\cite{Komargodski:2010rb}.
In our model, this problem is, however, absent.
The scalar $c$ and the fermion $\tilde{a}$ are automatically stabilized
by the F-term-induced mass $m_F$ in the superpotential (see Eq.~\eqref{eq:WhidAXY}).
To make this statement more precise, let us now review the mass
spectrum in the IYIT sector at low energies (see also~\cite{Harigaya:2015soa,Domcke:2014zqa}). 


The relevant superfields at low energies are the Polonyi field
$X = \left(x,\tilde{x}\right)$, the stabilizer field $Y = \left(y,\tilde{y}\right)$,
the Goldstone field $A = \left(a,\tilde{a}\right)$, and the vector field 
$V = \left(\lambda,A_\mu\right)$.
Here, $x$, $y$, and $a$ are complex scalars. 
The complex scalar $a$ contains the Goldstone boson and its scalar partner,
$a = 2^{-1/2}\left(c+i\varphi\right)$.
The fields $\tilde{x}$, $\tilde{y}$, $\tilde{a}$, and $\lambda$ are Weyl fermions,
where $\lambda$ denotes the $B$$-$$L$ gaugino.
$A_\mu$ is the $B$$-$$L$ vector boson.
To determine the mass spectrum for these fields, we work in the weakly gauged
limit where $\Delta m_{xy}$ in Eq.~\eqref{eq:mXY} is negligible.
A standard calculation in global supersymmetry then yields
\begin{align}
\label{eq:spectrum}
m_x^2 & = m_{1\ell}^2      \,, & m_y^2 & = m_F^2 \,, &
m_c^2 & = 2\,m_F^2 + m_V^2 \,, & m_\varphi & = 0 \,,
\\ \nonumber
m_{\tilde{x}}^2 & =0              \,, & m_{\tilde{y}}^2 & = m_F^2 + m_V^2 \,, &
m_{\tilde{a}}^2 & = m_F^2 + m_V^2 \,, & m_\lambda & = 0 \,.
\end{align}
Here, the indices refer to the respective mass eigenstates which
are not necessarily identical to the corresponding
``flavor'' eigenstates (see, e.g., $\tilde{y}$, $\tilde{a}$, and $\lambda$).
As expected, we find that the Goldstone boson $\varphi$ remains massless, $m_\varphi = 0$.
Similarly, also the fermionic component of the Polonyi field $X$ remains massless,
$m_{\tilde{x}} = 0$.
This is because $\tilde{x}$ corresponds to the goldstino field of spontaneous
supersymmetry breaking.
It is absorbed by the gravitino field in supergravity.
Finally, also the $B$$-$$L$ gaugino $\lambda$ remains massless
(however, see also Eq.~\eqref{eq:mfX}).
This is due to the following reason.
The original Weyl fermion $\lambda$ shares a Dirac mass with 
the axino, $\mathcal{L} \supset i\,m_V\lambda\,\tilde{a}$.
One would, thus, think that $\lambda$ ends up forming a Dirac fermion 
with the axino, i.e., the fermionic component of the $B$$-$$L$ Goldstone multiplet.
The axino, however, also shares a Dirac mass with the fermionic
component of the stabilizer field, $\mathcal{L} \supset m_F\,\tilde{a}\,\tilde{y}$,
that is parametrically larger, $m_F \gg m_V$, in the weakly gauged limit.
$\tilde{y}$ therefore ``steals'' the axino from the gaugino, so that
the gaugino no longer has a mass partner to form a Dirac fermion.
In more technical terms, this goes back to the fact that the
diagonalization of two Dirac masses for three Weyl fermions
necessarily results in one massive Dirac fermion and one massless Weyl fermion.
For our purposes, the most important lesson from Eq.~\eqref{eq:spectrum} is
that the superpartners of the Goldstone boson, the saxion $c$ and the axino $\tilde{a}$,
are indeed stabilized.
Both fields obtain masses from the gauge sector%
\footnote{This follows from the super-Higgs mechanism.
In consequence of spontaneous $B$$-$$L$ breaking, the massless vector multiplet $V$
(one Weyl fermion, one massless vector boson) absorbs an entire chiral multiplet
(one complex scalar, one Weyl fermion), so that it eventually contains the DOFs
of a massive vector multiplet (one real scalar, one Dirac fermion, one massive vector boson).
In the absence of supersymmetry breaking, all these DOFs would have a common mass $m_V$.}
as well as from the F-term-induced
mass in the superpotential.
This solves the modulus problem and assures us that $\xi$ is
a viable input for the construction of our inflation model.


Up to now, our discussion only dealt with the properties of the IYIT model
in global supersymmetry.
The next important step is to embed the IYIT model into
supergravity.
As discussed in Sec.~\ref{subsec:ingredients}, we especially intend to 
work in Jordan-frame supergravity with canonically normalized kinetic terms.
However, before going into any details, let us briefly discuss
the implications of supergravity for the low-energy mass spectrum.
On general grounds, we essentially expect two effects:
(i) In supergravity, $R$ symmetry is broken to ensure the vanishing of the
\textit{cosmological constant} (CC) in the true vacuum.
The order parameter of $R$ symmetry breaking is the gravitino mass $m_{3/2}$.
We therefore expect that supergravity leads to corrections to the various 
mass eigenvalues of $\mathcal{O}\left(m_{3/2}\right)$.
This should, in particular, also hold true if all other sectors
are sequestered from the hidden SUSY-breaking sector.
(ii) As a consequence of $R$ symmetry breaking, the Polonyi field $X$
acquires a nonzero VEV that is typically
parametrically larger than the gravitino mass,
$\left<X\right> \propto m_{3/2}/\left(c_{1\ell}^2\lambda^3\right)$
(see Sec.~\ref{subsec:polonyi}).
This induces an effective Majorana mass for the Goldstone multiplet $A$
in the superpotential (see Eq.~\eqref{eq:WhidAXY}),
\begin{align}
m_A = \lambda_X \left<X\right> \,.
\end{align}


The Majorana mass $m_A$ breaks some of the mass degeneracies in
Eq.~\eqref{eq:spectrum} and helps to make sure that the
$B$$-$$L$ gaugino $\lambda$ acquires a nonzero mass after all.
To illustrate the effect of nonzero $m_A$, we compute the fermion masses
in the limit of vanishing gravitino mass, $m_{3/2} \rightarrow 0$,
and small Majorana mass, $m_A \ll m_F$.
This exercise leads to the following mass eigenvalues (in the Jordan frame),
\begin{align}
\label{eq:mfX}
m_\lambda \simeq \frac{2\, m_V m_F}{m_F^2 + m_V^2}\, m_V \sin\theta  \,, \quad
m_{\tilde{a},{\tilde{y}}}^2 \simeq m_F^2 + m_V^2 \pm 
\left(m_A - m_\lambda\right)\left(m_F^2 + m_V^2\right)^{1/2} \,,
\end{align}
and again $m_{\tilde{x}} = 0$.
Now, the vector boson mass $m_V$ also receives contributions from
the Polonyi VEV,
\begin{align}
m_V \rightarrow m_V = \sqrt{2}\,gq_0\,v_A  \,, \quad 
v_A = \left(f_A^2 + \left<X^2\right>\right)^{1/2} \,,
\end{align}
where we parametrize the two contributions from $f_A$ and $\left<X\right>$
to $v_A$ in terms of a mixing angle $\theta$,
\begin{align}
\cos\theta = \frac{f_A}{v_A} \,, \quad \sin\theta = \frac{\left<X\right>}{v_A} \,.
\end{align}
Eq.~\eqref{eq:mfX} demonstrates that, for $\left<X\right> \neq 0$, the gaugino $\lambda$
indeed obtains a nonvanishing mass.
Moreover, we recognize that the Dirac fermion $\left(\tilde{y},\tilde{a}\right)$
splits into two nondegenerate Majorana fermions.
Making the same assumptions as before, we also compute the scalar
masses for a nonvanishing Polonyi VEV,
\begin{align}
m_{y_-}^2 = \frac{1}{2}\left(A-B^{1/2}\right) \,, \quad
m_{y_+}^2 = m_F^2 + m_A^2                     \,, \quad
m_c^2     = \frac{1}{2}\left(A+B^{1/2}\right) \,,
\end{align}
where
\begin{align}
A & = 3\, m_F^2 +  m_A^2 + m_V^2 \,,
\\ \nonumber
B & = m_F^4 + m_A^4 + m_V^4 
+ 6\, m_A^2 m_F^2 + 2 \, m_V^2 \left[\left(m_F^2 + m_A^2\right)
\cos\left(2\theta\right) 
- 2\, m_A\, m_F \sin\left(2\theta\right)\right]  \,,
\end{align}
and $m_x = m_{1\ell}$ and $m_\varphi = 0$.
For a dominant F-term-induced mass, $m_A, m_V \ll m_F$, this simplifies to
\begin{align}
m_{y_-}^2 \simeq m_F^2 - m_A^2 \,, \quad
m_{y_+}^2 =      m_F^2 + m_A^2                     \,, \quad
m_c^2     \simeq 2\,m_F^2 + 2\,m_A^2 + m_V^2 \,.
\end{align}
We, thus, find that the Majorana mass $m_A$ breaks the degeneracy among the two real scalar
components of $Y$ and shifts the saxion mass to even larger values.
This concludes our analysis of the IYIT mass spectrum.
We will now discuss the embedding into Jordan-frame supergravity in more detail.


\subsection{Effective Polonyi model and embedding into supergravity}
\label{subsec:polonyi}


In global supersymmetry, the Polonyi field $X$ is stabilized
at the origin in field space, $\left<X\right> = 0$, by its one-loop
effective mass $m_{1\ell}$ (see Eq.~\eqref{eq:m1l}).
Gravitational corrections in supergravity, however, threaten
to destabilize this vacuum solution.
We shall now explain why this is a serious problem.
At field values larger than some critical value,
$\left|x\right| \gtrsim x_c$, the one-loop effective potential
for the complex Polonyi field $x$ changes from a quadratic to a logarithmic
behavior (see, e.g., \cite{Schmitz:2016kyr} for an explicit calculation),
\begin{align}
\label{eq:V1lx}
V_{1\ell}\left(x\right) \simeq
\begin{cases}
m_{1\ell}^2    \left|x\right|^2               & ; \quad \left|x\right| \ll x_c \\
V_{1\ell}^0\, \ln \left|x/x_c\right| & ; \quad \left|x\right| \gg x_c                                 
\end{cases} \,, \quad
V_{1\ell}^0 \sim \frac{m_F^4}{16\pi^2} \,.
\end{align}
Here, $V_{1\ell}^0$ denotes the height of the logarithmic plateau at large field values.
The critical field value $x_c$ is reached once the
Polonyi field induces effective masses
for the quarks in the IYIT sector of $\mathcal{O}\left(\Lambda_{\rm dyn}\right)$.
The Yukawa interactions of the Polonyi field with the IYIT quarks
follow from Eqs.~\eqref{eq:WhidHE} and \eqref{eq:AMXY}, 
\begin{align}
W_{\rm hid} \supset \frac{X}{\sqrt{2}}
\left(\lambda_+ \Psi_1 \Psi_2 + \lambda_- \Psi_3 \Psi_4 \right) \,.
\end{align}
This allows us to estimate $x_c$.
For definiteness, we will work with $x_c = \sqrt{2}\,\Lambda_{\rm dyn}/\lambda$
in the following.
At $\left|x\right| \gtrsim x_c$, the IYIT quarks decouple
perturbatively, which gives rise to the logarithmic
corrections in Eq.~\eqref{eq:V1lx}.
Dangerous SUGRA corrections can shift
the Polonyi VEV from the origin in field space towards the
logarithmic plateau at large field values.
Once this happens, the Polonyi field is no longer stabilized
and the system settles into a completely different vacuum at field
values of $\mathcal{O}\left(M_{\rm Pl}\right)$.
In the following, we will illustrate where in parameter
space this unwanted conclusion can be avoided.
This will, in particular, provide us with a useful lower bound
on the Yukawa coupling $\lambda$ (see Eq.~\eqref{eq:lmin}).


For the purposes of this section, it will suffice if we
exclusively focus on the Polonyi field $X$ and integrate out
the heavier fields $A$ and $Y$.
The low-energy superpotential in Eq.~\eqref{eq:WhidAXY} then turns into
\begin{align}
\label{eq:Polonyi}
W_{\rm hid} = \mu^2\, X + w \,.
\end{align}
Here, $w$ denotes a constant contribution to the superpotential
that we added by hand.
$w$ is meaningless in global supersymmetry.
In supergravity, it accounts for the fact that $R$ symmetry must be broken
to ensure a vanishing CC in the true vacuum.
In the presence of the constant $w$, Eq.~\eqref{eq:Polonyi} is nothing but the
superpotential of the standard Polonyi model of spontaneous
supersymmetry breaking~\cite{Polonyi:1977pj}.
In this sense, the IYIT model can be regarded as a UV completion
of the Polonyi model that offers a dynamical explanation for the
origin of the SUSY-breaking scale $\mu$.
This is expected as the IYIT model is, after all, just a dynamical
realization of spontaneous supersymmetry breaking \`a la O'Raifeartaigh.
The IYIT model does not explain the UV origin of the constant $w$.
In this paper, we will not speculate about this issue any further.
That is, we do not have anything new to say about the CC problem.


Eq.~\eqref{eq:Polonyi} needs to be supplemented by
the following K\"ahler potential (see Eqs.~\eqref{eq:KF}
and \eqref{eq:F}),
\begin{align}
\label{eq:KX}
K_{\rm tot} \supset -3 M_{\rm Pl}^2\,\ln\left[1-\frac{F_{\rm hid} + F_{\rm inf}
}{3M_{\rm Pl}^2}\right] \,, \quad
F_{\rm hid} = X^\dagger X \,, \quad 
F_{\rm inf} = F \,,
\end{align}
where we set all terms to zero that are irrelevant for the present discussion.
The functions in Eq.~\eqref{eq:Polonyi} and \eqref{eq:KX} allow us to
calculate the total SUGRA potential for the Polonyi field
in the Jordan frame,
\begin{align}
\label{eq:VhidJ}
V_{\rm hid}^J\left(x\right) = \mu^4 + V_{1\ell}^J \left(x\right)
- \frac{\left|3\, w + 2\,\mu^2 x\right|^2}
{\left(1-f\right)3 M_{\rm Pl}^2} + m_R^2 \left|x\right|^2 \,.
\end{align}
Here, the second term on the right-hand side, i.e., the Jordan-frame
one-loop effective potential $V_{1\ell}^J$, is equivalent
to the global-SUSY expression $V_{1\ell}$ in Eq.~\eqref{eq:V1lx}.
The third term on the right-hand side corresponds to a tree-level
SUGRA correction in the Jordan frame, while the last term represents
the gravity-induced mass discussed around Eq.~\eqref{eq:mR}.
The function $f$ in Eq.~\eqref{eq:VhidJ} is defined as follows,
\begin{align}
f = \frac{1}{3 M_{\rm Pl}^2} \left(F - F_{S^{\vphantom{\dagger}}} F_{S^\dagger}\right) \,,
\end{align}
which may be regarded as a dimensionless measure for the amount of superconformal
symmetry breaking in the kinetic function of the inflaton field.
In the following, we will refer to $f$ as the \textit{reduced
kinetic function} of the inflaton field.
Given our choice for $F$ in Eq.~\eqref{eq:F}, the function $f$ evaluates to
\begin{align}
\label{eq:f}
f = \frac{1-2\chi}{3M_{\rm Pl}^2} \left[ \chi\,\sigma^2 - \left(1-\chi\right) \tau^2 \right] \,.
\end{align}
As expected, this expression vanishes for
$\chi \rightarrow \chi_{\rm CSS} = 1/2$.
The imaginary component of the complex inflaton field will be stabilized
during inflation, $\tau = 0$.
During inflation, we therefore have to deal with
\begin{align}
\label{eq:fz}
f = \left(1-2\chi\right)z  \,, \quad
z = \frac{\chi\sigma^2}{3M_{\rm Pl}^2} \,.
\end{align}


The constant $w$ controls the value of the CC.
To ensure that inflation ends in a Minkowski vacuum with vanishing CC,
we need to impose the following two conditions after inflation, i.e., for $f = 0$,
\begin{align}
\label{eq:vacuumconds}
V_{\rm hid}^J\left(x\right) = 0 \,, \quad
\frac{d}{dx}\, V_{\rm hid}^J\left(x\right) = 0 \qquad 
\textrm{for} \quad f = 0\,, \quad x = \left<X\right> \,, \quad w = w_0 \,.
\end{align}
These two conditions can be solved for the two unknowns $\left<X\right>$ and $w_0$.
Let us focus on $w_0$ for now,%
\footnote{The  K\"ahler potential in Eq.~\eqref{eq:KX} turns the
usual $e^{K/M_{\rm Pl}^2}$ term in the SUGRA potential into a simple
rational function (see Appendix~\ref{app:frames}).
Our Jordan-frame formulation of supergravity therefore allows us
to solve the two conditions in Eq.~\eqref{eq:vacuumconds} in a closed form.
Note that this is not possible for the Polonyi model in standard Einstein-frame supergravity.}
\begin{align}
\label{eq:w0}
w_0 = \left(1-\frac{4}{3k}\right)^{1/2} \frac{\mu^2 M_{\rm Pl}}{\sqrt{3}} \,.
\end{align}
Here, $k$ is a convenient measure for the relative importance
of the SUGRA corrections in Eq.~\eqref{eq:VhidJ},
\begin{align}
k = \frac{m_{1\ell}^2 + m_R^2}{\mu^4/M_{\rm Pl}^2} \,.
\end{align}
The definition of $k$ is chosen such that it mimics the effect of a higher-dimensional
operator in $F_{\rm hid}$,
\begin{align}
F_{\rm hid} \supset X^\dagger X - \frac{k}{\left(2!\right)^2}
\frac{\left(X^\dagger X\right)^2}{M_{\rm Pl}^2} \,,
\end{align}
that induces a mass
correction $\Delta m_k^2 = m_{1\ell}^2 + m_R^2$ in the SUGRA potential.
Large values of $k$, thus, indicate that the Polonyi field
is strongly stabilized by its one-loop effective mass, $\Delta m_k^2 \gg m_{3/2}^2$.


Given the result for the constant $w_0$ in Eq.~\eqref{eq:w0}, we can go
one step back and determine the time-dependent Polonyi VEV during inflation.
To do so, we just need to solve one equation,
\begin{align}
\label{eq:dVdx0}
\frac{d}{dx}\, V_{\rm hid}^J\left(x\right) = 0 \qquad 
\textrm{for} \quad f \neq 0 \,, \quad x = \left<X\right> \,, \quad w = w_0 \,.
\end{align}
Assuming that $\left<X\right>$ is located in the quadratic
part of the potential, Eq.~\eqref{eq:dVdx0} has the following solution,
\begin{align}
\label{eq:XVEV}
\left<X\right> = \frac{\left(k-4/3\right)^{1/2}}{\left[\left(1-f\right)k-4/3\right]k^{1/2}}
\frac{2 M_{\rm Pl}}{\sqrt{3}} \,.
\end{align}
Indeed, in the limit of a large one-loop effective mass,
$k \gg 1$, the Polonyi field remains stabilized close to the origin,
$\left<X\right> \ll M_{\rm Pl}$.
For $f=0$, Eq.~\eqref{eq:XVEV} turns into
the solution of Eq.~\eqref{eq:vacuumconds}.
Both the constant $w_0$ and the Polonyi VEV $\left<X\right>$ break
$R$ symmetry.
This is because both the superpotential $W$ as well as the IYIT singlets $Z_\pm$
carry $R$ charge $2$.
We can therefore use our results in Eqs.~\eqref{eq:w0} and \eqref{eq:XVEV}
to determine the order parameter of $R$ symmetry breaking, i.e.,
the gravitino mass in the Jordan frame,
\begin{align}
\label{eq:m32}
m_{3/2} = \frac{\left<W\right>}{M_{\rm Pl}^2} = 
\frac{\mu^2 \left<X\right> + w_0}{M_{\rm Pl}^2} = 
\left(1-\frac{4}{3k}\right)^{1/2}
\frac{\left(1-f\right)k +2/3}{\left(1-f\right)k-4/3}
\frac{\mu^2}{\sqrt{3} M_{\rm Pl}} \,.
\end{align}
This result allows us to write the total mass of the Polonyi field
in the Jordan frame, $m_x^J$, as follows,
\begin{align}
\left(m_x^J\right)^2 = \left[1 - \frac{4}{3\left(1-f\right)k}\right]
\left(m_{1\ell}^2 + m_R^2\right)
= \frac{3\,k}{\left(1-f\right)\left(k-4/3\right)}
\frac{\left[\left(1-f\right)k-4/3\right]^3}
{\left[\left(1-f\right)k+2/3\right]^2}\:m_{3/2}^2 \,.
\end{align}
In the following, we will restrict ourselves to the large-$k$ limit.
This is justified because $k$ is typically at least of
$\mathcal{O}\left(10\right)$ in the part of parameter space that we 
are interested in (see Eqs.~\eqref{eq:mu} and \eqref{eq:m1l}),
\begin{align}
k \approx \frac{m_{1\ell}^2}{\mu^4/M_{\rm Pl}^2} =
\frac{1}{2} \left(\frac{c_{1\ell}\,\lambda\,M_{\rm Pl}}{\Lambda}\right)^2\simeq
47\, \bigg(\frac{c_{1\ell}}{0.02}\bigg)^2 \left(\frac{\lambda}{1}\right)^2
\left(\frac{5\times10^{15}\,\textrm{GeV}}{\Lambda}\right)^2 \,.
\end{align}
In the large-$k$ limit, the Polonyi VEV in Eq.~\eqref{eq:XVEV} can
be approximately written as follows,
\begin{align}
\label{eq:Xm32}
\left<X\right> \approx \frac{2\sqrt{3}\, m_{3/2}^2}
{\left(1-f\right)\left(m_x^J\right)^2}\,M_{\rm Pl}\
\approx \frac{2\sqrt{2}}{1-f}\frac{m_{3/2}}{c_{1\ell}^2\lambda^3} \,,
\end{align}
confirming that $\left<X\right>$ is parametrically enhanced compared to
the gravitino mass, $\left<X\right> \propto m_{3/2}/\left(c_{1\ell}^2\lambda^3\right)$.
For large values of $k$, also the relations between $w_0$, $\mu$, and $m_{3/2}$
become significantly simpler,
\begin{align}
k \gg 1 \qquad\Rightarrow\qquad w_0 \approx \frac{\mu^2 M_{\rm Pl}}{\sqrt{3}} 
\approx m_{3/2}\,M_{\rm Pl}^2 \,.
\end{align}
In the remainder of this paper, we will restrict ourselves to working
with these approximate expressions.


We derived the result in Eq.~\eqref{eq:Xm32} under the assumption that
the leading term in the one-loop effective potential is an effective mass term,
$V_{1\ell}^J = m_{1\ell}^2 \left|x\right|^2 + \mathcal{O}\left(x^4\right)$.
The derivation of Eq.~\eqref{eq:Xm32} is therefore only self-consistent and valid
as long as $\left<X\right> \lesssim x_c$ (see Eq.~\eqref{eq:V1lx}).
This implies a lower bound on the Yukawa coupling $\lambda$ that depends
on the energy scale $\Lambda$ as well as on the hierarchy parameter $\rho$.
We plot this lower bound $\lambda_{\rm min}$ as a function of $\Lambda$ and $\rho$
in the right panel of Fig.~\ref{fig:bounds}.
The exact numerical result shown in this figure is well approximated
by the following analytical expression,
\begin{align}
\label{eq:lmin}
\lambda_{\rm min} \simeq \min\left\{
\frac{16\pi\sqrt{2}}{\sqrt{3}\left(2\ln2-1\right)\rho^6}
\frac{\Lambda}{M_{\rm Pl}},
\left[\frac{4\pi\sqrt{2}\,\lambda_0^2}{\sqrt{3}\left(2\ln2-1\right)}
\frac{\Lambda}{M_{\rm Pl}}\right]^{1/3}\right\} \,.
\end{align}
$\lambda_{\rm min}$ is sensitive to $\rho$ at large values
of $\rho$ where $n_M^{\rm eff}$ in Eq.~\eqref{eq:nMeff} is dominated
by $\rho$ rather than the neutral meson contribution $4\,\ell_0$.
At small $\rho$ values, where $n_M^{\rm eff} \approx 4\,\ell_0$, the $\rho$
dependence disappears.
In addition to the lower bound $\lambda_{\rm min}$, we also require that $\lambda$
must remain perturbative, $\lambda < \lambda_{\rm pert} = 4$.
In the following, we will eliminate $\lambda$ from our analysis and
set it to the following value, for simplicity,
\begin{align}
\bar{\lambda} = \left(\lambda_{\rm min} \lambda_{\rm pert}\right)^{1/2} \,.
\end{align}
That is, we fix $\lambda$ just at the central value of the allowed range of values,
$\lambda_{\rm min} < \lambda < \lambda_{\rm pert}$.
Together with our choice for the hierarchy parameter, $\rho = \bar{\rho} \simeq 0.80$,
this removes all dimensionless parameters from the IYIT sector,
so that the only remaining free parameter is the scale $\Lambda$.
We then obtain for $\bar{\lambda}$,
\begin{align}
\label{eq:lbar}
\bar{\lambda} \simeq 1.78 \left(\frac{\Lambda}{5\times10^{15}\,\textrm{GeV}}\right)^{1/2} \,.
\end{align}
At the same time, $\lambda$ must not be larger than $\lambda \simeq 2.41$,
since otherwise $\lambda_+$ or $\lambda_-$ will exceed $\lambda_{\rm pert}$
(see Eq.~\eqref{eq:lmax}).
This results in an upper bound on the dynamical scale, $\Lambda \lesssim 10^{16}\,\textrm{GeV}$.
The relation in Eq.~\eqref{eq:lbar} enables us to express
all mass scales in the IYIT sector in terms of $\Lambda$.
For $\lambda = \bar{\lambda}$ and $\rho = \bar{\rho}$ and making
use of Eqs.~\eqref{eq:mu}, \eqref{eq:mF}, and \eqref{eq:m32}, we find
(in addition to the relations in Eq.~\eqref{eq:xifaL})
\begin{align}
\label{eq:mumfm32}
\mu & \simeq 7.94 \times 10^{15}\,\textrm{GeV}
\left(\frac{\Lambda}{5 \times 10^{15}\,\textrm{GeV}}\right)^{5/4} \,, 
\\ \nonumber
m_F & \simeq 7.15 \times 10^{15}\,\textrm{GeV}
\left(\frac{\Lambda}{5 \times 10^{15}\,\textrm{GeV}}\right)^{3/2} \,, 
\\ \nonumber
m_{3/2} & \simeq 1.51 \times 10^{13}\,\textrm{GeV}
\left(\frac{\Lambda}{5 \times 10^{15}\,\textrm{GeV}}\right)^{5/2} \,.
\end{align}
The relations in Eqs.~\eqref{eq:xifaL} and \eqref{eq:mumfm32}
are the main result of this chapter.
They demonstrate how the dynamical breaking of supersymmetry 
in the IYIT sector generates all mass scales relevant for our model
via dimensional transmutation.
Our model, thus, does not require any hard dimensionful input parameters.
Moreover, Eq.~\eqref{eq:mumfm32} illustrates that supersymmetry breaking
close to the GUT scale results in a large gravitino mass.
As expected, our model therefore predicts a very heavy sparticle spectrum
(see Sec.~\ref{subsec:spectrum}).
Eqs.~\eqref{eq:xifaL} and \eqref{eq:mumfm32} now set the stage for our model 
of D-term inflation.


\subsection{Scalar potential in the inflaton sector}
\label{subsec:potential}


We now have all ingredients at our disposal to construct our inflationary model.
In a first step, we add the usual superpotential of D-term inflation (see Eq.~\eqref{eq:FHIDHI})
to the Polonyi superpotential in Eq.~\eqref{eq:Polonyi}, 
\begin{align}
\label{eq:Wtot}
W_{\rm tot} \supset \kappa\, S\, \Phi \bar{\Phi} +  \mu^2\, X + w_0 \,.
\end{align}
The inflaton $S$ as well as the waterfall fields $\Phi$ and $\bar{\Phi}$
belong to a separate sector that is sequestered from the IYIT sector.
The relevant K\"ahler potential is of the following form
(see Eqs.~\eqref{eq:KF} and \eqref{eq:F}),
\begin{align}
\label{eq:Ktot}
K_{\rm tot} \supset -3 M_{\rm Pl}^2\,\ln\left[1-\frac{F_{\rm hid} + F_{\rm inf}
}{3M_{\rm Pl}^2}\right] \,, \quad
F_{\rm hid} = X^\dagger X \,, \quad 
F_{\rm inf} = F + \Phi^\dagger \Phi + \bar{\Phi}^\dagger \bar{\Phi} \,.
\end{align}
The waterfall fields $\Phi$ and $\bar{\Phi}$ carry $B$$-$$L$ charges $+q$
and $-q$, respectively.
They, thus, appear in the D-term scalar potential, together with the dynamically
generated FI parameter $\xi$ (see Eq.~\eqref{eq:xiIYIT}),
\begin{align}
\label{eq:VDJ}
V_D^J = \frac{g^2}{2} \left[q_0\, \xi
- q\left(\left|\phi\right|^2 - \left|\bar{\phi}\right|^2 \right)\right]^2 \,.
\end{align}
$V_D^J$ denotes the D-term scalar potential in the Jordan frame which
is identical to the D-term scalar potential in global supersymmetry, $V_D^0$.
In addition, the total tree-level scalar potential $V_{\rm tree}^J$ also
receives an F-term contribution $V_F^J$ which can be computed
by making use of Eqs.~\eqref{eq:Wtot} and \eqref{eq:Ktot},
\begin{align}
V_F^J = V_F^0 + \Delta V_F^J \,.
\end{align}
Here, the first contribution, $V_F^0$,
denotes the usual F-term scalar potential in global supersymmetry,
\begin{align}
\label{eq:VF0}
V_F^0 = \mu^4 + m_{\rm eff}^2 \left(\left|\phi\right|^2 +
\left|\bar{\phi}\right|^2 \right) +
\kappa^2 \left|\phi\right|^2 \left|\bar{\phi}\right|^2 \,,
\end{align}
while the second term, $\Delta V_F^J$, corresponds to the tree-level SUGRA
correction in the Jordan frame,
\begin{align}
\label{eq:DeltaVJ}
\Delta V_F^J = - \frac{\mu^4}{1-f}
+ \left(\delta m_{\rm eff}^2\right)^* \phi \bar{\phi} + \delta m_{\rm eff}^2\, \phi^*\bar{\phi}^*
+ \delta\kappa^2 \left|\phi\right|^2 \left|\bar{\phi}\right|^2 \,.
\end{align}
For more details on the computation of $\Delta V_F^J$, see Appendix~\ref{app:frames}.
In Eqs.~\eqref{eq:VF0} and \eqref{eq:DeltaVJ}, we introduced
the masses squared $m_{\rm eff}^2$ and $\delta m_{\rm eff}^2$ as well as the quartic coupling
$\delta\kappa^2$.
These are defined as follows,
\begin{align}
\label{eq:meff}
m_{\rm eff}^2 = \kappa^2\left|s\right|^2 \,, \quad 
\delta m_{\rm eff}^2 =  - \frac{1-2\chi}{1-f}\,m_{3/2}\,\kappa s \,, \quad
\delta\kappa^2 = - \frac{\left(1-2\chi\right)^2}{1-f}
\frac{\kappa^2\left|s\right|^2}{3M_{\rm Pl}^2} \,.
\end{align}
Note that all three parameters are field-dependent.
The real mass parameter $m_{\rm eff}^2$ denotes the effective inflaton-dependent mass
that stabilizes the waterfall fields during inflation.
The complex mass parameter $\delta m_{\rm eff}^2$ is a bilinear mass that originates from
the interference between the supersymmetric mass of the waterfall fields in
the superpotential, $\kappa \left<S\right>$, and the gravitino mass $m_{3/2}$.
This so-called B term is, hence, a consequence of $R$ symmetry breaking in the superpotential. 
It is only generated for the scalar waterfall fields and not for the corresponding
fermions, which is why it breaks supersymmetry.
Just like the usual A terms in models of broken supersymmetry,
the B term results in a soft breaking of supersymmetry.
The coupling $\delta\kappa^2$ will be irrelevant for our purposes
as it constitutes just a small correction to $\kappa^2$, i.e., the quartic
coupling of the waterfall fields in global supersymmetry.


During inflation, the waterfall fields are stabilized at their origin,
$\big<\phi\big> = \big<\bar{\phi}\big> = 0$.
Along the inflationary trajectory, the tree-level scalar potential
in the Jordan frame therefore reads as follows,
\begin{align}
\label{eq:VtreeJ}
V_{\rm tree}^J = \frac{1}{2} D_0^2
- \frac{f}{1-f}\,F_0^2 \,, \quad D_0 = g q_0 \xi \,, \quad F_0 = \mu^2 \,.
\end{align}
Here, $D_0^2/2$ denotes the contribution to the vacuum energy
density from the D-term scalar potential. 
The large F-term contribution, $V_F^0 \supset +F_0^2$, is canceled by
the contribution from $R$ symmetry breaking that is contained in the
SUGRA correction to the scalar potential,
$\Delta V_F^J \supset -F_0^2$.
This explains why the vacuum energy density during inflation is
dominated by the D-term contribution in our model.
During the $B$$-$$L$ phase transition at the end of inflation,
the D term is absorbed by the VEV of one of the waterfall fields.
In the true vacuum after inflation, all contributions to the CC
therefore approximately cancel.
In the next section, we will perform a standard slow-roll analysis
of our inflationary model.
This is most easily done in terms of the usual slow-roll parameters
in the Einstein frame.
To compute these parameters, we need to convert the potential in
Eq.~\eqref{eq:VtreeJ} from the Jordan frame to the Einstein frame.
This is achieved by rescaling $V_{\rm tree}^J$
by the fourth power of the conformal factor $\mathcal{C}$ (see Eq.~\eqref{eq:Weyl}),
\begin{align}
\label{eq:Vtree}
V_{\rm tree} = \mathcal{C}^4\,V_{\rm tree}^J = 
\left(-\frac{3M_{\rm Pl}^2}{\Omega}\right)^2
\left[\frac{1}{2} D_0^2 - \frac{f}{1-f}\,F_0^2 \right] \,.
\end{align}
On the inflationary trajectory, this can be rewritten as
a function of the parameter $z$ (see Eq.~\eqref{eq:fz}), 
\begin{align}
V_{\rm tree} = \frac{1}{\left(1-z\right)^{2}}
\left[\frac{1}{2} D_0^2 - \frac{\left(1-2\chi\right)z}
{1-\left(1-2\chi\right)z}\,F_0^2 \right] \,, \quad 
z = \frac{\chi\sigma^2}{3M_{\rm Pl}^2} \,.
\end{align}
The second derivative of $V_{\rm tree}$ w.r.t.\ the field $\sigma$ provides us 
with the mass parameter $m_\sigma^2$ (see Eq.~\eqref{eq:msigma}),
\begin{align}
\label{eq:msigmaDF}
m_\sigma^2 = 2\chi\left[\frac{D_0^2}{3 M_{\rm Pl}^2} -
\left(1-2\chi\right) \frac{F_0^2}{3 M_{\rm Pl}^2} \right] \,, \quad
\frac{D_0^2}{3 M_{\rm Pl}^2} \approx m_R^2  \,, \quad
\frac{F_0^2}{3 M_{\rm Pl}^2} \approx m_{3/2}^2 \,.
\end{align}
As anticipated, $m_\sigma$ is suppressed by the shift-symmetry-breaking
parameter $\chi$.
By appropriately choosing $\chi$, we will therefore be able to adjust
the scalar spectral index $n_s$ so that it matches the observed best-fit value.
We also note that the mass parameter $m_\sigma$ is not a physical mass eigenvalue
in the actual sense.
This is because the scalar inflaton field $\sigma$ is not properly normalized.
The mass of the canonically normalized inflaton field $\hat{\sigma}$
is instead given in terms of the slow-roll parameter $\eta$ in the Einstein frame,
$m_{\hat{\sigma}}^2  = 3\,\eta H^2 \sim m_\sigma^2$ (see Sec.~\ref{subsec:analysis}).
Moreover, $m_\sigma$ is just a tree-level parameter, whereas the scalar 
potential also receives important contributions at the one-loop level.


The radiative corrections are encoded in the one-loop
effective Coleman-Weinberg potential~\cite{Coleman:1973jx},
\begin{align}
\label{eq:V1lJ}
V_{1\ell}^J = \frac{Q_J^4}{64\pi^2} \,\textrm{STr}\left[
\frac{\mathcal{M}_J^4}{Q_J^4}\left(\ln \frac{\mathcal{M}_J^2}{Q_J^2}-C\right)\right] \,, \quad 
C = \frac{3}{2} \,.
\end{align}
Here, $\textrm{STr}\left[\cdot\right]$ stands for
the supertrace over a (matrix-valued) function of the total tree-level
mass matrix squared $\mathcal{M}_J^2$ in the Jordan frame.
We evaluate $V_{1\ell}^J$ in the {\footnotesize$\overline{\textrm{MS}}$}
renormalization scheme and only consider contributions from scalars and fermions.
This fixes the numerical constant $C$ in Eq.~\eqref{eq:V1lJ}
to $C = 3/2$ (see, e.g., \cite{Martin:2001vx}).
The energy scale $Q_J$ denotes the {\footnotesize$\overline{\textrm{MS}}$}
renormalization scale.
To determine the radiative corrections to the inflaton potential,
it is sufficient to focus on the inflaton-dependent masses
in the waterfall sector.
From Eqs.~\eqref{eq:Snonmin}, \eqref{eq:VDJ},
\eqref{eq:VF0}, and \eqref{eq:DeltaVJ}, we find for the scalar fields,
\begin{align}
\label{eq:Mphi2}
V_{\rm tree}^J \supset
\frac{1}{2} \,
\begin{pmatrix}\phi^* \\ \bar{\phi}^* \\ \phi \\ \bar{\phi} \end{pmatrix}^T
\begin{pmatrix}
m_\phi^2 & 0 & 0 & \delta m_{\rm eff}^2 \\
0 & m_{\bar{\phi}}^2 & \delta m_{\rm eff}^2 & 0 \\
0 & \left(\delta m_{\rm eff}^2\right)^*  & m_\phi^2 & 0 \\
\left(\delta m_{\rm eff}^2\right)^* & 0 & 0 & m_{\bar{\phi}}^2
\end{pmatrix}
\begin{pmatrix}\phi \\ \bar{\phi} \\ \phi^* \\ \bar{\phi}^* \end{pmatrix} \,, \quad
m_{\phi,\bar{\phi}}^2  = m_{\rm eff}^2 + m_R^2 \mp q\, m_D^2 \,.
\end{align}
where the mass parameters $m_D^2$, $m_R^2$, $m_{\rm eff}^2$, and $\delta m_{\rm eff}^2$
are respectively defined below Eq.~\eqref{eq:VDMSSM} as well in Eqs.~\eqref{eq:mR}
and \eqref{eq:meff}.
In addition to these tree-level masses, the scalar waterfall fields also obtain
gauge-mediated masses at the loop level.
This is because supersymmetry breaking in the IYIT sector results
in a mass splitting among the components of the massive $B$$-$$L$
vector multiplet (see Eq.~\eqref{eq:spectrum})~\cite{Intriligator:2010be}. 
Including these one-loop masses in Eq.~\eqref{eq:V1lJ} would result 
in a two-loop contribution to the effective potential.
For this reason, we will ignore the effect of
\textit{gauge-mediated supersymmetry breaking} (GMSB) for now.
We will come back to this issue in Sec.~\ref{subsec:spectrum}.
In the next step, we diagonalize the mass matrix in Eq.~\eqref{eq:Mphi2}.
This results in two complex mass eigenstates $\phi_\pm$ with inflaton-dependent
mass eigenvalues,
\begin{align}
\label{eq:mpm}
m_\pm^2 = m_{\rm eff}^2 + m_R^2 \pm \left(1 + \delta^4\right)^{1/2} q\, m_D^2 \,, \quad
\delta^4 = \frac{\left|\delta m_{\rm eff}\right|^4}{q^2m_D^4} \,.
\end{align}
Here, $\delta$ is related to the rotation angle $\beta_{\phi\bar{\phi}}$ that diagonalizes the
scalar mass matrix, $\delta^2 = \tan\left(2\,\beta_{\phi\bar{\phi}}\right)$.
We note that the parameter $\delta$ depends on the inflaton field value, which makes it
a time-dependent quantity. 
For this reason, the
mass eigenstates $\phi_\pm$ do not coincide with the charge eigenstates
$\phi,\bar{\phi}$  during inflation.
The mass parameters $m_R^2$, $m_D^2$, and $\delta m_{\rm eff}^2$ in Eq.~\eqref{eq:mpm}
arise from various effects in the scalar sector of our model: the conformal
coupling to the Ricci scalar in Eq.~\eqref{eq:Snonmin}, the D-term scalar potential
in Eq.~\eqref{eq:VDJ}, and the soft B term in Eq.~\eqref{eq:DeltaVJ}.
None of these effects are relevant for the fermions in the waterfall sector.
The two Weyl fermions in $\Phi$ and $\bar{\Phi}$
simply form a Dirac fermion $\tilde{\phi}$ with
Jordan-frame mass $m_{\tilde{\phi}} = m_{\rm eff}$.
No further SUGRA corrections arise in the Jordan frame~\cite{Ferrara:2010yw}.


We are now ready to evaluate $V_{1\ell}^J$ in Eq.~\eqref{eq:V1lJ}.
Our final result can be written as follows,
\begin{align}
\label{eq:V1lJL}
V_{1\ell}^J = \frac{Q_J^4}{16\pi^2}\,L\left(x,\alpha\right) \,, \quad
L\left(x,\alpha\right) = \frac{1}{4}\,\textrm{STr}\left[
\frac{\mathcal{M}_J^4}{Q_J^4}\left(\ln \frac{\mathcal{M}_J^2}{Q_J^2}-\frac{3}{2}\right)\right] \,,
\end{align}
where $L$ is a one-loop function that takes the same form in the Jordan
frame as in the Einstein frame,
\begin{align}
\label{eq:L}
L\left(x,\alpha\right) = \frac{1}{2}\sum_\pm \left(x \pm 1\right)^2\left[
\ln\left(x \pm 1\right) - \frac{3}{2}\right]
- \left(x-\alpha\right)^2\left[\ln\left(x-\alpha\right) - \frac{3}{2}\right] \,,
\end{align}
and where the variable $x$, the parameter $\alpha$, and the renormalization scale $Q_J$
are introduced such that
\begin{align}
\label{eq:xaQJ}
x = \frac{m_{\rm eff}^2 + m_R^2}{Q_J^2} \,, \quad
\alpha = \frac{m_R^2}{Q_J^2} \,, \quad 
Q_J^2 = \left(1 + \delta^4\right)^{1/2}q\,m_D^2 \,.
\end{align}
All of these quantities depend on the inflaton field value
by virtue of the parameters $m_{\rm eff}^2$, $m_R^2$, and $\delta$.
In the following, we will, however, neglect the field dependence of $m_R^2$
and approximate it instead by the constant expression in Eq.~\eqref{eq:msigmaDF}.
The fact that $Q_J$ is field-dependent does not pose any problem for our model.
Recall that one usually encounters a field-independent renormalization scale $Q_J$
in the Jordan frame and a field-dependent renormalization scale $Q = \mathcal{C}\,Q_J$
in the Einstein frame\,---\,or vice versa.
In our model, the renormalization scale is, by contrast, field-dependent in both frames.
This is \textit{a priori} a perfectly valid choice. 
Independent of whether $Q_J$ is field-dependent or not, we merely have to make sure
that our final results are not overly sensitive to our particular choice for $Q_J$.
This is a requirement that we will have to check \textit{a posteriori} as part of our
slow-roll analysis (see Sec.~\ref{sec:inflation}).
The main field dependence of Eq.~\eqref{eq:V1lJL} is encoded in $x$
which may also be written as follows,
\begin{align}
\label{eq:x}
x = \frac{m_+^2 + m_-^2}{m_+^2 - m_-^2} = \frac{2\,m_{\rm eff}^2}{m_+^2 - m_-^2} + \alpha \,.
\end{align}
$x = 1$ therefore corresponds to the critical
point along the inflationary trajectory at which $m_-$ vanishes,
\begin{align}
x = 1 \qquad\Leftrightarrow\qquad \sigma = \sigma_c \qquad\Leftrightarrow\qquad m_- = 0 \,.
\end{align}
At this point in field space, the mass eigenstate $\phi_-$ becomes tachyonically
unstable which triggers the $B$$-$$L$ waterfall transition.
We also note that Eq.~\eqref{eq:x} illustrates the physical meaning of $\alpha$.
The parameter $\alpha$ represents a shift in the field variable $x$ compared
to the situation in global supersymmetry where one simply
has $x_0 = 2\,m_{\rm eff}^2 / \left(m_+^2 - m_-^2\right)$.
As evident from Eq.~\eqref{eq:xaQJ}, this shift results from the fact that the 
waterfall scalars obtain a gravity-induced mass $m_R$, while the waterfall fermion does not.


To convert Eq.~\eqref{eq:V1lJL} into the one-loop effective Coleman-Weinberg
potential in the Einstein frame, we need to multiply again by $\mathcal{C}^4$,
just like in the case of the tree-level scalar potential (see Eq.~\eqref{eq:Vtree}), 
\begin{align}
\label{eq:V1l}
V_{1\ell} = \mathcal{C}^4\,V_{1\ell}^J = \frac{Q^4}{16\pi^2}\,L\left(x,\alpha\right) \,, \quad
Q = \mathcal{C}\,Q_J \,. 
\end{align}
The Weyl transformation from the Jordan frame to the Einstein frame
therefore corresponds to nothing but a rescaling of
the $Q_J^4$ factor in Eq.~\eqref{eq:V1lJL}.
The loop function $L$ remains unchanged.
This is consistent with the fact that the Weyl transformation in Eq.~\eqref{eq:Weyl}
only affects dimensionful parameters.
A mass scale $m_J$ in the Jordan frame is, e.g., mapped onto $m = \mathcal{C}\,m_J$
in the Einstein frame.
Dimensionless ratios of mass parameters, thus, remain
invariant under the Weyl transformation~\cite{Postma:2014vaa,Kamenshchik:2014waa}.
In passing, we also mention that the effective scalar
potential $V_{1\ell}$ in Eq.~\eqref{eq:V1l} cannot be
derived from the effective K\"ahler potential $K_{1\ell}$ in Eq.~\eqref{eq:K1l}.
The reason for this is that, in D-term inflation, the effective K\"ahler potential $K_{1\ell}$
can enter into the total scalar potential only via the D-term scalar potential,
\begin{align}
K_{\rm tot} \rightarrow K_{\rm tot} + K_{1\ell} \qquad\Rightarrow\qquad
V_D \supset - gq\,D_0 \left(\phi\,\frac{\partial}{\partial\phi}\,K_{1\ell}-
\bar{\phi}\,\frac{\partial}{\partial\bar{\phi}}\,K_{1\ell}\right) \,.
\end{align}
This, however, only constitutes a contribution to the one-loop effective potential
for the waterfall fields which we are not interested in.
The one-loop effective potential for the inflaton field in Eq.~\eqref{eq:V1l}
has a different origin.
This can be explicitly seen in the superspace formulation of global supersymmetry.
There, Eq.~\eqref{eq:V1l} does not follow
from the effective potential for the chiral multiplets $S$, $\Phi$, and $\bar{\Phi}$, i.e.,
from the effective K\"ahler potential $K_{1\ell}$, but from the effective potential
for the auxiliary $D$ component of the vector field $V$~\cite{Buchbinder:1994iw}.
This quantity is discussed less often in the literature.
Alternatively, Eq.~\eqref{eq:V1l} can also be derived in a superspace language that applies
to models of softly broken global supersymmetry~\cite{Nibbelink:2006si}.
In this approach, one first integrates out the heavy vector multiplet $V$ such that
supersymmetry is softly (and explicitly) broken in the effective theory at low energies.
Then, one calculates the radiative corrections to the soft SUSY-breaking terms
in the Lagrangian.
This allows one to recover Eq.~\eqref{eq:V1l} as the one-loop renormalization
of the so-called soft K\"ahler potential $\widetilde{K}$.
In an explicit calculation, we convince ourselves that Eq.~\eqref{eq:V1l}
indeed satisfies the relation $V_{1\ell} = - \widetilde{K}_{1\ell}$.
It is interesting to note that this result differs from the situation
in F-term inflation.
There, the effective K\"ahler potential directly contributes to the
effective inflaton potential via the F-term scalar potential.


The total inflaton potential follows from the
combination of our results in Eqs.~\eqref{eq:Vtree} and \eqref{eq:V1l},
\begin{align}
\label{eq:V}
\boxed{
V = \frac{V^J}{\left(1-z\right)^{2}} \,, \quad
V^J = \left[\frac{1}{2} + \left(1 + \delta^4\right)
\frac{g^2q^2}{16\pi^2}\,L\left(x,\alpha\right)\right] D_0^2 -
\frac{\left(1-2\chi\right)z} {1-\left(1-2\chi\right)z}\,F_0^2 \,. }
\end{align}
The individual parameters and functions appearing in this potential can be
found in the following equations:
$z$ in Eq.~\eqref{eq:fz}, $D_0$ and $F_0$ in Eq.~\eqref{eq:VtreeJ},
$\delta$ in Eq.~\eqref{eq:mpm}, and $L$ in Eq.~\eqref{eq:L}.
The potential in Eq.~\eqref{eq:V} is one of the main results in this
paper and the starting point of our phenomenological study of the
inflationary dynamics (see Sec.~\ref{sec:inflation}).
In conclusion, let us summarize the main differences
between Eq.~\eqref{eq:V} and the inflaton potential of ordinary D-term inflation
in global supersymmetry,
\begin{align}
V_0 = \left[\frac{1}{2} + \frac{g^2q^2}{16\pi^2}\,L\left(x_0,0\right)\right] D_0^2 \,, \quad
x_0 = \frac{m_{\rm eff}^2}{q\,m_D^2} \,.
\label{eq:V01l}
\end{align}
Compared to this potential, our total inflaton potential $V$ receives four
different SUGRA corrections:
(i) The total potential is rescaled by $\mathcal{C}^4$ to account for the transition
from the Jordan frame to the Einstein frame.
(ii) The approximate shift symmetry in the inflaton kinetic function in combination 
with F-term SUSY breaking in the IYIT sector results in a small contribution
from the F-term scalar potential.
(iii) The soft B term mass $\delta m_{\rm eff}^2$ modifies the prefactor of
the one-loop effective potential as well as the definition of the field variable $x$.
(iv) The gravity-induced mass $m_R$ gives rise to the parameter $\alpha$.
All of these effects vanish in the global-SUSY limit, such that $V \rightarrow V_0$.
One of our main claims is that the SUGRA corrections 
in Eq.~\eqref{eq:V} are instrumental in realizing a viable scenario
of D-term inflation that is in agreement with all theoretical and
phenomenological constraints.




\section{Phenomenology: A viable scenario of D-term hybrid inflation}
\label{sec:inflation}


In the previous section, we introduced a supergravity embedding of the IYIT model of dynamical supersymmetry breaking with the following properties: (i) by promoting a $U(1)$ flavor symmetry of the DSB model to the gauged $U(1)_{B-L}$ symmetry, we connect the scales of supersymmetry and $B$$-$$L$ breaking and simultaneously generate an effective FI term for the $U(1)_{B-L}$ symmetry. 
We demonstrated that all mass scales, including the FI term, the $B$$-$$L$ breaking scale and the supersymmetry breaking scale, are set by the dynamical scale $\Lambda$, see Eqs.~\eqref{eq:xifaL} and \eqref{eq:mumfm32}.
(ii) The effective FI term generates a D-term scalar potential which can be used to implement DHI. Besides the usual one-loop contribution from integrating out the waterfall fields in the limit $M_{\rm Pl} \rightarrow \infty$, our construction entails several (calculable) supergravity corrections. The final result for the  one-loop effective scalar potential is given in Eq.~\eqref{eq:V}.
(iii) The requirement of perturbativity and the necessity to stabilize the Polonyi field lead to constraints on the parameters of the DSB model, see Fig.~\ref{fig:bounds}.

In this section we turn to the phenomenology of the resulting DHI model, outlined in Sec.~\ref{subsec:HI}. This will essentially fix the only remaining free parameter of our DSB sector, the dynamical scale $\Lambda$. After briefly reviewing the standard picture in global supersymmetry, we proceed to an analytical study of the parameter space, highlighting the most important effects of the different contributions to the scalar potential calculated in the previous section. We then present a full numerical study of the relevant parameter space, supplemented by a discussion on the initial conditions in different parts of the parameter space.


\subsection{D-term inflation in global supersymmetry}
\label{subsec:globalDHI}


The key ingredients of globally supersymmetric DHI are the superpotential and D-term potential given in Eq.~\eqref{eq:FHIDHI}. The waterfall fields $\Phi, \bar \Phi$ obtain masses which depend on the scalar component $s$ of the chiral multiplet $S$, which stabilize them for values of the inflaton field above the critical value
$|s^\text{glob}_c|^2 = g^2 q q_0 \xi /\kappa^2 $.
These field-dependent masses result in a Coleman-Weinberg one-loop contribution to the effective potential of the inflaton, so that the scalar potential for the inflaton  above the critical field value is given by Eq.~\eqref{eq:V01l}.
At the critical field value (corresponding to $x_0 = 1$) one of the waterfall fields acquires a nonvanishing vacuum expectation value, absorbing the FI term~$\xi$. 

Identifying the inflaton field as the radial component of $s$, $\rho =\sqrt{2} \, |s|$, its classical evolution  during inflation is described by the slow-roll equation,
\begin{equation}
 V(\rho) \, \rho'(N) = M_\text{Pl}^2 \, V'(\rho) \,,
 \label{eq:slowroll_global}
\end{equation}
where $N = - \int H dt$ counts the number of remaining e-folds until the end of inflation (with  $N = 0$ at the end of inflation).  At field values much larger than the critical field value $\rho_c$, the scalar potential~\eqref{eq:V01l} can be approximated as 
\begin{equation}
 V_0 \simeq \left[ \frac{1}{2} + \frac{g^2 q^2}{16 \pi^2} \ln x_0 \right] D_0^2 \,.
\end{equation}
Eq.~\eqref{eq:slowroll_global} is an accurate description of the inflationary dynamics as long as the slow-roll parameters,
\begin{equation}
\varepsilon(\rho) = \frac{M_{\rm Pl}^2}{2} \left( \frac{V'(\rho)}{V(\rho)} \right)^2 \simeq \frac{g^4 q^4}{32 \pi^4 } \left( \frac{ M_\text{Pl}}{\rho} \right)^2 \,, \quad \eta(\rho) = M_{\rm Pl}^2 \frac{V''(\rho)}{V(\rho)} \simeq - \frac{g^2 q^2 }{4 \pi^2 }  \left( \frac{ M_\text{Pl}}{\rho} \right)^2\,,
\label{eq:slowroll_parameters}
\end{equation}
are much smaller than one. The CMB observables, describing the statistical properties of quantum vacuum fluctuations during inflation, can be expressed in terms of these variables as
\begin{equation}
A_s = \frac{V}{24 \pi^2 \varepsilon M_{\rm Pl}^4} \,, \quad n_s = 1 - 6 \,\varepsilon + 2 \, \eta \,, \quad r = 16 \, \varepsilon \,,
\label{eq:infpred}
\end{equation}
evaluated at $N_* \simeq 55$ e-folds before the end of inflation.

DHI ends at the critical field value $\rho_c$ or  even earlier, when the second slow-roll parameter $\eta$ becomes large,
$\rho_\eta =  g q M_{\rm Pl} /(2 \pi)$,
depending on the size of $\rho_\eta / \rho_c$. 
For  $\rho_\eta / \rho_c \gg 1$, i.e., if the slow-roll condition is violated before the critical point, the value of $\rho$ at $N_*$ e-folds before the end of inflation is given by
$\rho_*^2 \simeq (g^2 q^2  N_* M_{\rm Pl}^2 )/(2 \pi^2) $.
The amplitude of the scalar spectrum is mainly controlled by the FI parameter,
\begin{equation}
A^0_s \simeq  \frac{N_* q_0^2}{3 \, q^2} \left( \frac{\xi}{ M_{\rm Pl}^2} \right)^2  \qquad \text{(large-$\kappa$ regime)} \,,
\end{equation}
and its spectral index, governed by the second slow-roll parameter $\eta$, is obtained as $n_s \simeq 1 - 1/N_* \simeq 0.98$
in the limit of $g q \ll 4 \pi$. For values of $\xi$ around the GUT scale, this yields the correct scalar amplitude, albeit with a somewhat too large spectral index, disfavored by about $2\sigma$ by the current data~\cite{Ade:2015lrj}.

On the other hand, if the slow-roll conditions are satisfied all the way down to the critical field value, we find $\rho_* \simeq \rho_c$. The small value of the inflaton coupling $\kappa$ in this region of parameter space implies that the field excursion during $N_* \simeq 55$ e-folds of inflation is typically small compared to the field value at the end point of inflation $\rho_c$. The observed value of the scalar amplitude fixes 
\begin{equation}
 A^0_s = \frac{4 \pi^2 q_0^3}{3 \, q^3 \kappa^2}  \left( \frac{\xi}{ M_{\rm Pl}^2} \right)^3
 \qquad \text{(small-$\kappa$ regime)} \,,
\end{equation} 
enabling lower values of $\xi$ for smaller values of $\kappa$. The  spectral index in this region is found to be $n_s \simeq 1$, excluded at more than 5$\sigma$ by the PLANCK data~\cite{Ade:2015lrj}. 

Despite its simplicity and obvious connection to particle physics, this model has several major shortcomings, as discussed in Sec.~\ref{subsec:ingredients}. These are connected to the origin of the FI mass scale in supergravity, the stability of scalar fields during inflation, gravitational corrections to the inflaton trajectory in supergravity, and phenomenological constraints from CMB observations. In the following, we demonstrate how all these shortcomings can simultaneously be overcome in our setup.


\subsection{Analytical description of the inflationary phase}
\label{subsec:analysis}


\subsubsection*{Inflationary dynamics in SUGRA}


In the following we implement DHI with the dynamically generated FI term of Sec.~\ref{subsec:iyit}, supplemented by the assumption of an approximate shift symmetry in the direction of the inflaton field. As discussed in Section~\ref{subsec:ingredients}, this shift symmetry is broken by one-loop effects in the scalar potential and K\"ahler potential. The interplay of these two small contributions will enable us to identify regions in parameter space which comply with all experimental constraints.\footnote{This includes the nonobservation of cosmic strings, as will be demonstrated in Sec.~\ref{subsec:strings}.}

The dynamics of inflation is determined by the scalar potential~\eqref{eq:V}, which contains all relevant supergravity and one-loop contributions. Our choice of kinetic function $F$ (see Eq.~\eqref{eq:F}) with $\chi \ll 1$ ensures that $\sigma$, the real part of the complex scalar $s$, plays the role of the inflaton. The supergravity version of Eq.~\eqref{eq:slowroll_global} in the Einstein frame reads
\begin{equation}
{\cal K}_{s s^*}(\sigma) \, V(\sigma) \, \sigma'(N) = M_{\rm Pl}^2 \, V'(\sigma) \,.
\label{eq:slowroll}
\end{equation}
where $V = {\mathcal C}^4 \, V^J$ is the Einstein-frame scalar potential and ${\cal K}_{s s^*} = \partial^2 K/(\partial s \partial s^*)$ is the prefactor of the kinetic term for the inflaton.  
The initial condition (i.e., the end of inflation) is given by $\sigma(N = 0) = \text{max}(\sigma_c, \sigma_\eta)$. 
 This enables us to evaluate the (Einstein-frame) slow-roll parameters $\varepsilon(\hat \sigma)$ and $\eta(\hat \sigma)$ 
and hence the CMB observables at $N_* = 55$ e-folds before the end of inflation. Evaluating the slow-roll parameters requires derivatives of the scalar potential w.r.t.\ the canonically normalized field $\hat \sigma$, which we perform by exploiting $\partial \hat \sigma / \partial \sigma = \sqrt{{\cal K}_{s s^*}}$, as follows from the canonical normalization of the kinetic terms in the Einstein frame,
\begin{equation}
\frac{1}{2} {\cal K}_{s s^*} \partial_\mu \sigma \partial^\mu \sigma \simeq  \frac{1}{2} {\cal C}^4 (1 - f) \partial_\mu \sigma \partial^\mu \sigma = \frac{1}{2} \partial_\mu \hat \sigma \partial^\mu \hat \sigma \,.
\label{eq:kinetic}
\end{equation}
For convenience, we recall here a few key quantities (evaluated along the inflationary trajectory) introduced earlier (see Eqs.~\eqref{eq:Weyl}, \eqref{eq:OmegaPhi}, \eqref{eq:F}, and \eqref{eq:fz})
\begin{align}
 {\cal C}^2 =  - \frac{3 M_\text{Pl}^2}{\Omega} \,, \quad \Omega = - 3 M_\text{Pl}^2 + F \,, \quad F = \chi \sigma^2 \,, \quad  f = (1 - 2 \chi) \, z \,,  \quad z = \frac{\chi \sigma^2}{3 M_\text{Pl}^2} \,.
 \label{eq:recap}
\end{align} 
For more details on translating between the Einstein and Jordan frames, see App.~\ref{app:frames}.

The results of the numerical analysis are shown in Fig.~\ref{fig:parameterscan}. Before discussing them in detail, we will give an analytical analysis of the parameter space in the vicinity of the globally supersymmetric limit. This will prove instructive for interpreting the numerical results.


\subsubsection*{Slow-roll parameters}


The slow-roll parameters in the Einstein frame can be expressed in terms of derivatives of the scalar potential and of the kinetic function in the Jordan frame as (see Appendix~\ref{app:slowroll_parameters}),
\begin{align}
\varepsilon & = \frac{1}{{\cal N}^2} \left(\varepsilon_J^{1/2} -2\, \xi_J^{1/2}\right)^2 \,,  \label{eq:epsjordan}\\
\eta  & = \frac{1}{{\cal N}^2} \left(\eta_J +
12\,\xi_J -  \, 8\left(\varepsilon_J\,\xi_J\right)^{1/2} - 2\,\zeta_J  - 
2\, \nu^{1/2} \left( \varepsilon_J^{1/2} - 2\,\xi_J^{1/2} \right) \right)\,,
\label{eq:etajordan}
\end{align}
with ${\cal N }  \equiv  {\cal K}_{ss^*}^{1/2} = {\cal C}^2 (1 - f)^{1/2}$ and
\begin{align}
\varepsilon_J  &\equiv \frac{M_{\rm Pl}^2}{2}\left(\frac{V_\sigma^J}{V^J}\right)^2 \,, \quad
\eta_J \equiv M_{\rm Pl}^2\,\frac{V_{\sigma\sigma}^J}{V^J} \,, \\
\xi_J^{1/2} &\equiv \frac{M_{\rm Pl}}{\sqrt{2}}\,\frac{\Omega_\sigma}{\Omega} \,, \quad
\zeta_J \equiv M_{\rm Pl}^2\,\frac{\Omega_{\sigma\sigma}}{\Omega}  \,, \quad \nu \equiv \frac{M_\textrm{Pl}^2}{2} \left( \frac{{\cal N}_\sigma}{{\cal N}} \right)^2 \,. \nonumber
\end{align}
These expressions are equivalent to those found in Appendix~A of Ref.~\cite{Chiba:2008ia}. In the following, we will use Eqs.~\eqref{eq:epsjordan} and \eqref{eq:etajordan} to analyze the inflationary predictions, since this format enables us to nicely disentangle the different contributions in various parts of the parameter space.

With the definitions above, simplified expressions for the \textit{Jordan-frame slow-roll parameters} $\varepsilon_J$ and $\eta_J$  can be obtained by approximating the Coleman-Weinberg one-loop potential for $x \gg 1$ and $\alpha \ll 1$ as 
\begin{align}
V_{1\ell}^J = \frac{Q_J^4}{16\pi^2} \left[
\vphantom{\frac{1}{4}} \ln x + \mathcal{O}\left(\alpha x \right)\right] \,,
\label{eq:CWapp}
\end{align}
with $x, \, \alpha$ and $Q_J$ given in Eq.~\eqref{eq:xaQJ}. As in the globally supersymmetric case, $x = 1$ denotes the critical point. Eq.~\eqref{eq:CWapp} is a good approximation as long as $1 \ll x \ll 1/\alpha$, which will hold in most of the parameter regime of interest.
 We then find
\begin{align}
\varepsilon_J & \simeq \phantom{-}\left(\frac{M_{\rm Pl}}{\sigma/\sqrt{2}}\right)^2 \left[
\left(1+\delta_\varepsilon^4\right)\frac{q^2g^2}{16\pi^2}\frac{D_0^2}{V^J} 
- \frac{f}{\left(1-f\right)^2} \frac{F_0^2}{V^J}\right]^2 \,, \label{eq:epsJ0}
\\ 
\eta_J & \simeq - \left(\frac{M_{\rm Pl}}{\sigma/\sqrt{2}}\right)^2 \left[
\left(1-\delta_\eta^4\right)\frac{q^2g^2}{16\pi^2}\frac{D_0^2}{V^J} 
+ \frac{f\left(1+3f\right)}{\left(1-f\right)^3} \frac{F_0^2}{V^J} \right] \,,
\label{eq:etaJ0}
\end{align}
with $f$ given in Eq.~\eqref{eq:fz} and
\begin{align}
\delta_\varepsilon \equiv \left(\ln x+\frac{1}{2}\right)^{1/4} \delta  \,, \quad
\delta_\eta \equiv \left(\ln x+\frac{1}{2} + \frac{2+\delta^4}{1+\delta^4}\right)^{1/4} \delta \,,
\label{eq:deltas}
\end{align}
where $\delta$ was introduced in Eq.~\eqref{eq:mpm} and $F_0$ and $D_0$ denote the F- and D-term contributions from global supersymmetry, respectively.
For both $\varepsilon_J$ and $\eta_J$, the term proportional to $D_0^2$ stems from the Coleman-Weinberg one-loop potential whereas the term proportional to $F_0^2$ is a supergravity effect, induced by the noncanonical terms in the K\"ahler potential. Moreover, we note that the Coleman-Weinberg term splits into the expression familiar from global supersymmetry (indicated by the ``1'' in the parentheses) and the supergravity contribution to the waterfall field sector, parametrized by $\delta_{\varepsilon,\eta}^4$. Note that, as in global supersymmetry, $\varepsilon_J$ is suppressed compared to $\eta_J$.

Eqs.~\eqref{eq:epsJ0} and \eqref{eq:etaJ0} illustrate the main effect of the F-term SUGRA corrections in our model. 
For $f \ll 1$ and $V^J \simeq D_0^2/2$, the second term in Eq.~\eqref{eq:etaJ0} yields
\begin{equation}
\Delta \eta_J \simeq - \frac{4}{3} (1 - 2 \chi) \chi \frac{F_0^2}{D_0^2} \,,
\label{eq:DetaJ}
\end{equation}
indicating that the supergravity contributions from the tree-level F-term scalar potential can induce the desired lowering of the spectral index  ($\Delta \eta \sim - 0.01$) compared to the result of global supersymmetry if $\chi \simeq 7.5 \times 10^{-3} \, D_0^2/F_0^2$.\footnote{For $\chi = 1/2$, we have $\Delta \eta_J = 0$, which corresponds to the absence of a $m_{3/2}^2$ term for the inflaton field due to the sequestering K\"ahler potential. One might thus expect a second solution, $\chi = 1/2 - |\delta \chi|$ with $|\delta \chi| \ll 1$, to produce $\Delta \eta \sim - 0.01$. However in this case the corresponding gravity-induced mass for the inflaton, see Eq.~\eqref{eq:msigma_H0}, is too large.}  
This small value of $\chi$ indicates the presence of an approximate shift symmetry.

Turning to the effects of a small, positive value for $\chi$ on the first slow-roll parameter, see Eq.~\eqref{eq:epsJ0}, we note that a positive $\chi$ will lead to a decrease in $\varepsilon$. The one-loop and SUGRA contributions may even cancel each other, indicating the presence of a hilltop or a saddle point in the scalar potential. We will come back to the consequences of such a scenario below.


The explicit expressions for the remaining auxiliary Jordan-frame slow-roll parameters are 
\begin{align}
\xi_J^{1/2} = - \left(\frac{M_{\rm Pl}}{\sigma/\sqrt{2}}\right) \frac{z}{1-z} \,, \quad
\zeta_J = - \left(\frac{M_{\rm Pl}}{\sigma/\sqrt{2}}\right)^2 \frac{z}{1-z} \,, \quad
\nu^{1/2} = \left(\frac{M_{\rm Pl}}{\sigma/\sqrt{2}}\right)
\frac{\left(1+2\chi-f\right)z}{2\left(1-z\right)\left(1-f\right)} \,.
\end{align}
Here, $\zeta_J$ captures the gravity-induced mass of the inflaton and emphasizes once more the need for an approximate shift symmetry ($\chi \ll 1$) for the inflaton field: for $\chi \simeq 1/2$, the inflaton picks up a gravity-induced mass just as all the other scalars do,
\begin{align}
\chi \simeq \frac{1}{2} \quad\Rightarrow\quad
m_\sigma^2 \simeq m_R^2 \,.
\label{eq:msigma_H0}
\end{align}
with $m_R^2 \simeq D_0^2/(3 M_{\rm Pl}^2)$.
This implies a contribution to the slow-roll parameter $\eta$
of $\Delta\eta_\zeta \simeq 2/3$ (see Eq.~\eqref{eq:eta23}) and, hence, leads to an $\eta$ problem.
A purely canonical term in the inflaton kinetic function, $F \simeq S^\dagger S$,
is therefore not viable in our model. The approximate shift symmetry resolves the problem, suppressing this contribution as
$\Delta \eta_\zeta \simeq 4 \chi/3$. Of course, Eq.~\eqref{eq:DetaJ} and \eqref{eq:msigma_H0} directly correspond to the second and first terms of Eq.~\eqref{eq:msigmaDF}, respectively. Due to $F_0 > D_0$, the contribution from Eq.~\eqref{eq:DetaJ} will always dominate for $\chi \ll 1$.

Inserting these results into Eqs.~\eqref{eq:epsjordan} and \eqref{eq:etajordan}, we find to leading order ($z \ll 1$,  $D_0/F_0 \ll 1$):
\begin{align}
\varepsilon & \simeq
\left(\frac{M_{\rm Pl}}{\sigma/\sqrt{2}}\right)^2 \left[
\left(1 +\delta_\varepsilon^4\right)\frac{q^2g^2}{16\pi^2}\frac{D_0^2}{V^J}
- f \frac{F_0^2}{V^J}
 \right]^2 \,, \label{eq:epsfinal} \\
\eta & \simeq - \left(\frac{M_{\rm Pl}}{\sigma/\sqrt{2}}\right)^2 \left[
\left(1-\delta_\eta^4\right)\frac{q^2g^2}{16\pi^2}\frac{D_0^2}{V^J}
+ f\, \frac{F_0^2}{V^J} \right] \,. \label{eq:etafinal}
\end{align}
%


\subsubsection*{Viable parameter space}


Starting from Eqs.~\eqref{eq:epsfinal} and \eqref{eq:etafinal} and the results of D-term hybrid inflation in global supersymmetry, we can analytically assess the viable parameter space in the vicinity of the globally supersymmetric results. 
Comparing the global-SUSY DHI results with the observed value for $n_s$, we conclude that the supergravity contributions must enhance $|\eta|$ by (at least) an ${\cal O}(1)$ factor. At the same time, requiring $V'(\hat \sigma) > 0$ implies an upper bound on these contributions (see Eq.~\eqref{eq:epsfinal}):
\begin{equation}
\left(1-\delta_\eta^4\right)\frac{q^2g^2}{16\pi^2}\frac{D_0^2}{V^J} \, \lesssim  \,  f \frac{F_0^2}{V^J} \, < \,  \left(1+\delta_\varepsilon^4\right)\frac{q^2g^2}{16\pi^2}\frac{D_0^2}{V^J} \,,
\end{equation}
for $\sigma = \sigma_*$.
This implies a lower bound on $\delta$ (governing the sizes of both $\delta_\varepsilon$ and $\delta_\eta$). 
At the same time, $\delta_\eta$ yields a positive contribution to $\eta$ and a too large value will lead to an enhancement of the spectral index $n_s$. 
Together this implies $\delta \sim {\cal O}(0.1 \cdots 1)$,
where to leading order
$\delta^4 \simeq \kappa^2 \sigma^2 m_{3/2}^2/(2 q_0^2 q^2 g^4 \xi^2)$. 
We can estimate $\delta$ by exploiting the analytical results for $\sigma_*$ in globally supersymmetric D-term hybrid inflation, see Sec.~\ref{subsec:globalDHI}.
In addition, we note that the requirement that the tree-level supergravity term contributes $\Delta \eta \sim - 0.01$ implies 
\begin{equation}
- \Delta \eta \simeq  \frac{2 \chi}{3} \left( \frac{m_{3/2}}{H_J} \right)^2  \simeq 0.01 \quad \rightarrow \quad
\chi \simeq  10^{-4} \, \left( \frac{g}{0.1}\right)^2 \left(\frac{10^{15}~\text{GeV}}{\sqrt{\xi}} \right) \,.
\label{eq:chiconstr}
\end{equation}
where we have inserted the relations~\eqref{eq:xifaL} and \eqref{eq:mumfm32}.

In the regime of large $\kappa$ (and taking for simplicity $|q| \sim |q_0| \sim 1$), the constraints on $\delta$ thus roughly fix the parameter combination $\kappa^2 \sqrt{\xi}/g^2$. 
Using Eq.~\eqref{eq:chiconstr} we immediately see that the correct spectral index can be obtained for
\begin{equation}
\chi \simeq 3.5 \times 10^{-4} \left(\frac{0.1}{\delta^4}\right) \left( \frac{\kappa}{0.1} \right)^2 \left( \frac{N_*}{55} \right)\,.
\label{eq:chi_largek}
\end{equation}
The amplitude of the scalar power spectrum $A_s$ is mainly dependent on $\xi$, see Sec.~\ref{subsec:globalDHI}. This essentially fixes $\xi \simeq 10^{-5} M_{\rm Pl}^2$, and determines the preferred range of the gauge coupling, e.g.\ $g \sim 0.1$ for $\kappa = 0.1$.
Note that this constraint can be circumvented if one allows inflation to begin very close to the hilltop of the scalar potential, $\varepsilon_* \simeq 0$, which can be obtained by tuning the contributions in Eq.~\eqref{eq:epsfinal}. From Eq.~\eqref{eq:infpred} we see that in this case, we can in principle arbitrarily lower $\xi$. However, this corresponds to a very tuned situation and we will not focus on this regime of the parameter space. 

In the regime of small $\kappa$ we note from the expressions of the globally supersymmetric limit that $\sigma_* \simeq \sigma_c$ and $n_s \simeq 1$. This indicates that (i) the leading-order term in the expansion of $V_{\text{CW}}$ in $1/x$ becomes a poor approximation and (ii) to obtain the correct spectral index, we must rely nearly exclusively on the supergravity terms in $\eta$. As a result of the first point, the lower bound on $\delta$ in fact becomes irrelevant when using the full expression for the one-loop potential. We are thus left with
$\xi^{3/2}/g^2\lesssim 8.5 \times 10^{-3} \, M_{\rm Pl}^3/2 \, \delta^4$.
Imposing the observed value for $A_s$ and approximating $\sigma_*$ by the corresponding expression in globally supersymmetric DHI, this yields
\begin{equation}
\xi \simeq 5.5 \times 10^{-6} \, M_{\rm Pl}^2 \, \left( \frac{\kappa}{10^{-3}} \right)^{2/3} \,.
\end{equation}
Inserting this into Eq.~\eqref{eq:chiconstr}, we find
\begin{equation}
\chi \simeq 0.1 \, g^2 \left( \frac{\kappa}{10^{-3}} \right)^{-1/3} \,.
\label{eq:chi_smallk}
\end{equation}
{Both Eq.~\eqref{eq:chi_largek} and Eq.~\eqref{eq:chi_smallk} point to small values of $\chi$, i.e., an approximate shift symmetry. For $\kappa = 0.1$, Eq.~\eqref{eq:chi_largek} indicates a value of $\chi$ that is larger than that obtained by only integrating out the waterfall fields, $\chi_{1\ell} \simeq \kappa^2/(16 \pi^2) = 6 \times 10^{-5} \, (\kappa/0.1)^2$, see Eq.~\eqref{eq:K1l}. On the other hand, for smaller values of $\kappa$, the symmetry breaking induced by the waterfall fields can be sufficient to generate the correct spectral index for accordingly small values of the gauge coupling $g$, see Eq.~\eqref{eq:chi_smallk}. We will confirm these results with a dedicated numerical analysis below.}

\subsection{Scan of parameter space and numerical results}
\label{subsec:scan}

\begin{figure}
\centering
\includegraphics[width=0.48\textwidth]{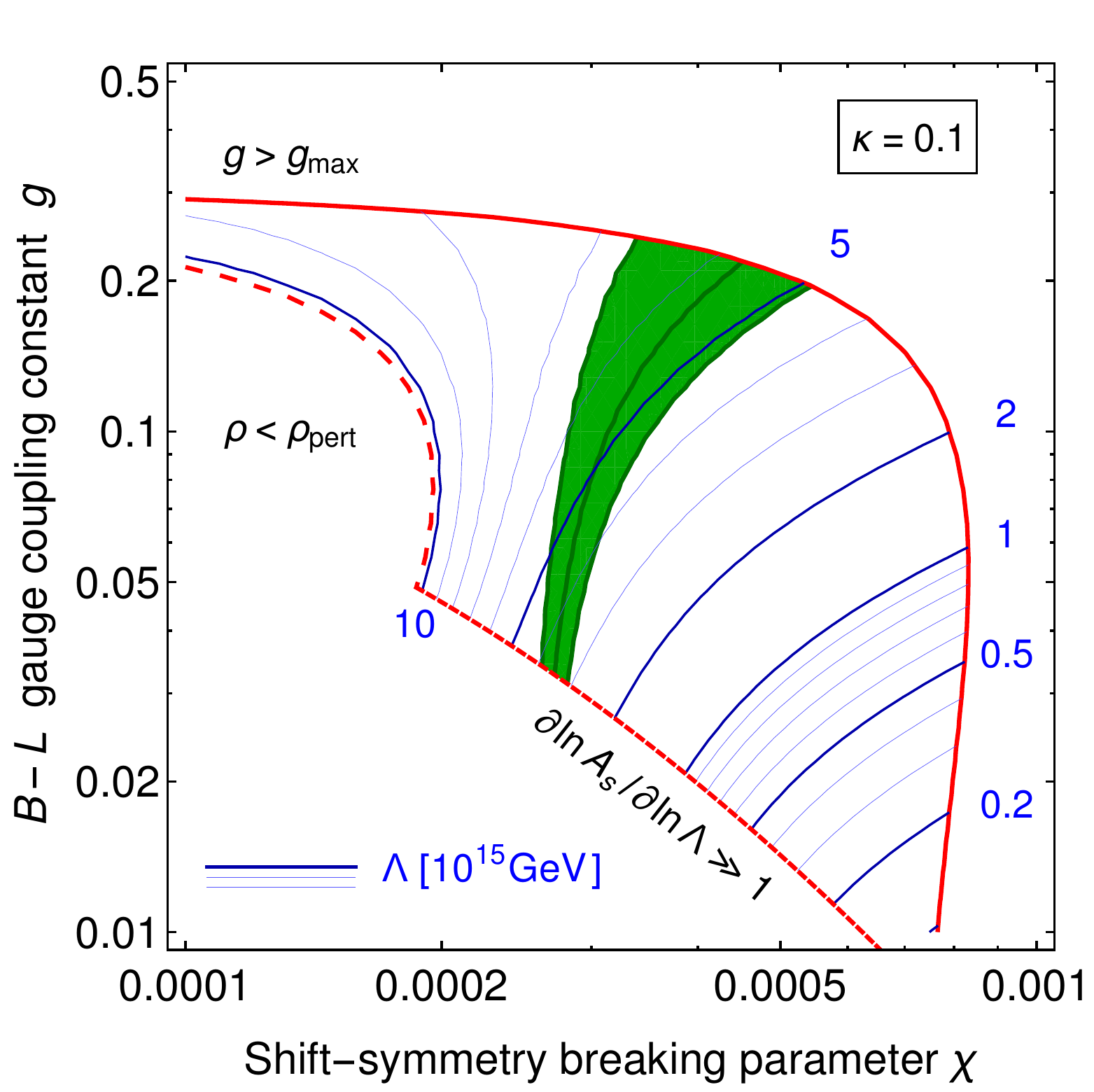}\hfill
\includegraphics[width=0.48\textwidth]{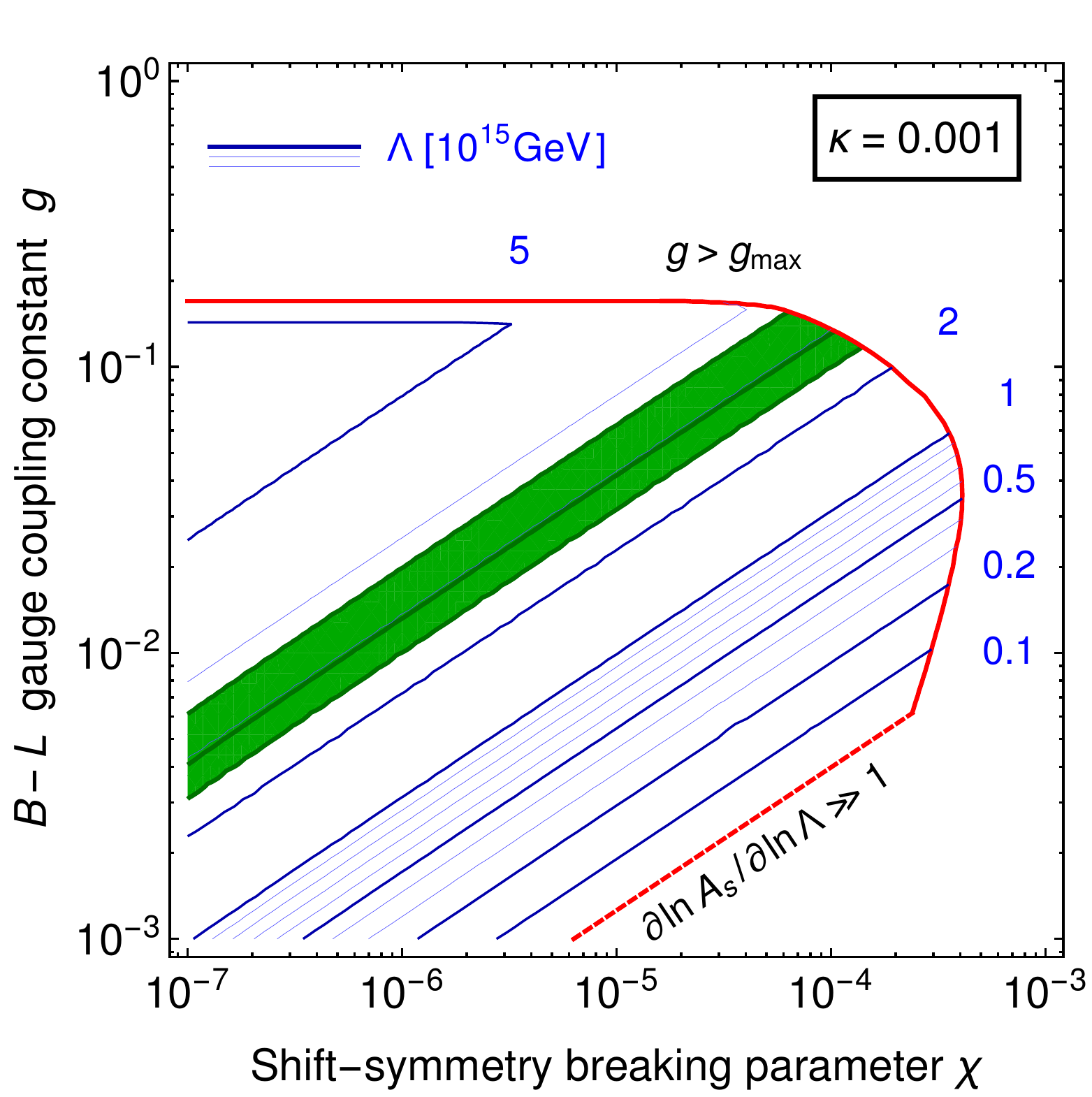}
\caption{CMB observables and viable parameter space for two representative, fixed values of the superpotential coupling, $\kappa = 0.1$ and $\kappa = 0.001$. The red boundaries show constraints enforcing perturbativity ($\rho > \rho_\text{pert}$, see Eq.~\eqref{eq:rhopert}), the stabilization of the Polonyi field ($g < g_\text{max}$, see Eq.~\eqref{eq:gmax}) and limiting the amount of fine-tuning in the model parameters (we disregard parameter values for which $\partial \ln A_s/\partial \ln \Lambda \gtrsim 30$, see the discussion below Eq.~\eqref{eq:chi_largek}). Blue lines indicate  contour lines for the dynamical scale $\Lambda$ that reproduce the observed amplitude of the scalar spectrum. The green band indicates values of the spectral index in agreement at 95$\%$~CL with the current best-fit value, $n_s = 0.9677 \pm 0.006$~\cite{Ade:2015lrj}.
}
\label{fig:parameterscan}
\end{figure}

In this section we present our results for a numerical scan of parameter space, focusing on the regions identified analytically in the previous section. Starting from the full scalar potential~\eqref{eq:V} and the equation of motion~\eqref{eq:slowroll}, we determine the slow-roll parameters~\eqref{eq:slowroll_parameters} and consequently the inflationary observables~\eqref{eq:infpred} at $N_* = 55$ e-folds before the end of inflation.
Fixing the charges $|q_0| = 1$ and $|q| = 2$, for each parameter point in the $(\kappa, \, g, \, \chi, \, \xi)$ plane we (i) determine the end point of inflation, (ii) solve the slow-roll equation of motion and (iii) determine for fixed $(\kappa, \, g, \, \chi)$ the value of $\xi$ that reproduces the observed amplitude of the scalar power spectrum. We also explicitly check that our results are not sensitive to the precise choice of the renormalization scale $Q_J$.

\begin{figure}[t]
\centering
\includegraphics[width=0.48\textwidth]{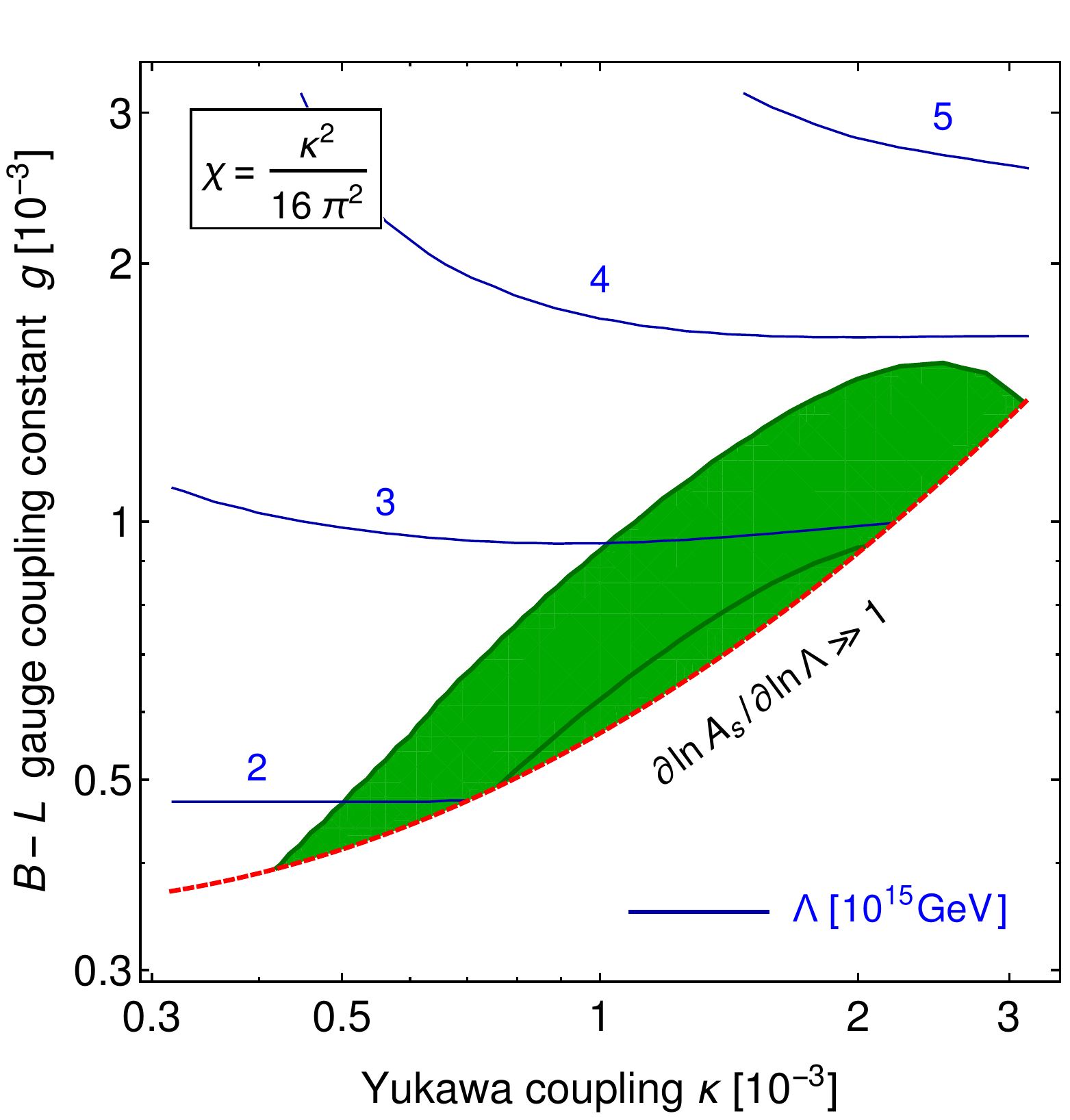}
\caption{CMB observables and viable parameter space for shift symmetry breaking exclusively through the superpotential coupling of the waterfall fields, $\chi = \chi_{1 \ell} =  \kappa^2/(16 \pi^2)$. The color code is the same as in Fig.~\ref{fig:parameterscan}.
}
\label{fig:parameterscan2}
\end{figure}

Our results are depicted in Fig.~\ref{fig:parameterscan}, for $\kappa = 0.1$ (left panel) and $\kappa = 10^{-3}$ (right panel), {as well as in Fig.~\ref{fig:parameterscan2} where we have imposed the additional relation $\chi =  \chi_{1 \ell} = \kappa^2/(16 \pi^2)$, see Eq.~\eqref{eq:K1l}.} In all figures, the green band indicates the region of parameter space in accordance with all constraints.
In the parameter space of interest, we find values for the tensor-to-scalar ratio $r$ of ${\cal O}(10^{-6} \cdots 10^{-4})$, far below the current bounds. The numerical results for the inflationary observables excellently agree with the results obtained from our analytical expressions for the slow-roll parameters, Eq.~\eqref{eq:epsfinal} and Eq.~\eqref{eq:etafinal}, as well as with our estimates for the shift-symmetry-breaking parameter $\chi$ in Eqs.~\eqref{eq:chi_largek} and \eqref{eq:chi_smallk}. This underlines that although our numerical analysis takes into account all contributions to the scalar potential, the most relevant contribution to lower the spectral index is the shift-symmetry-suppressed soft mass for the inflaton, leading to Eq.~\eqref{eq:DetaJ}.

Our choices for the coupling $\kappa$ are designed to cover the relevant aspects of the parameter space, while focusing on particularly interesting benchmark points. As it is responsible for explicit shift symmetry breaking in the superpotential, we expect $\kappa \lesssim 1$. In the left panel of Fig.~\ref{fig:parameterscan}, we consider $\kappa = 0.1$. This enables us to reproduce the observed CMB observables (in particular $n_s$) with loop and SUGRA contributions of comparable size, leading to $\chi \sim 10^{-4}$. 
In global-SUSY DHI the parameter space splits into two regimes, characterized by the size of $|\sigma_\eta/\sigma_c|$ and consequently by different parameter dependencies of $s_*$, see Sec.~\ref{subsec:globalDHI}. The value of $\kappa = 0.1$ falls into the regime of $\sigma_\eta \gg \sigma_c$. %
In the right panel of Fig.~\ref{fig:parameterscan} we turn to the opposite regime, $\sigma_\eta \ll \sigma_c$.
To reproduce the observed spectral index, we here need to require the SUGRA contributions to clearly dominate over the one-loop contributions. 
Note that for even smaller values of $\kappa$, the critical value $\sigma_c$ can take super-Planckian values, enabling a phase of ``subcritical hybrid inflation'' after the inflation field has passed $\sigma_c$~\cite{Buchmuller:2014rfa,Buchmuller:2014dda}.

{In Fig.~\ref{fig:parameterscan2} we impose $\chi = \chi_{1\ell} = \kappa^2/(16 \pi^2)$ (Eq.~\eqref{eq:K1l}), thus reducing the parameter space of our model by one dimension. Interestingly and nontrivially, we find solutions which obey all constraints if $ \kappa \sim g = {\cal O}(10^{-3})$. Hence the minimal model setup with shift symmetry breaking only through the coupling of the waterfall fields in the superpotential can reproduce the observed CMB observables, thereby essentially determining all model parameters: the superpotential coupling $\kappa$, the gauge coupling $g$ and the mass scale $\Lambda$. The value of the gauge coupling $g$ is small compared to the expectation in GUTs, but interestingly, the relation $\kappa = \sqrt{2}g$ is precisely the relation predicted in $\mathcal{N} = 2$ supersymmetric hybrid inflation~\cite{Watari:2000jh}.


\subsection{Initial conditions}
\label{subsec:iniconds}


In the viable parameter space, inflation occurs either near a hilltop (i.e., a local maximum in the scalar potential) or near an inflection point, depending on the exact values of $\chi$ and $g$. In Fig.~\ref{fig:potentials} we depict these two possibilities (for $\kappa = 0.1$), together with the decomposition of the total scalar potential into its dominant components. 
\begin{figure}
\centering
\includegraphics[width=0.48\textwidth]{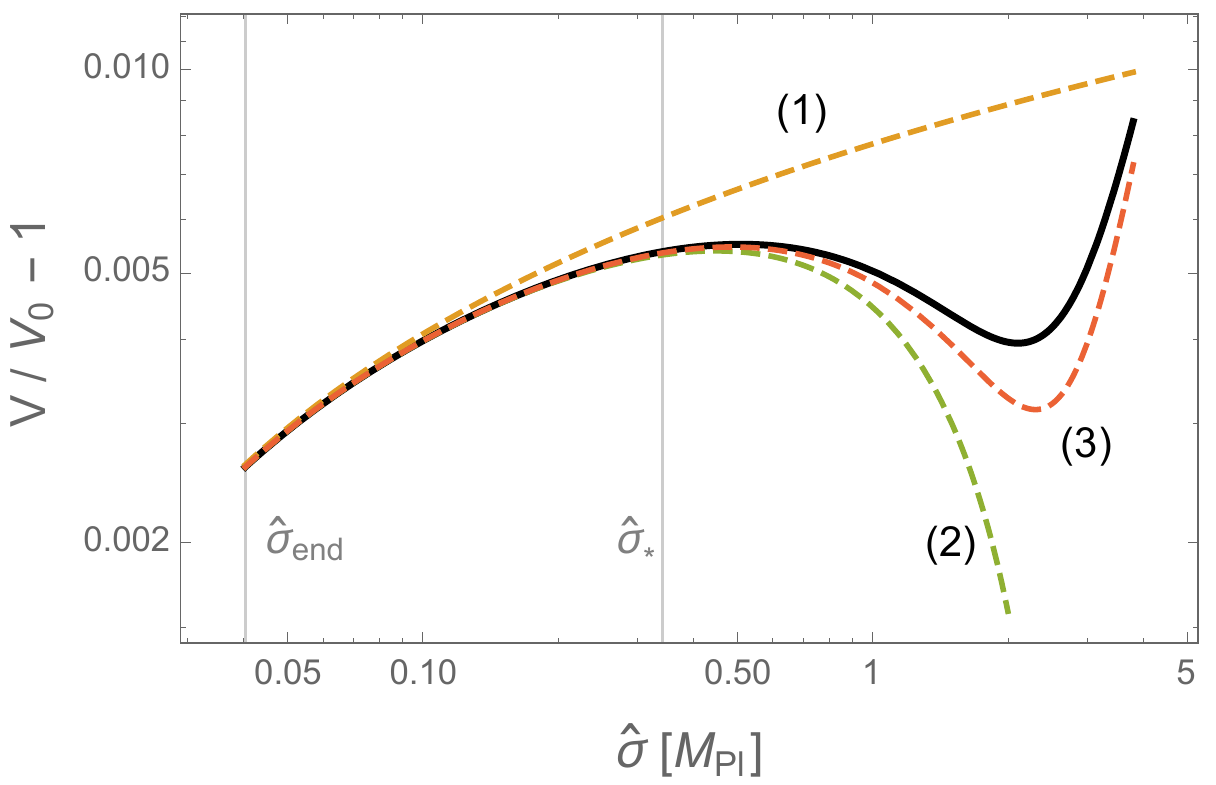}\hfill
\includegraphics[width=0.48\textwidth]{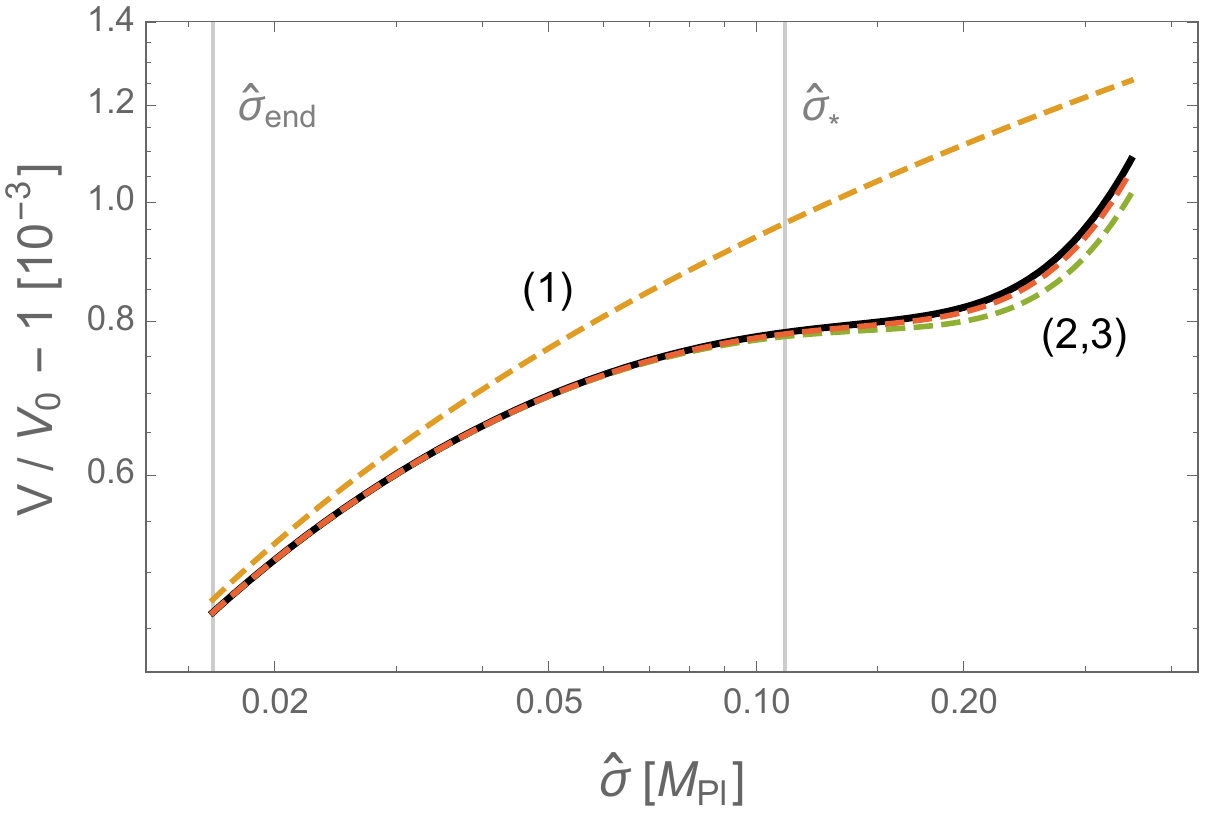}
\caption{Contributions to the scalar potential (for $\kappa = 0.1$). The total scalar potential is shown in solid black, while the dashed curves mark the following contributions: (1) global-SUSY one-loop potential, (2) + leading SUGRA effects, and (3) + gravity-induced effects in the one-loop potential. For details see text. The parameter values are chosen as $g = 0.13$, $\chi = 3.5 \times 10^{-4}$, $\xi = 10^{-5} M_{\rm Pl}^2$ \textbf{(left panel)} and $g = 0.05$, $\chi = 2.8 \times 10^{-4}$, $\xi = 6.3 \times 10^{-6} M_{\rm Pl}^2$ \textbf{(right panel)}.}
\label{fig:potentials}
\end{figure}
The solid line shows the full scalar potential, while the labeled dashed lines indicate the following components: (1) leading-order term in the Coleman-Weinberg potential in the global-SUSY limit, (2) supplemented with the leading supergravity terms to the inflaton F-term potential and to the waterfall mass spectrum and (3) in addition supplemented with the next-to-leading-order term in the expansion of the Coleman-Weinberg potential (see Eq.~\eqref{eq:CWapp}), $V_{1\ell}^{J,NLO} = Q_J^4/(8 \pi^2) \alpha x \ln(x)$. The latter term becomes relevant as $x(\sigma)$ increases to $x \sim 1/\alpha$.
The remaining discrepancy compared to the full scalar potential (in the left panel) is mainly due to the $\sigma$ dependence of the D-term potential in the Einstein frame, induced by the conformal factor.

Implementing inflation in the left panel of Fig.~\ref{fig:potentials} requires some fine-tuning in the initial conditions, to ensure the correct vacuum is reached. However, we point out two further observations: (i) in the entire parameter space of interest, we find $H_* / (2 \pi) \lll (\sigma_\text{max} - \sigma_*)$, i.e., if (by accepting some tuning), the initial conditions are in the desired regime, they are at least stable against quantum fluctuations. (ii) lowering the $B$$-$$L$ gauge coupling $g$, the energy level of the false minimum is raised compared to $V(\sigma_*)$. An interesting (albeit fine-tuned) situation arises if the vacuum energy density of the false minimum lies just a tiny bit above $V(\sigma_*)$, allowing for a phase of eternal inflation, followed by $N_*$ e-folds of inflation arising once the inflaton field tunnels through the potential barrier.

On the other hand, in the right panel of Fig.~\ref{fig:potentials} inflation can start at large field values, avoiding an initial conditions problem. There is however some degree of tuning required in the model parameters to ensure this shape of the potential. For $\kappa = 0.1$, this becomes particularly relevant for small values of $g$, when large SUGRA contributions to the slow-roll parameters need to be carefully balanced. This results in the `fine-tuning' constraint on the parameter space in Fig.~\ref{fig:parameterscan}. An overview of these different regions in parameter space is given in Fig.~\ref{fig:InitialConditions} {for the cases of $\kappa = 0.1$ and $\chi = \kappa^2/(16 \pi^2)$, in both cases focusing on the region of parameter space which reproduces the correct CMB observables.}

Note that for very large field values, $\sigma^2 \sim 3 M_\text{Pl}^2/\chi \sim 10^4 M_\text{Pl}^2$, both the conformal factor ${\cal C}$ and the F-term potential exhibit a pole (see Eqs.~\eqref{eq:chi_largek} and \eqref{eq:Vtree}):
\begin{align}
\Omega  = 0  \; \rightarrow  \; (\sigma^\infty_C)^2  \simeq \frac{1}{\chi}(3 M_{\rm Pl}^2) \,, \qquad 
f  = 1 \; \rightarrow \; (\sigma^\infty_F)^2 = \frac{3 M_{\rm Pl}^2}{\chi (1 - 2 \chi)} \,.
\label{eq:sigmaF}
\end{align}
After canonical normalization of the inflaton field, the pole in the conformal factor will be pushed to infinity. The pole in the F-Term potential is always at larger field values and is hence never reached.\footnote{The pole in the conformal factor implies that in the field space of the canonically normalized field $\hat \sigma$, the scalar potential asymptotes to $\lim\limits_{\hat \sigma \rightarrow \infty} V(\hat \sigma) =  V(\sigma_C^\infty)$ and is thus always bounded from below.} However, for $\chi \ll 1$, $\sigma^\infty_F$ approaches $\sigma^\infty_C$ and the F-term potential begins to dominate the tree-level potential already for $\sigma < \sigma^\infty_C$. This can generate a false, negative-valued vacuum at large field values. {In the regions of parameter space where the inflaton field reaches values of order $M_{\rm Pl}$, this can impact the vacuum structure.} However at these large field values, $\sigma \gg M_{\rm Pl}$, higher-order operators may significantly modify the scalar potential.

\begin{figure}
\centering
\includegraphics[height=0.48\textwidth]{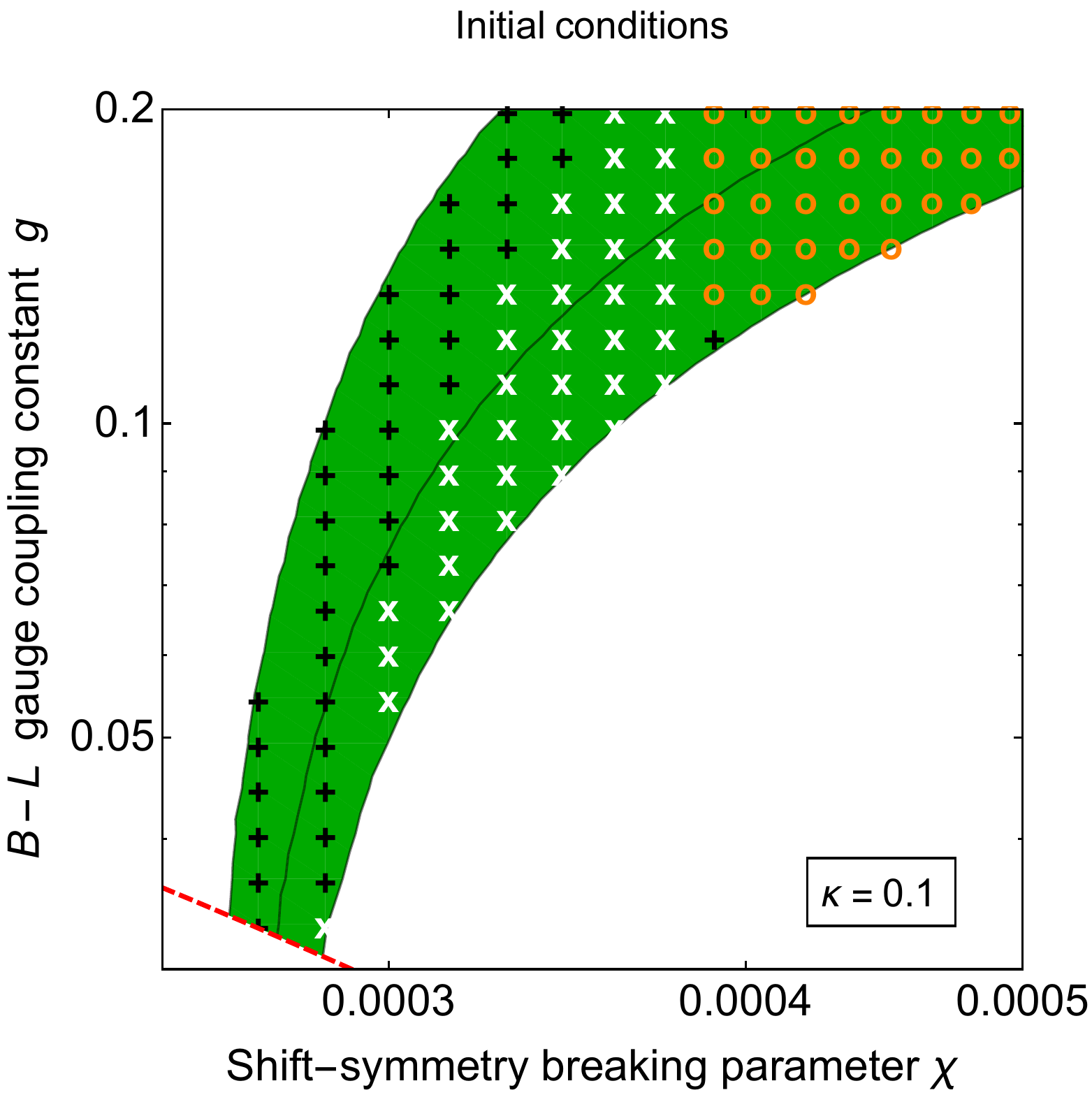}\hfill
\includegraphics[height=0.48\textwidth]{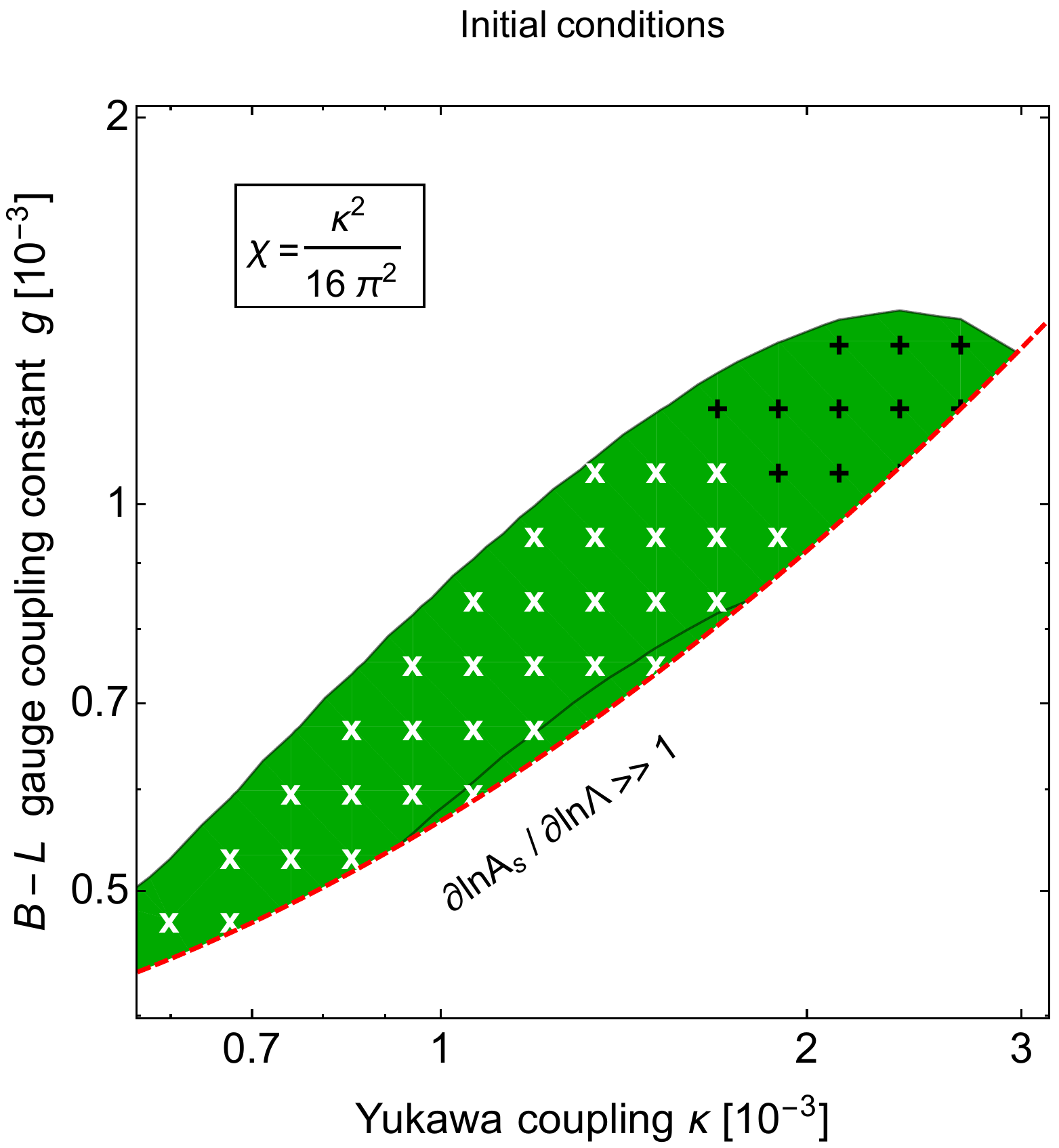}
\caption{Initial conditions for $\kappa = 0.1$ \textbf{(left panel)} and $\chi = \kappa^2/(16 \pi^2)$ \textbf{(right panel)}. The three different possibilities for the structure of the scalar potential are (i) inflection point inflation (black ``$+$'' symbol, same situation as in the right panel of Fig.~\ref{fig:potentials}), (ii) hilltop inflation with a local minimum at $\sigma \gg \sigma_*$ (white ``x'' symbol, same situation as in the left panel of Fig.~\ref{fig:potentials}) and (iii) hilltop inflation without a local minimum  (red ``o'' symbol). This last situation arises due to the effect described below Eq.~\eqref{eq:sigmaF}, and requires tuning the initial conditions. As in Figs.~\ref{fig:parameterscan} and \ref{fig:parameterscan2}, the green band indicates that the scalar spectral index lies within the $2 \sigma$ band.}
\label{fig:InitialConditions}
\end{figure}

For $\kappa = 10^{-3}$, we find that inflation typically occurs in a small field region in the vicinity of a hilltop, accompanied by a false, often negative-valued vacuum at large field values. The amplitude of this vacuum is lifted as $g$ and $\chi$ are decreased, until at values of $g \lesssim 5 \times 10^{-4}$ inflation occurs close to an inflection point. The shape of the potential is well described by the D-term potential, ${\cal C}^4 D_0^2/2$, supplemented by the full globally supersymmetric one-loop potential (not truncated at ${\cal O}(1/x)$) and the leading-order SUGRA contribution from the F-term potential.



\section{Discussion: Particle spectrum and cosmology after inflation}
\label{sec:discussion}


During inflation, supersymmetry is broken through F- and D-term contributions. After inflation, when the D-term is absorbed into the VEV of the waterfall field, only F-term supersymmetry breaking remains. This is communicated to the particles of the MSSM through (i) higher-dimensional terms in the K\"ahler potential (\textit{Planck-scale-mediated supersymmetry breaking}, PMSB) (ii) \textit{anomaly-mediated supersymmetry breaking} (AMSB) and (iii) couplings to the $B$$-$$L$ multiplet which receives a supersymmetry-breaking mass splitting at tree level (\textit{gauge-mediated supersymmetry breaking}, GMSB), see e.g.~\cite{Martin:1997ns} for an overview. While PMSB will play a crucial role for the SM squarks and sleptons and for the mass parameters of the Higgs sector, the standard model gauginos will only receive a loop-suppressed AMSB contribution. At low energies, the particle spectrum thus resembles the results obtained in pure gravity mediation (PGM)~\cite{Ibe:2006de,Ibe:2011aa,ArkaniHamed:2012gw,Ibe:2012hu},\footnote{See also Ref.~\cite{Babu:2015xba} for a related discussion.} with the overall scale of the spectrum determined by the inflationary observables (which determine the value of the FI parameter $\xi$), see Eqs.~\eqref{eq:xifaL} and \eqref{eq:mumfm32}. In this section, we discuss the mass spectrum of our model during and after inflation, and discuss consequences for early-Universe cosmology, including reheating, leptogenesis, dark matter and the production of topological defects.


\subsection{Particle spectrum}
\label{subsec:spectrum}


\subsubsection*{Stabilization of squarks and sleptons during and after inflation}


The total scalar mass of an MSSM matter field $Q_i$ with gauge charge $q_i$ is given by 
\begin{align}
m_{0,i}^2 = \mathcal{C}^2\left[ 
-q_i\,m_D^2 - a_{ii}\, \Delta m_0^2 + q_i^2 m_{\textrm{gm}}^2 + m_R^2 \right] \,,
\label{eq:mssm_masses}
\end{align}
where the first term denotes the D-term-induced mass present only during inflation, the second term is a tree-level supergravity contribution induced by higher-dimensional operators in the K\"ahler potential, the third term is the gauge-mediated SUSY-breaking contribution and the fourth term is the gravity-induced mass (present only during inflation). Assuming that the MSSM sector couples to the supersymmetry-breaking sector via higher-dimensional operators in the Jordan frame (see Eq.~\eqref{eq:Ftot}),
\begin{align}
\Omega \supset X^\dagger X + Q_i^\dagger Q_i + 
\frac{a_{ii}}{M_*^2}\, Q_i^\dagger Q_i\, X^\dagger X +
\mathcal{O}\left(M_{*}^{-4}\right) \,,
\label{eq:Kmssm}
\end{align}
yields
\begin{align}
V_F^J \supset 
- a_{ii}\,\frac{\mu^4}{M_{*}^2} |\tilde{q}_i|^2  =
- 3\,a_{ii}\, \frac{m_{3/2}^2 M_\text{Pl}^2}{M_*^2} \left|\tilde{q}_i\right|^2  \,,
\end{align}
and hence $\Delta m_0^2 = 3\,m_{3/2}^2 M_\text{Pl}^2/M_*^2$. The quantities $m_D$ and $m_R$ have been introduced in Eqs.~\eqref{eq:VDMSSM} and Eq.~\eqref{eq:mR}, respectively. 
$m_\text{gm}$ denotes the mass contribution obtained through gauge mediation, with the dominant
(one-loop) effect arising due to the mass splitting within the $B$$-$$L$ gauge multiplet~\cite{Intriligator:2010be}
\begin{equation}
 m_\text{gm}^2 \simeq  \frac{g^2}{32 \pi^2} \, m_V^2 \ln \left( \frac{m_c^2 m_V^6}{m_{\tilde a}^8}\right)\,,
\end{equation}
with $m_V$, $m_c$ and $m_{\tilde a}$ given in section~\ref{subsec:iyit}. This contribution is clearly subdominant compared to the tree-level D-term contribution during inflation.
The stabilization of the MSSM scalars during inflation ($m_{0,i}^2 \gg H^2$) requires $a_{ii} < 0$ and
\begin{align}
|a_{ii}| \, \Delta m_0^2 & \gg  q_i\,m_D^2 - q_i^2 m_{\textrm{gm}}^2 -  H_J^2 \simeq q_i \, m_D^2 \,,
\end{align}
which for $|q_i| \simeq |q_0| \simeq  |a_{ii}| \simeq 1$ implies
\begin{equation}
 M_* \ll \frac{m_{3/2} M_{\rm Pl}}{m_D} \sim 10^{16}~\text{GeV} \left( \frac{0.1}{g} \right) \,,
\end{equation}
a value relatively close to the dynamical scale~$\Lambda$.

After the end of inflation, only the second and third terms in Eq.~\eqref{eq:mssm_masses} remain, leading to very heavy MSSM squarks and sleptons,
\begin{equation}
m_{0,i} \simeq (- a_{ii})^{1/2} \, \Delta m_0 > 10^{14}~\text{GeV} \, \left(\frac{g}{0.1} \right) \left(\frac{\sqrt{\xi}}{10^{15}~\text{GeV}} \right) \,. 
\end{equation}


\subsubsection*{MSSM gauginos}
\label{sec:gauginos}


Similar to the squarks and sleptons of the previous subsection, the $\mu$ and $B$ parameters of the MSSM receive tree-level supergravity contributions. Consequently, the heavy Higgs scalars and the Higgsinos obtain masses of ${\cal O}(m_{3/2})$. For the MSSM gauginos on the other hand tree-level supergravity contributions are strongly suppressed (since our supersymmetry-breaking field $X$ is not a total SM singlet)~\cite{Ibe:2006de,Ibe:2011aa,ArkaniHamed:2012gw,Ibe:2012hu}. The dominant contributions are thus obtained at one loop trough anomaly mediation and (in the case of binos and winos) through Higgsino threshold effects. As detailed in Ref.~\cite{Ibe:2012hu} in the context of PGM, these are generically both of the same order $( \sim g_a^2 \, m_{3/2} /(16 \pi^2) )$, where $g_a$ is the respective SM gauge coupling. Depending on the size of the Higgsino threshold effects, either the wino or the bino can take the place of the lightest neutral MSSM particle\footnote{Due to their larger gauge coupling at low energy scales, the gluinos are typically significantly heavier.}\,---\,and hence of the dark matter candidate. In the following we will focus on the case where the Higgsino threshold contributions do not dominate over the AMSB contribution, rendering the wino the lightest supersymmetric particle (LSP). The motivation for this is twofold. First, this is the more likely scenario in PGM~\cite{Ibe:2012hu}. Second, due to its smaller annihilation cross section, a thermal bino LSP population with $m_{\tilde b} \gtrsim 300~\text{GeV}$ leads to the overproduction of dark matter~\cite{Olive:1989jg,Griest:1989zh}. In combination with the bounds set by ATLAS~\cite{Aaboud:2017bac} and CMS~\cite{Sirunyan:2017cwe}, this excludes a thermal bino as a viable dark matter candidate.

A particularly interesting situation arises if the two contributions to the wino mass are tuned to very similar values, leading to a cancellation of these two terms and hence to a wino mass which can be arranged to be much lighter than its generic mass scale $m_{3/2} /(16 \pi^2)$. For example, in the notation of Ref.~\cite{Ibe:2012hu} this is achieved for $-\mu_H \simeq - 2 B \simeq m_{3/2}$ (for $\tan \beta = v_u/v_d \simeq 1$). Such a fine-tuning might be  justified from the anthropological requirement of dark matter, see Sec.~\ref{sec:DM}. Additional contributions to the gaugino masses may arise from threshold and anomaly-mediated corrections from heavy vector matter multiplets charged under $SU(2)_L$ and/or threshold corrections from the F terms of flat directions in KSVZ-type axion models~\cite{Harigaya:2013asa}. In this case, these contributions would also play a role in tuning
the wino mass.


\subsubsection*{Particles beyond the MSSM}
\label{subsec:BMSSM}


The masses of all remaining degrees of freedom are set by the dynamical scale $\Lambda$,
effectively decoupling these particles from low-energy physics.


In the supersymmetry-breaking sector, the only degree of freedom  which is present in the low-energy effective theory of the IYIT model is the pseudomodulus $X$, which acts as the Polonyi field of SUSY breaking, see Sec.~\ref{subsec:polonyi}. The dominant mass contribution after the end of inflation arises from its one-loop effective  potential,
\begin{equation}
m_X \simeq 0.02 \, \lambda^2 \Lambda \sim 10^{14}~\text{GeV} \,.
\end{equation}


The masses of the scalar fields of the inflation sector can be obtained from the scalar potential in Eq.~\eqref{eq:V}. For $q/(q_0 \xi) > 0$, the field $\phi$ obtains a VEV of $\langle | \phi|^2 \rangle = q_0/q \, \xi$ after the end of inflation.\footnote{{If $q$ or $q_0 \xi$ has opposite sign, the roles of $\phi$ and $\bar \phi$ are inverted.}} In this vacuum, the masses of the scalar degrees of freedom are given as
\begin{align}
m_{\bar\phi} = m_s & = 10^{14}~\text{GeV} \, \left(\frac{q_0}{q}\right)^{1/2} \, \left( \frac{\kappa}{0.1} \right) \left( \frac{\sqrt{\xi}}{10^{15}~\text{GeV}} \right), \\
 m_{| \phi|} & = 10^{14}~\text{GeV} \, (q_0 \, q)^{1/2} \, \left( \frac{g}{0.1} \right) \left( \frac{\sqrt{\xi}}{10^{15}~\text{GeV}} \right) \,,
\end{align}
where $m_{| \phi|}$ denotes the mass of the radial degree of freedom of $ \phi$ in the true vacuum. The fermionic DOFs contained in $S$ and $\bar \Phi$ form a Dirac fermion with a mass term $W_{\bar \phi s} = \kappa \langle \phi \rangle$. Similarly, the fermionic DOF of $\Phi$ pairs up with the fermionic DOF $\lambda$ from the $B$$-$$L$ gauge multiplet to form a Dirac gaugino of mass $\sqrt{2}\,q g\langle \phi \rangle$. Hence, after inflation, all components of the $B$$-$$L$ gauge multiplet obtain large masses of order $m_F$ or $m_V$, see also the discussion towards the end of Sec.~\ref{subsec:iyit}.

If the IYIT sector is fully sequestered from the inflaton sector, the angular degree of freedom arg($\phi$) remains massless. The reason for this is that although $B$$-$$L$ is spontaneously broken by $\langle M_\pm \rangle \neq 0$, there is a remaining accidental global $U(1)$ symmetry in the superpotential, associated with $ \phi \mapsto \exp(i \alpha)  \phi$, $\bar \phi \mapsto \exp(- i \alpha) \bar \phi$. This global symmetry is spontaneously broken at the end of inflation when $\phi$ acquires a VEV. We then have two spontaneously broken $U(1)$ symmetries, this global $U(1)$ symmetry and the gauged $U(1)_{B-L}$. The angular degree of freedom arg($\phi$) and the field $\varphi$ in the IYIT sector (see discussion around Eq.~\eqref{eq:Stuckel}) form two massless scalar modes. One linear combination of these couples to $A_\mu$. This is the $B$$-$$L$ Goldstone boson which is eaten by the $B$$-$$L$ gauge multiplet and is gauged away in unitary gauge. The orthogonal linear combination is the Goldstone boson of the global $U(1)$ symmetry. Assuming a full sequestering between the IYIT sector and the inflation sector, the $B$$-$$L$ breaking contained in the IYIT sector is not communicated to the inflaton sector and this Goldstone boson remains massless.
The spontaneous breaking of the global $U(1)$ symmetry at the end of inflation would lead to the formation of cosmic strings (see Sec.~\ref{subsec:strings}), in contradiction with observations. The discussion above, however, immediately reveals how to resolve this issue. If we drop the assumption of a complete sequestering between the IYIT and the inflaton sector,  $B$$-$$L$ breaking can be communicated to the $\Phi$ multiplet, rendering the Goldstone boson of the global $U(1)$ symmetry massive. In Sec.~\ref{subsec:strings} we will achieve this by introducing higher-dimensional operators in the K\"ahler potential coupling $\Phi$ to $Z_-$ and/or $M_-$. These operators will come with small coefficients (respecting the level of sequestering necessary to prevent the waterfall fields from obtaining too large masses) and will explicitly break the global $U(1)$ symmetry (in agreement with general arguments that no exact global symmetries should exist in any theory of quantum gravity~\cite{Banks:2010zn}). As we will see below, this leads to a mass for the Goldstone boson of the order of the Hubble scale during inflation.

Note that this situation is crucially different than in standard DHI, where the VEV of the waterfall field $\phi$ spontaneously breaks a local $U(1)$ symmetry at the end of inflation, triggering the super-Higgs mechanism: there, the complex phase is ``eaten'' by the $U(1)$ gauge boson, providing the longitudinal degree of freedom for the massive vector field. In our case, however, the $U(1)_{B-L}$ vector boson is already massive, having absorbed the corresponding degree of freedom from the meson multiplets. 


\subsection[The cosmology of a \texorpdfstring{$B$$-$$L$}{B-L} phase transition]
{The cosmology of a \texorpdfstring{\boldmath{$B$$-$$L$}}{B-L} phase transition}
\label{subsec:phasetransition}


D-term hybrid inflation ends in a phase transition in which the waterfall field $ \phi$, charged under $B$$-$$L$, obtains a vacuum expectation value. Assigning $B$$-$$L$ charge $q = -2$ to $\Phi$, it can couple to the right-handed neutrinos $N_i$ (which carry $B$$-$$L$ charge +1) in the seesaw extension of the MSSM, $W \supset \frac{1}{2}\, h_{ij}\, \Phi\, N_i N_j$. In fact, when gauging $B$$-$$L$, the introduction of three right-handed neutrinos is the simplest way to ensure anomaly cancellation. Once the waterfall field $ \phi$ obtains a VEV, this generates the Majorana mass matrix for the right-handed neutrinos, $M_{ij} = h_{ij} \langle \phi \rangle$.

A very similar cosmological phase transition was studied in the case of F-term hybrid inflation in Refs.~\cite{Buchmuller:2010yy,Buchmuller:2011mw,Buchmuller:2012wn,Buchmuller:2013dja}. It was shown that this phase transition can set the initial conditions for the hot early Universe: With the energy initially stored in oscillations of the waterfall field (as well as in degrees of freedom created in tachyonic preheating), this energy is transferred to the thermal bath through the decay into right-handed neutrinos (which obtain their mass from the coupling to the $B$$-$$L$ breaking waterfall field). In the course of this process, both thermal and nonthermal processes generate a lepton asymmetry, which, after conversion into a baryon asymmetry through sphaleron processes, can explain the baryon asymmetry observed today. Using a coupled set of Boltzmann equations, Refs.~\cite{Buchmuller:2010yy,Buchmuller:2011mw,Buchmuller:2012wn,Buchmuller:2013dja} provide a time-resolved picture of the entire reheating and leptogenesis process.

We expect this overall picture to also hold in our model. There are, however, a few differences in the details of the phase transition. Contrary to~\cite{Buchmuller:2010yy,Buchmuller:2011mw,Buchmuller:2012wn,Buchmuller:2013dja}, in the DHI model presented here (i)  $B$$-$$L$ is broken already during inflation, (ii) there is a tree-level mass splitting in the $B$$-$$L$ multiplet, (iii) supersymmetry is broken in the true vacuum with $m_{3/2} \sim 10^{12}\cdots10^{13}~\text{GeV}$ (iv) there is an additional pseudoscalar degree of freedom in the waterfall sector,\footnote{This relatively light degree of freedom  in the waterfall sector has a decay rate into right-handed neutrinos of $\Gamma \simeq h^2/(4 \pi)  H^2 = {\cal O}(\text{GeV})$ for typical values of the Yukawa coupling of $h \simeq 10^{-5}$. It thus decays into the SM thermal bath before the onset of BBN and does not create any cosmological problems.\smallskip} 
which in the analysis of \cite{Buchmuller:2012wn} plays the role of the $B$$-$$L$ Goldstone boson and (v) we allow here for smaller values of the $B$$-$$L$ gauge coupling. Consequently, no (local) cosmic strings are formed at the end of inflation, the $B$$-$$L$ multiplet is not produced in tachyonic preheating and the gravitino is too heavy to be a dark matter candidate as in~\cite{Buchmuller:2012wn}. Hence, while we expect the same sequence of events as in~\cite{Buchmuller:2010yy,Buchmuller:2011mw,Buchmuller:2012wn,Buchmuller:2013dja}, leading to successful leptogenesis for a mass of the lightest right-handed neutrino above about $10^{10}$~GeV and a reheating temperature of about $T_{RH} \gtrsim 10^8$~GeV, the above-mentioned differences require a detailed study to verify these expectations. This is beyond the scope of the current paper. 


\subsection{Particle candidates for dark matter}
\label{sec:DM}


The high reheating temperature expected in our model (see above) implies an abundant production of gravitinos~\cite{Jeong:2012en}. Since these gravitinos are very heavy, their decay temperature
\begin{equation}
T_{3/2} \simeq 1.5 \times 10^8~\text{GeV}\,  \left( \frac{g_*(T_{RH})}{80} \right)^{-1/4} \left( \frac{m_{3/2}}{10^{12}~\text{GeV}} \right)^{3/2} \,,
\end{equation}
is much larger than the temperature of BBN ($\sim$~MeV). The gravitinos will thus decay into MSSM gauginos (the lightest particles in our MSSM spectrum) before the onset of BBN, a well-known solution to the classical gravitino problem~\cite{Weinberg:1982zq,Gherghetta:1999sw,Ibe:2004tg}. If one of the gauginos (in our setup the wino, see Sec.~\ref{sec:gauginos}) is sufficiently light, so that its freeze-out temperature, $T_f \sim m_{\tilde w}/28$, is lower than the gravitino decay temperature $T_{3/2}$, then its relic abundance will be set by the usual thermal freeze-out contribution.\footnote{On the other hand, if $T_f > T_{3/2}$ the LSP abundance will be dominated by the nonthermal contribution from gravitino decay. The gravitino abundance in turn receives contributions from thermal production, production from the decay of the Polonyi field and production through oscillations of a field in the inflaton sector~\cite{Jeong:2012en}. In particular the thermal production~\cite{Bolz:2000fu} and the decay of $ \phi$ to two gravitinos through a supergravity coupling present when $\langle K_{\phi} \rangle \simeq \langle \phi \rangle \neq 0$~\cite{Kawasaki:2006gs} yield large gravitino abundances and thus overclose the Universe for the large LSP masses consistent with $T_f > T_{3/2}$.}
This occurs for $m_{\tilde w} \lesssim 4 \times 10^9~\text{GeV} \, (m_{3/2}/(10^{12}~\text{GeV}))^{3/2}$, 
with the correct relic density  obtained for $m_{\tilde w} \simeq 2.7$~TeV~\cite{Hisano:2006nn,Cirelli:2007xd}. This value is much smaller than the generic gaugino mass scale $m_{3/2}/(16 \pi^2)$, but may be achieved by fine-tuning the anomaly mediation and Higgsino threshold corrections (see Sec.~\ref{sec:gauginos}). Without such fine-tuning, the strongly enhanced LSP abundance would lead to an overclosure of the Universe. Such a fine-tuning may therefore be justified by anthropological arguments. {Alternatively, one can take the gauginos to be at their natural scale $m_{3/2}/(16 \pi^2)$ and invoke R-parity breaking to ensure a sufficiently fast decay of the LSP into the SM degrees of freedom~\cite{Barbier:2004ez}. In this case, the question of the nature of dark matter remains open and may, e.g., be addressed by the QCD axion.}


\subsection{Topological defects}
\label{subsec:strings}


In standard DHI, the angular degree of freedom of the waterfall field $ \phi$ is massless, protected by the $U(1)$ gauge symmetry of DHI. Consequently, cosmic strings are formed at the end of standard DHI. This is known as the cosmic string problem of DHI, since the nonobservation of cosmic strings in the CMB~\cite{Ade:2013xla}, together with constraints on the spectral index, essentially exclude the entire parameter space. The setup we propose here is crucially different. The $U(1)_{B-L}$ symmetry is already broken during inflation by the meson VEVs $\langle M_\pm \rangle \neq 0$, and no local cosmic strings are formed at the end of inflation. There is instead an accidental global symmetry which is not expected to be exact, see Sec.~\ref{subsec:BMSSM}. {We can express this by adding higher-dimensional operators\footnote{Here, we assumed $q_0 = -1$. Similar terms (also involving the inflaton field) can be written down for $q_0 = -2$.\smallskip} in the K\"ahler potential of the type $M_{\rm Pl} K \supset \epsilon^K_{ZZ} Z_-^2 \Phi$, $\epsilon^K_{MM} M_-^2 \Phi$, $\epsilon^K_{MZ} M_- Z_- \Phi$, all supplemented by their complex conjugate. Here the parameters $\epsilon^K_{IJ}$ are expected to be exponentially small, respecting the sequestering between the inflation and the IYIT sector. By means of a K\"ahler transformation these holomorphic terms can be equivalently considered as terms in the superpotential, $W \supset  W_0 \, \epsilon^K_{ZZ} \, Z_-^2  \Phi/M_{\rm Pl}^3$, etc. Taking into account the vacuum expectation values for the scalar and auxiliary (F-term) components in $Z_\pm, M_\pm, \bar \Phi, \Phi$ and $S$, this leads to linear terms in the scalar potential for the waterfall fields. Schematically,
\begin{equation}
V = V_0 - c \left( \phi +  \phi^* \right) + \tilde m^2 \left|\phi\right|^2 + \frac{\tilde \lambda}{4} \left| \phi \right|^4\,,
\label{eq:Vcs}
\end{equation}
where $\tilde \lambda = 2 g^2 q^2$ denotes the self-coupling of the waterfall field, $\tilde m^2$ is its (inflaton-dependent) mass and $c \sim \epsilon_{IJ}^K  \Lambda^3 m_{3/2}/M_{\rm Pl}$ is determined by the higher-dimensional operators mentioned above.

To study cosmic string formation,\footnote{We thank the authors of Ref.~\cite{Evans:2017bjs} for very helpful discussions on this point.} 
we consider the system close to the end of inflation, just when Eq.~\eqref{eq:Vcs} develops a local maximum. At this point, the local minimum of the potential is given by $| \phi| = 2 (c/\tilde \lambda)^{1/3}$. The phase of the local minimum is set by the phase of $c$ and we will take it to be zero in the following. The mass of the canonically normalized radial degree of freedom $\alpha$ in this local minimum is given by $m_\alpha^2 =\frac{1}{2} (c^2 \tilde \lambda)^{1/3}$. 

To avoid the production of cosmic strings, we require that quantum fluctuations of the angular degree of freedom (see e.g.~\cite{Mukhanov:1990me})  cannot overcome the barrier at $\alpha/(\sqrt{2} \langle  \phi \rangle) = \pi$, i.e.,
\begin{equation}
\langle \delta a^2 \rangle = \frac{H}{3 m_\alpha} \left( \frac{H}{2 \pi}\right)^2 \ll 2 \langle \phi \rangle^2 \pi^2\,.
\end{equation}
This leads to 
\begin{equation}
c^{1/3} \gg 0.05 \, \tilde \lambda^{1/6} \, H_\text{inf} \,, 
\label{eq:boundc}
\end{equation}
with $H_\text{inf}$ denoting the Hubble scale at the end of inflation.
In this paper, we will consider
\begin{equation}
\tilde c \equiv c/H_\text{inf}^3 \gtrsim 1\,,
\end{equation}
safely satisfying Eq.~\eqref{eq:boundc} but also ensuring that the fluctuations in the radial direction are small compared to the position of the local minimum and that the decay rate of the angular component $\alpha$ in the true vacuum is not significantly smaller than the decay rate of the radial component. For the couplings in the K\"ahler and/or superpotential, this implies
\begin{equation}
\epsilon^K_{IJ} \gtrsim \left(\frac{H}{\Lambda} \right)^3 \sim \left(\frac{g \Lambda}{M_{\rm Pl}} \right)^3 \sim g^3 \, 10^{-9} \,, \qquad  W_0 \, \epsilon^K_{IJ}/M_{\rm Pl}^3  \sim  \epsilon^K_{IJ} \left( \frac{m_{3/2}}{M_{\rm Pl}} \right) \sim g^3 \, 10^{-16} \,,
\end{equation}
in good agreement with our sequestering ansatz.}

However, scenarios with a richer phenomenology are possible. For example, imagine that a term such as $K \supset  \Phi Z_- Z_-/M_{\rm Pl}$ is forbidden by an additional discrete symmetry, which in turn is explicitly broken by Planck-suppressed operators of even higher dimension. If this explicit breaking is of a suitable size, unstable domain walls will form~\cite{Vilenkin:1981zs,Gelmini:1988sf,Larsson:1996sp}. A similar situation has been discussed for the QCD axion~\cite{Sikivie:1982qv,Chang:1998tb}, see also \cite{Choi:2009jt,Dias:2014osa,Ringwald:2015dsf}. The decaying domain walls will emit energy in the form of gravitational waves. The resulting gravitational-wave spectrum depends mainly on two parameters, the tension $\sigma$ of the domain walls and their annihilation temperature $T_\text{ann}$. For the high energy scales present in our model, the resulting stochastic gravitational-wave background might be within the sensitivity reach of upcoming advanced LIGO runs~\cite{Nakayama:2016gxi,Saikawa:2017hiv}, depending on the details of the discrete symmetry (breaking).


\section{Conclusions: A unified model of the early Universe}
\label{sec:conclusions}


In this paper, we constructed a phenomenologically viable SUGRA model
of hybrid inflation in which reheating proceeds
via the $B$$-$$L$ phase transition.
We focused on the case of D-term inflation to avoid the notorious
complications associated with the inflaton tadpole term in F-term inflation
(see Eq.~\eqref{eq:tadpole}).
This tadpole term turns F-term inflation into a two-field model,
potentially spoils the slow-roll motion of the inflaton field,
and creates a false vacuum at large field values.
D-term inflation does not, by contrast, involve any inflaton tadpole term,
which prevents one from running into these problems.


The first part of our paper contains the details of our model-building effort
(see Sec.~\ref{sec:model}).
To meet all theoretical and phenomenological constraints,
our model combines the following three features:


(i) The vacuum energy driving D-term inflation is provided
by a \textit{Fayet-Iliopoulos} (FI) D term.
We assume that this D term is dynamically generated in the
hidden SUSY-breaking sector~\cite{Domcke:2014zqa}.
Our construction involves two steps.
First, we suppose that SUSY breaking in the hidden sector is
accomplished by the dynamics of a strongly coupled supersymmetric gauge theory.
To be specific, we employ the \textit{Izawa-Yanagida-Intriligator-Thomas}
(IYIT) model~\cite{Izawa:1996pk,Intriligator:1996pu}, which represents the
simplest vector-like model of dynamical SUSY breaking.
Thanks to the strong interactions in the IYIT sector,
our model does not require any hard dimensionful input scales.
All mass scales (including the SUSY breaking scale itself) turn out to be
related to the dynamical scale in the IYIT sector, $\Lambda_{\rm dyn}$.
The dynamical scale $\Lambda_{\rm dyn}$ is in turn generated via the quantum
effect of dimensional transmutation, just like the confinement scale of QCD.
The second step in our construction consists in promoting a global
axial $U(1)_A$ flavor symmetry in the IYIT sector to a weakly gauged local
$U(1)_{B-L}$ symmetry.
The SUSY-breaking dynamics in the IYIT sector then result in an effective
FI parameter $\xi$ that is determined by the vacuum expectation values of
certain moduli in the effective theory at low energies.


Our dynamically generated FI term has a number of
interesting properties.
(1) Being an effective field-dependent FI parameter, it can be consistently
coupled to supergravity.
In this sense, it differs from genuinely constant
FI parameters whose coupling to supergravity always requires
an exact global continuous symmetry.
(2) The generation of field-dependent FI parameters typically results
in dangerous flat directions in the scalar potential.
In our case, all moduli are, however, automatically stabilized by a
large mass term in the superpotential that is induced by the SUSY-breaking F term.
(3) The generation of effective FI parameters is always accompanied by
the spontaneous breaking of the underlying Abelian gauge symmetry.
This is also the case in our model where the generation of $\xi$ spontaneously
breaks $B$$-$$L$ in the IYIT sector.
We use this fact to our advantage and communicate the breaking of $B$$-$$L$
to the visible sector via marginal couplings in the K\"ahler potential.
This allows us to prevent the formation of dangerous cosmic strings during
the $B$$-$$L$ phase transition at the end of inflation.
(4) The magnitude of the FI parameter and, hence, the energy scale
of inflation are related to the SUSY breaking scale.
This unifies the dynamics of inflation and SUSY breaking.
The energy scales of both phenomena derive from the dynamical
scale, $\Lambda_{\rm dyn} = e^{-S_{\rm inst}/b_{\rm hid}}M_{\rm Pl}$, which is generated
via nonperturbative dynamics in the infrared.%
\footnote{Here, $S_{\rm inst} = 8\pi^2 / g_{\rm hid}^2$ denotes the
nonperturbative instanton action in the IYIT sector (see Eq.~\eqref{eq:Lambdadyn}).
$g_{\rm hid}$ and $b_{\rm hid}$ stand for the gauge
coupling constant and the beta function coefficient of the IYIT gauge group,
respectively.}
This explains the exponential hierarchy between the energy scales
of inflation and SUSY breaking on the one hand and the Planck scale 
on the other hand.


(ii) We assume that the natural SUGRA description of our model
corresponds to an embedding into the standard Jordan frame with canonically
normalized kinetic terms for all complex scalar fields.
From the Einstein-frame perspective, this corresponds to a noncanonical K\"ahler
geometry based on a K\"ahler potential of the sequestering type.
This sequestering structure allows us to control the soft 
masses in the visible MSSM sector independently of the corresponding soft
masses in the inflation sector.
In fact, thanks to our choice of the K\"ahler potential,
the inflation sector sequesters from the IYIT sector such that none
of the fields in the inflation sector obtains a large soft mass. 
This is a crucial requirement for a successful $B$$-$$L$ phase transition.
Otherwise, i.e., without sequestering, large soft masses in the inflation sector
would keep the waterfall fields stabilized at the origin and, thus, remove
the tachyonic instability in the scalar potential.
At the same time, we introduce a higher-dimensional coupling between the
visible MSSM sector and the IYIT sector in the K\"ahler potential to stabilize
all MSSM sfermions during and after inflation.
Again, this is an important ingredient of our model.
Without any extra stabilization mechanism, the MSSM sfermions would destabilize
the FI term in the D-term scalar potential and inflation would prematurely end in
the wrong vacuum.
For the purposes of this paper, we do not specify the high-energy 
origin of the additional coupling between the visible MSSM sector
and the IYIT sector in the K\"ahler potential. 
However, it would be interesting to study different scenarios for the
possible origin of these operators in future work.


We caution that one should not attribute too much
meaning to our choice to work in Jordan-frame supergravity.
The formulation of our model in the language of Jordan-frame supergravity
should rather be regarded as a placeholder for a hypothetical completion of 
our model at high energies.
Possible candidates for an ultraviolet completion of our model that feature
an appropriate K\"ahler geometry include models of
extra dimensions, strongly coupled conformal field theories, no-scale supergravity,
and string theory.
Again, any further speculations into this direction are left for future work.


(iii) Our third and final assumption consists in an approximate shift symmetry in
the direction of the inflaton field in the K\"ahler potential.
Such an approximate shift symmetry is a popular tool in many SUGRA models
of inflation.
As usual, it helps us to suppress dangerously large
SUGRA corrections to the inflaton mass and, hence, solve the SUGRA eta problem.
In our case, the most dangerous such correction, $m_R^2 = R_J/6$,
stems from the nonminimal coupling between the inflaton field to the
Ricci scalar in the Jordan frame, $R_J$.
This effect can be completely suppressed by an exact
shift symmetry\,---\,which is, however, not feasible in our model, since
the superpotential of D-term inflation inherently breaks any shift symmetry.
But this is not a problem.
As we are able to show, also an approximate shift symmetry
manages to adequately suppress all dangerous SUGRA corrections.
On top of that, we can use the fact that the inflaton shift symmetry
must be slightly broken to adjust our prediction for the scalar spectral
index $n_s$.
The amount of shift symmetry breaking in the K\"ahler potential
is quantified by a parameter $\chi$. 
By choosing this additional parameter appropriately, we can reach
agreement between our prediction for $n_s$ and the current best-fit
value reported by PLANCK.
Here, an interesting special case arises if $\chi$ is zero
at tree level and only radiatively generated because of the shift-symmetry-breaking
Yukawa coupling in the superpotential,
$\chi = \chi_{1\ell} = \kappa^2/\left(16\pi^2\right)$.
We are able to demonstrate that even this minimal scenario
allows to successfully reproduce the CMB data.
In this case, the constraints $A_s \simeq A_s^{\rm obs}$ and $n_s \simeq n_s^{\rm obs}$
fix all free parameters of our model (see Fig.~\ref{fig:parameterscan2}),
\begin{align}
\Lambda \sim 3 \times 10^{15}\,\textrm{GeV} \,, \quad
\kappa \sim 10^{-3} \,, \quad g \sim 10^{-3} \,, \quad
\chi_{1\ell} \sim 10^{-8} \,.
\end{align}


In summary, we conclude that the above three assumptions allow us
to solve five problems of $B$$-$$L$ D-term inflation:
(i)   Our FI term can be consistently coupled to supergravity;
(ii)  we avoid the formation of cosmic strings at the end of inflation;
(iii) all MSSM sfermions are sufficiently stabilized during and after inflation;
(iv)  we do not encounter any SUGRA eta problem; and 
(v)   our prediction for $n_s$ is in agreement with the PLANCK data.
This is a highly nontrivial success of our model.


A further outcome of our model is a unified picture of the
early Universe (see Secs.~\ref{sec:inflation} and \ref{sec:discussion}).
Provided that we include the right couplings in the superpotential,
the $B$$-$$L$ phase transition at the end of inflation generates large 
Majorana masses for a number of right-handed neutrinos.
This sets the stage for baryogenesis via leptogenesis as well as for the
generation of small standard model neutrino masses via the seesaw mechanism.
In the end, our model therefore unifies the scales of dynamical SUSY breaking,
inflation, and spontaneous $B$$-$$L$ breaking.
All of these scales derive from the dynamical scale $\Lambda_{\rm dyn} = 4\pi\Lambda$
in the IYIT sector.
We also find that, in order to reproduce the amplitude of the
scalar power spectrum, the reduced dynamical scale $\Lambda$ must take a value
close to the GUT scale,
\begin{align}
\label{eq:scales}
\Lambda_{\rm SUSY} \sim \Lambda_{\rm inf} \sim \Lambda_{B-L} \sim \Lambda
\sim 5\times 10^{15}\,\textrm{GeV} \sim \Lambda_{\rm GUT} \,.
\end{align}
This is another highly nontrivial result of our analysis.
Before confronting our model with the experimental CMB data,
we did not need to make any assumption about the numerical value of $\Lambda$.
All of the above scales only become fixed once we require that our model 
yields the correct value for the scalar spectral amplitude.
This leads to the interesting physical question of which scale in Eq.~\eqref{eq:scales}
actually corresponds to a fundamental scale and which scale is only a derived quantity.
Is there, e.g., an anthropic reason for the observed scalar spectral amplitude
which then determines the SUSY breaking scale?
Or is the SUSY breaking scale rather determined by the scale of $R$ symmetry breaking
in the superpotential and the requirement of a nearly vanishing cosmological constant? 
Or should one instead regard the GUT scale as the most fundamental
scale which then fixes all other scales?
All of these questions are beyond the scope of this work.
But we feel that our model provides an interesting starting point for further 
studies in this direction.
An important task would be to embed our model into a full-fledged GUT scenario
that explains the occurrence of the GUT scale in Eq.~\eqref{eq:scales}.


A central prediction of our model is that supersymmetry is broken
at a high energy scale.
The naturalness of the electroweak scale is therefore lost.
This sacrifice is, however,
compensated for by the unification of the dynamics of SUSY breaking and inflation.
One of our key messages therefore is that pushing the SUSY breaking scale
to very high values is not necessarily just a loss.
A high SUSY breaking scale also represents an opportunity for novel
ideas such as those presented in this paper.
In the end, supersymmetry might play a different role in nature 
than previously expected.
Following the arguments presented this paper, it is conceivable that supersymmetry's actual
purpose is not to ensure the stability of the electroweak scale,
but to provide the right conditions for successful inflation!


In this paper, we only touched upon the implications of a
high SUSY breaking scale for the particle spectrum of the
MSSM and more work in this direction is certainly needed.
In particular, one should reevaluate in more detail how the running of
the standard model coupling constants can be matched with the coupling
constants in the MSSM provided that supersymmetry is broken at energies
close to the GUT scale.
This matching of the low-energy parameters with their counterparts at high
energies is sensitive to important experimental input data, such as the top
quark mass $m_t$ and the strong coupling constant $\alpha_s$. 
Given the current experimental uncertainty in these observables,
we expect that it should actually not pose any problem to successfully match the
standard model to our high-scale scenario.
On top of that, large threshold corrections due to nonuniversal
soft masses at high energies may help us to achieve a successful matching
(see~\cite{Ellis:2017erg} for a recent analysis).
In fact, given our treatment of the MSSM soft masses
(see Eq.~\eqref{eq:mssm_masses}), large nondegeneracies
in the sparticle mass spectrum at high energies are quite likely.
Moreover, one should reevaluate in more detail under which conditions
our high-scale scenario is compatible with the idea of gauge coupling
unification. 
Again, such an analysis would be sensitive to the experimental input data
at low energies.
In addition, it would also depend on the details of the anticipated unification
scenario. 
We are, however, confident on general grounds that it should be feasible
to realize gauge coupling unification in our model.
After all, unification is also possible in entirely nonsupersymmetric scenarios.
We therefore expect that supersymmetry, despite the large value of its breaking
scale, will only help in achieving gauge coupling unification~\cite{Ellis:2015jwa}.


In conclusion, we find that our model provides a consistent
cosmological scenario that unifies five different phenomena:
(i)   dynamical supersymmetry breaking at a high energy scale,
(ii)  viable D-term hybrid inflation in supergravity,
(iii) spontaneous $B$$-$$L$ breaking at the GUT scale,
(iv)  baryogenesis via leptogenesis, and
(v)   standard model neutrino masses due to the type-I seesaw mechanism.
Our model is built around a strongly coupled hidden
sector, which puts it on a sound theoretical footing.
We do not need to make any \textit{ad hoc} assumptions about
the dimensionful parameters in our model.
Instead, all important mass scales are related to the dynamical scale of
the strong interactions in the hidden sector.
Thanks to its precise parameter relations, our model is
therefore well suited to be used as a basis for further explicit calculations.
It would, e.g., be worthwhile to study the reheating process after inflation
in greater detail and determine the corresponding implications for the
spectrum of gravitational waves. 
Similarly, a more comprehensive study of the MSSM particle spectrum 
and its consequences for dark matter would be desirable. 
The analysis in the present paper should only be regarded as a first step.
It served the purpose to illustrate our main point: \textit{the fact
that SUSY breaking close to the GUT scale might be the key to a
unified picture of particle physics and cosmology}.
This is a fascinating observation and we are excited to see where it will
lead us in the future.
One possibility is that it will eventually cause a paradigm shift in our understanding
of SUSY's role in the physics of the early Universe.
High-scale SUSY breaking might be the driving force behind inflation!




\subsubsection*{Acknowledgements}


The authors wish to thank T.~T.~Yanagida
for inspiring discussions at Kavli IPMU at the University of Tokyo
in the early stages of this project in 2014.
The authors are grateful to W.~Buchm\"uller, J.~Evans,
B.~v.~Harling, M.~Peloso and L.~Witkowski for helpful
remarks and discussions at various stages of this project.
This project has received support from the European Union's Horizon 2020
research and innovation programme under the Marie Sk\l{}odowska-Curie grant
agreement No.\ 674896 (K.\,S.).



\appendix



\section{Technicalities: Supergravity in the Einstein/Jordan frame}
\label{app:frames}


Our model is based on a particular embedding into supergravity.
We assume that the coupling to gravity is most naturally described in a Jordan frame
where all scalar kinetic terms are canonically normalized (see Sec.~\ref{sec:model}).
At the same time, we wish to perform a standard slow-roll analysis of the inflationary
dynamics (see Sec.~\ref{sec:inflation}), which requires a reformulation
of our model in the Einstein frame.
To facilitate the transition between these two different frames, this appendix
provides a dictionary that allows one to translate back and forth between
the two different formulations of our model.


\subsection{Bosonic action}


In the usual Einstein frame, the purely bosonic action
of our model takes the following form,%
\footnote{Some authors in the literature distinguish between
Einstein-frame and Jordan-frame quantities by labeling them with indices
$E$ and $J$, respectively.
We will, by contrast, not use any particular label for quantities in the Einstein frame
and merely label quantities in the Jordan frame with an index $J$.
This will slightly simplify our notation.}
\begin{align}
\label{eq:Sbos}
S_{\rm bos} = \int d^4 x\, \sqrt{-g} \left[\frac{1}{2}\,M_{\rm Pl}^2\,R -
\mathcal{K}_{\bar{\imath}j}\,g^{\mu\nu} D_\mu \phi_{\bar{\imath}}^*\, D_\nu \phi_j
- \frac{1}{4}\,F_{\mu\nu}F^{\mu\nu} - V\right] \,.
\end{align}
Here, $M_{\rm Pl}$ denotes the reduced Planck Mass,
$M_{\rm Pl} \simeq 2.44 \times 10^{18}\,\textrm{GeV}$;
$g$ is the determinant of the Einstein-frame spacetime metric $g_{\mu\nu}$;
$R$ is the Ricci scalar constructed from $g_{\mu\nu}$;
$g^{\mu\nu}$ stands for the inverse of the metric $g_{\mu\nu}$;
the fields $\phi_i$ represent the complex scalar fields in our model;
$D_\mu$ denotes the usual gauge-covariant derivative;
$F_{\mu\nu}$ is the field strength tensor of the Abelian $B$$-$$L$ vector field; and
$V$ represents the total scalar potential in the Einstein frame.
As evident from Eq.~\eqref{eq:Sbos}, the scalar fields $\phi_i$ couple
to gravity only via the inverse spacetime metric $g^{\mu\nu}$.
This corresponds to the case of minimal coupling.
At the same time, the scalar fields exhibit a nontrivial (K\"ahler)
geometry in field space.
This is accounted for by the K\"ahler metric $\mathcal{K}$
which multiplies the scalar kinetic terms in Eq.~\eqref{eq:Sbos}.
The K\"ahler metric $\mathcal{K}$ is defined as
the Hessian of the real-valued K\"ahler potential $K$,
\begin{align}
\mathcal{K}_{\bar{\imath}j} =
\frac{\partial^2 K}{\partial \phi_{\bar{\imath}}^* \partial \phi_j} \,.
\end{align}
In our model, the K\"ahler potential $K$ is not canonical, such that
$\mathcal{K}_{\bar{\imath}j} \neq \delta_{\bar{\imath}j}$.
The scalar kinetic terms (and, hence, the scalar fields
themselves) are, thus, not canonically normalized in the Einstein frame.


To obtain the equivalent of Eq.~\eqref{eq:Sbos} in the Jordan frame,
we need to perform a Weyl rescaling,
\begin{align}
\label{eq:gmnJ}
g_{\mu\nu}^J = \mathcal{C}^2\,g_{\mu\nu} \,, \quad
g^{\mu\nu}_J = \mathcal{C}^{-2}\,g^{\mu\nu} \,, \quad
\sqrt{-g_J} = \mathcal{C}^4 \sqrt{-g} \,, \quad
\mathcal{C} = \left(-\frac{3 M_{\rm Pl}^2}{\Omega}\right)^{1/2} \,,
\end{align}
where $\mathcal{C}$ is known as the conformal factor.
The frame function $\Omega$ is an arbitrary real negative function of the complex scalars,
$\Omega = \Omega\left(\phi_i,\phi_{\bar{\imath}}^*\right) < 0$.
Each choice for $\Omega$ defines a separate Jordan frame.
In Sec.~\ref{sec:model}, we make a particular choice for $\Omega$,
demanding the following relation to the K\"ahler potential,
\begin{align}
\label{eq:Omega}
\Omega = -3M_{\rm Pl}^2\, \exp\left(-\frac{K}{3M_{\rm Pl}^2}\right) 
\qquad\Leftrightarrow\qquad 
K = -3M_{\rm Pl}^2 \,\ln\left(-\frac{\Omega}{3M_{\rm Pl}^2}\right) \,.
\end{align}
This results in what may be regarded as the standard Jordan frame.
In light of the relation in Eq.~\eqref{eq:Omega}, the frame function 
$\Omega$ has two possible interpretations.
In the curved superspace approach to old minimal
supergravity~\cite{Stelle:1978ye,Ferrara:1978em}, $\Omega$ can be
identified as the generalized kinetic energy on curved
superspace, while in the superconformal approach to old minimal
supergravity~\cite{Cremmer:1982en,Kugo:1982mr}, $\Omega$ can be
identified as the prefactor of the kinetic term of the chiral compensator superfield.
Eq.~\eqref{eq:Omega} allows to relate the partial derivatives of the K\"ahler
potential to the partial derivatives of the frame function,
\begin{align}
\label{eq:KiOi}
K_i & = \frac{\partial K}{\partial \phi_i} =
\mathcal{C}^2\,\frac{\partial \Omega}{\partial \phi_i} = 
\mathcal{C}^2\,\Omega_i \,, \quad
K_{\bar{\imath}} = \frac{\partial K}{\partial \phi_{\bar{\imath}}^*} =
\mathcal{C}^2\,\frac{\partial \Omega}{\partial \phi_{\bar{\imath}}^*} = 
\mathcal{C}^2\,\Omega_{\bar{\imath}} \,.
\end{align}
Similarly, we are able to express the K\"ahler metric $\mathcal{K}$
in terms of derivatives of the frame function,
\begin{align}
\label{eq:Komega}
\mathcal{K}_{\bar{\imath}j} = \mathcal{C}^2\,\omega_{\bar{\imath}j} \,, \quad
\omega_{\bar{\imath}j} = \Omega_{\bar{\imath}j}
- \frac{\Omega_{\bar{\imath}}\,\Omega_j}{\Omega} \,, \quad
\Omega_{\bar{\imath}j} =
\frac{\partial^2 \Omega}{\partial \phi_{\bar{\imath}}^* \partial \phi_j} \,.
\end{align}
Here, we introduced $\omega$ as a rescaled field-space metric
that is determined by the derivatives of the frame function and that is
equivalent to the K\"ahler metric up to the rescaling factor $\mathcal{C}^2$.
Eq.~\eqref{eq:Komega} automatically implies a similar relation between
the respective inverse metrics, $\mathcal{K}^{-1}$ and $\omega^{-1}$,
\begin{align}
\left(\mathcal{K}^{-1}\right)_{i\bar{\jmath}} = 
\mathcal{C}^{-2}\left(\omega^{-1}\right)_{i\bar{\jmath}} \,.
\end{align}


We now apply the Weyl transformation in Eq.~\eqref{eq:gmnJ} to the Einstein-frame
action in Eq.~\eqref{eq:Sbos}.
This yields the purely bosonic action of our
model in the Jordan frame (see \cite{Ferrara:2010yw} for more details),
\begin{align}
\label{eq:SbosJ}
S_{\rm bos}^J = \int d^4 x\, \sqrt{-g_J} \left[\frac{1}{2}\left(-\frac{\Omega}{3}\right)R_J -
\Omega_{\bar{\imath}j}\,g_J^{\mu\nu} D_\mu \phi_{\bar{\imath}}^*\, D_\nu \phi_j
+ \Omega\,\mathcal{A}_\mu \mathcal{A}^\mu - \frac{1}{4}\,F_{\mu\nu}F^{\mu\nu} - V^J\right] \,.
\end{align}
In view of this action, several comments are in order:


(i) The frame function $\Omega$ depends on the scalar fields of our model.
The Einstein-Hilbert term (i.e., the kinetic term for the metric that is proportional
to the Ricci scalar $R_J$) therefore becomes field-dependent. 
Or in other words, the scalar fields  are now nonminimally coupled to gravity.


(ii) The Planck mass squared in Eq.~\eqref{eq:Sbos} is now replaced by $-\Omega/3$.
This indicates that the square root of $-\Omega/3$ should be interpreted
as the effective field-dependent Planck mass in the Jordan frame,
\begin{align}
\label{eq:MPlJ}
M_{\rm Pl}^J = \left(-\frac{\Omega}{3}\right)^{1/2} \qquad\Leftrightarrow\qquad
M_{\rm Pl} = \mathcal{C}\,M_{\rm Pl}^J\,,
\end{align}
which is consistent with the fact that all Jordan-frame mass scales $m_J$ pick
a factor $\mathcal{C}$ when transforming from the Jordan frame
to the Einstein frame, $m = \mathcal{C}\,m_J$.
From this perspective, the conformal factor $\mathcal{C}$
turns out to be nothing but the ratio of the two respective
Planck masses, $\mathcal{C} = M_{\rm Pl}/M_{\rm Pl}^J$.


(iii) The K\"ahler metric $\mathcal{K}$ in Eq.~\eqref{eq:Sbos} is now replaced
by the Hessian of $\Omega$.
In our model, we choose $\Omega$ such that it only contains canonical
as well as purely holomorphic/antiholomorphic terms,
\begin{align}
\label{eq:OAnsatz}
\Omega = -3 M_{\rm Pl}^2 + \delta_{\bar{\imath}j}\, \phi_{\bar{\imath}}^* \phi_j +
\left[J\left(\phi_i\right) + \textrm{h.c.}\right] \,,
\end{align}
where $J$ is an arbitrary holomorphic function.
The canonical terms,
$\Omega \supset \delta_{\bar{\imath}j}\, \phi_{\bar{\imath}}^* \phi_j$,
lead to nonminimal couplings between the complex scalars and
the Ricci scalar $R_J$ that are invariant under a classical conformal symmetry.
These conformal couplings can be disturbed by a nonzero function $J$ which 
explicitly breaks the conformal symmetry.
Irrespective of whether $J = 0$ or $J \neq 0$,
Eq.~\eqref{eq:OAnsatz} leads to
\begin{align}
\label{eq:Odelta}
\Omega_{\bar{\imath}j} = \delta_{\bar{\imath}j} \,.
\end{align}
In this case, we obtain the following expressions for the rescaled field-space
metric $\omega$ and its inverse,
\begin{align}
\omega_{\bar{\imath}j} = \delta_{\bar{\imath}j}
- \frac{\Omega_{\bar{\imath}}\,\Omega_j}{\Omega} \,, \quad
\left(\omega^{-1}\right)_{i\bar{\jmath}} = \delta_{i\bar{\jmath}}
+ \frac{1}{\varrho}\frac{\Omega_{\bar{\imath}}\,\Omega_j}{\Omega} \,, \quad
\varrho = 1 - \frac{\Omega_{\bar{k}} \Omega_{\vphantom{\bar{k}}k}}{\Omega} \,.
\end{align}
Here, the dimensionless parameter $\varrho$ functions as a measure
for the amount of conformal symmetry breaking in the frame function $\Omega$. 
In the conformal limit, $J \rightarrow 0$, it simply reduces to the conformal
factor squared, $\rho \rightarrow \mathcal{C}^2$.
In this sense, $\varrho$ plays a similar role as the reduced kinetic function of 
the inflation field, $f$, defined in Eq.~\eqref{eq:f}.
In fact, in our concrete model, one can show that $\varrho = \mathcal{C}^2\left(1-f\right)$.


(iv) In the Jordan frame, the scalar kinetic terms receive additional contributions
from the bosonic part of the auxiliary SUGRA gauge field $\mathcal{A}_\mu$.
This is accounted for by the third term on the right-hand
side of Eq.~\eqref{eq:SbosJ}.
The auxiliary field $\mathcal{A}_\mu$ can be eliminated after solving
its equation of motion,
\begin{align}
\mathcal{A}_\mu = \frac{1}{\Omega}\,\textrm{Im}\left\{\Omega_i\,D_\mu \phi_i\right\} \,.
\end{align}
This solution illustrates that the $\mathcal{A}^2$ term in Eq.~\eqref{eq:SbosJ}
is only relevant as long as we are interested in the dynamics of angular degrees
of freedom, i.e., the complex phases of the complex scalars $\phi_i$.
This is, however, not the case.
In our model, inflation occurs along the real direction
of the complex inflation field $s$. 
The auxiliary field $\mathcal{A}_\mu$ therefore vanishes and 
the $\mathcal{A}^2$ term in Eq.~\eqref{eq:SbosJ} can be ignored.
Together with Eq.~\eqref{eq:Odelta}, $\mathcal{A}_\mu = 0$ leads to canonically normalized
kinetic terms for all complex scalar fields in our model.
This is an important result and the main motivation for our ansatz in 
Eq.~\eqref{eq:OAnsatz}.


Combining our above results, the action in Eq.~\eqref{eq:SbosJ}
can be simplified to the following expression,
\begin{align}
S_{\rm bos}^J = \int d^4 x\, \sqrt{-g_J} \left[\frac{1}{2}\left(M_{\rm Pl}^J\right)^2 R_J -
\delta_{\bar{\imath}j}\,g_J^{\mu\nu} D_\mu \phi_{\bar{\imath}}^*\, D_\nu \phi_j
- \frac{1}{4}\,F_{\mu\nu}F^{\mu\nu} - V^J\right] \,.
\end{align}
This is the starting point for the analysis of our model in the Jordan frame.
Thus far, we have not commented on the relation between the
potentials $V$ and $V^J$.
We will do this now in the next section.


\subsection{Scalar potential}


The total scalar potential in the Einstein frame, $V$, has mass dimension four.
On general grounds, this implies the following universal relation
to the total scalar potential in the Jordan frame, $V^J$,
\begin{align}
\label{eq:VVJ}
V = \mathcal{C}^4\, V^J \,, \quad V^J = \mathcal{C}^{-4}\, V \,,
\end{align}
which holds at tree level as well as at the loop level.
Eq.~\eqref{eq:VVJ} implies the following useful relations,
\begin{align}
\label{eq:VijVJij}
V_i & = \frac{\partial V}{\partial \phi_i} =
\mathcal{C}^4 \left(V_i^J - \frac{2}{\Omega}\, \Omega_i\, V^J\right) \,, \quad
V_i^J = \frac{\partial V^J}{\partial \phi_i} =
\mathcal{C}^{-4} \left(V_i + \frac{2}{\Omega}\, \Omega_i\, V\right) \,,
\\ \nonumber
V_{\bar{\imath}j} & = \frac{\partial^2 V}{\partial \phi_{\bar{\imath}}^* \partial \phi_j} =
\mathcal{C}^4 \:\:\,\left[V_{\bar{\imath}j}^J - \frac{2}{\Omega}
\left(\Omega_{\bar{\imath}} V_j^J + \Omega_j V_{\bar{\imath}}^J\right) - \frac{2}{\Omega}
\left(\omega_{\bar{\imath}j} - 2\,\frac{\Omega_{\bar{\imath}}\Omega_j}{\Omega}\right) V^J\right]
\,, \\ \nonumber
V_{\bar{\imath}j}^J & = \frac{\partial^2 V^J}{\partial \phi_{\bar{\imath}}^* \partial \phi_j} =
\mathcal{C}^{-4} \left[V_{\bar{\imath}j}^{\phantom{J}} + \frac{2}{\Omega}
\left(\Omega_{\bar{\imath}} V_j^{\phantom{J}} + \Omega_j V_{\bar{\imath}}^{\phantom{J}}\right)
+ \frac{2}{\Omega} \left(\omega_{\bar{\imath}j} + 2\,\frac{\Omega_{\bar{\imath}}\Omega_j}{\Omega}
\right) V^{\phantom{J}}\right] \,.
\end{align}
Together, Eqs.~\eqref{eq:VVJ} and \eqref{eq:VijVJij} illustrate
that a Minkowski vacuum in the Einstein frame ($V = V_i = 0$) also corresponds to a Minkowski
vacuum in the Jordan frame ($V^J = V_i^J = 0$), and vice versa.


In the next step, we shall become more specific and discuss
the individual contributions to $V$ and $V^J$, respectively.
The Einstein-frame potential consists of the usual F-term
and D-term contributions,
\begin{align}
V = V_F + V_D \,.
\end{align}
To begin with, let us focus on the F-term scalar potential
(see~\cite{Wess:1992cp} for more details),
\begin{align}
\label{eq:VFFF}
V_F = F_{\bar{\imath}}^* \mathcal{K}_{\bar{\imath}j} F_j 
- 3 \exp\left(\frac{K}{M_{\rm Pl}^2}\right) \frac{\left|W\right|^2}{M_{\rm Pl}^2} \,.
\end{align}
Here, the $F_i$ and $F_{\bar{\imath}}^*$ stand for the generalized F terms
in supergravity and their complex conjugates,
\begin{align}
F_i = - \exp\left[\frac{K}{2\,M_{\rm Pl}^2}\right]
\left(\mathcal{K}^{-1}\right)_{i\bar{\jmath}}
\left(\mathcal{D}W\right)_{\bar{\jmath}}^* \,, \quad
F_{\bar{\imath}}^* = - \exp\left[\frac{K}{2\,M_{\rm Pl}^2}\right]
\left(\mathcal{K}^{-1}\right)_{\bar{\imath}j}^*
\left(\mathcal{D}W\right)_j \,,
\end{align}
where $\mathcal{D}W$ denotes the K\"ahler-covariant
derivative of the superpotential on the K\"ahler manifold,
\begin{align}
\left(\mathcal{D}W\right)_i =
\frac{\partial W}{\partial \phi_i} + \frac{\partial K}{\partial \phi_i}
\frac{W}{M_{\rm Pl}^2} \,.
\end{align}
The F-term potential in the Jordan frame is
given as $V_F^J = \mathcal{C}^{-4}\,V_F$.
This can be rewritten as follows,
\begin{align}
\label{eq:VFJ}
V_F^J = \left(W_i - 3\,W \frac{\Omega_i}{\Omega}\right) \left(\omega^{-1}\right)_{i\bar{\jmath}}
\left(W_{\bar{\jmath}}^* - 3\,W^* \frac{\Omega_{\bar{\jmath}}}{\Omega}\right) +
\frac{9}{\Omega} \left|W\right|^2\,.
\end{align}
which underlines the similarity between the role of the inverse metric $\omega^{-1}$
in the Jordan frame and the role of the inverse K\"ahler
metric $\mathcal{K}^{-1}$ in the Einstein frame.
With our ansatz for the frame function $\Omega$ in Eq.~\eqref{eq:OAnsatz},
$V_F^J$ can be further simplified to the following
compact expression (see, e.g., \cite{Buchmuller:2012ex}),
\begin{align}
\label{eq:VFJVJ0DVJ}
V_F^J =  V_F^0 + \Delta V_F^J \,, \quad
V_F^0 = W_i W_{\bar{\imath}}^* \,, \quad
\Delta V_F^J =  \frac{1}{\varrho\,\Omega} \left|W_i\, \Omega_{\bar{\imath}} - 3\, W\right|^2 \,.
\end{align}


Eq.~\eqref{eq:VFJVJ0DVJ} illustrates a remarkable effect.
Provided that the scalar kinetic terms in the Jordan frame are canonically normalized,
$V_F^J$ splits into two separate contributions\,---\,where the first
contribution, $V_F^0$, is nothing but the ordinary F-term scalar potential
in global supersymmetry and the second contribution, $\Delta V_F^J$,
represents an additive SUGRA correction.
Eq.~\eqref{eq:VFJVJ0DVJ} provides the basis for the calculation
of the F-term scalar potential in our model (see Eqs.~\eqref{eq:VF0} and \eqref{eq:DeltaVJ}).
We first calculate the F-term scalar potential in the Jordan frame according
to Eq.~\eqref{eq:VFJVJ0DVJ}.
Then, we convert our result from the Jordan frame to the Einstein frame
making use of the general relation in Eq.~\eqref{eq:VVJ},
\begin{align}
V_F = \mathcal{C}^4 \left(W_i W_{\bar{\imath}}^* + 
\frac{1}{\varrho\,\Omega} \left|W_i\, \Omega_{\bar{\imath}} - 3\, W\right|^2 \right) \,.
\end{align}
Computing $V_F$ via this detour is considerably easier than a direct calculation
starting with Eq.~\eqref{eq:VFFF}.


The SUGRA correction $\Delta V_F^J$ in Eq.~\eqref{eq:VFJVJ0DVJ} is directly
proportional to the mass scales that appear in the superpotential. 
In our model, these mass scales correspond to the F-term SUSY breaking scale $\mu$,
the effective inflaton-dependent mass of the waterfall fields, $\kappa\left<S\right>$,
and the $R$-symmetry-breaking constant $w_0$ (see Eq.~\eqref{eq:Wtot}).
All of these mass scales are responsible for the explicit breaking of
superconformal symmetry.
Conversely, this means that, if the superpotential does not exhibit
any explicit mass scales, the SUGRA correction $\Delta V_F^J$ must vanish.
This is exactly what happens in the class of \textit{canonical superconformal
supergravity} (CSS) models studied in~\cite{Ferrara:2010in}.
These models are based on the frame function in Eq.~\eqref{eq:OAnsatz}
with the holomorphic function $J$ set to zero.
Moreover, they exhibit a purely cubic superpotential, such that
$\Delta V_F^J = 0$.
One generic feature of CSS models therefore is that their Jordan-frame scalar
potential coincides with the scalar potential in global supersymmetry,
\begin{align}
\textrm{CSS models:} \qquad
W = \frac{1}{3}\,\lambda_{ijk}\, \Phi_i \Phi_j \Phi_k \,, \quad
V_F^J = V_F^0 \,, \quad V_D^J = V_D^0 \,, \quad \Delta V_F^J = 0 \,.
\end{align}
Note that this statement applies to the total scalar potential,
including the D-term scalar potential.


In our model, the D-term scalar potential in the Einstein frame, $V_D$, is
given by~\cite{Wess:1992cp}
\begin{align}
\label{eq:VDD}
V_D = \frac{1}{2}\,D^2 \,,
\end{align}
where $D$ denotes the auxiliary component of the $B$$-$$L$ vector multiplet $V$.
In writing down Eq.~\eqref{eq:VDD}, we assumed a canonical gauge-kinetic
function for the $B$$-$$L$ vector field, $f_V = 1$, and absorbed the gauge coupling
constant $g$ into the definition of $D$.
On-shell, the auxiliary $D$ field can be replaced by
\begin{align}
\label{eq:D}
D = - g\,q_i\,K_i\, \phi_i \,.
\end{align}
In the language of K\"ahler geometry, this is equivalent to the Killing potential
of the $U(1)_{B-L}$ isometry of our K\"ahler manifold.
Together, Eqs.~\eqref{eq:VDD} and \eqref{eq:D} result
in the following expression for $V_D$,
\begin{align}
V_D = \frac{g^2}{2} \left(q_i\,K_i\,\phi_i\right)^2 \,.
\end{align}
This result can be easily translated into the Jordan frame by making use 
of Eqs.~\eqref{eq:KiOi} and \eqref{eq:VVJ},
\begin{align}
V_D^J = \mathcal{C}^{-4}\,V_D = \frac{g^2}{2} \left(q_i\,\Omega_i\,\phi_i\right)^2 \,.
\end{align}
In our model, all complex scalars with nonzero gauge charge $q_i$ appear with
a canonical term in $\Omega$.
Just like in the class of CSS models, $V_D^J$ therefore
obtains the same form as in global supersymmetry,
\begin{align}
q_i\,\Omega_i = q_i\,\phi_{\bar{\imath}}^* \qquad\Rightarrow\qquad
V_D^J = \frac{g^2}{2} \left(q_i \left|\phi_i\right|^2 \right)^2 = V_D^0 \,.
\end{align}
This means in turn that the D-term scalar potential in the Einstein frame can
be written as
\begin{align}
V_D = \frac{g^2}{2}\,\mathcal{C}^4\left(q_i \left|\phi_i\right|^2 \right)^2 \,.
\end{align}


Combining all of our above results, we conclude that $V$ and $V^J$ are given as follows
in our model,
\begin{align}
\label{eq:Vfinal}
V = \mathcal{C}^4\,V^J \,, \quad
V^J = W_i W_{\bar{\imath}}^* + \frac{g^2}{2} \left(q_i \left|\phi_i\right|^2 \right)^2 + 
\frac{1}{\varrho\,\Omega} \left|W_i\, \Omega_{\bar{\imath}} - 3\, W\right|^2 \,.
\end{align}
This result is the starting point for our calculation of the inflaton potential
in Sec.~\ref{subsec:potential}.


\subsection{Scalar mass parameters}


Eq.~\eqref{eq:Vfinal} allows us to derive
useful expressions for the scalar mass parameters $m_{ab}^J$ in the Jordan frame.%
\footnote{Here, $a$ and $b$ represent collective indices that encompass all scalar
fields $\phi_i$ as well as their complex conjugates $\phi_{\bar{\imath}}^*$.
In the following, the symbol $z_a$ will therefore either denote the field $\phi_i$
for some index $i$ or the field $\phi_{\bar{\imath}}^*$ for some index $\bar{\imath}$.}
The scalar mass matrix is given by the Hessian of
the scalar potential, $\left(m_{ab}^J\right)^2 = V_{ab}^J$.
As a consequence of the simple structure of $V^J$, the scalar masses, thus,
split into two contributions: the ordinary masses
in global supersymmetry, $m_{ab}^0$, as well as
additive corrections in supergravity, $\Delta m_{ab}$,
\begin{align}
\left(m_{ab}^J\right)^2 & =
\left(m_{ab}^0\right)^2 + \Delta m_{ab}^2 \,, \quad
\left(m_{ab}^0\right)^2 = \frac{\partial^2}{\partial z_a \partial z_b}
\left(V_F^0 + V_D^0 \right) \,, \quad 
\Delta m_{ab}^2 = \frac{\partial^2}{\partial z_a \partial z_b}\,\Delta V_F^J \,.
\end{align}
In the case of scalar fields that only appear with a canonical term in the frame function,
$\Omega \supset \left|\phi_i\right|^2$, the corrections $\Delta m_{ab}^2$
take a particularly simple form.
Based on our result in Eq.~\eqref{eq:VFJVJ0DVJ}, we find
\begin{align}
\Delta m_{\bar{\imath}j}^2 = 
\frac{\left(W_{j k\vphantom{\bar{k}}} \Omega_{\bar{k}} - 2\, W_j\right)
\left(W_{\bar{\imath}\bar{\ell}}^* \Omega_{\ell\vphantom{\bar{\ell}}}^{\phantom{*}}
- 2\, W_{\bar{\imath}}^*\right)}{\varrho\,\Omega} \,, \quad
\Delta m_{ij}^2 = \frac{
\big(W_{i j k\vphantom{\bar{k}}} \Omega_{\bar{k}} - W_{ij}\big)
\big(W_{\bar{\ell}}^* \Omega_{\ell\vphantom{\bar{\ell}}}^{\phantom{*}} - 3\, W^*\big)}
{\varrho\,\Omega} \,,
\end{align}
and similarly for the respective conjugate parameters,
$\Delta m_{i\bar{\jmath}}^2 = \big(\Delta m_{\bar{\imath}j}^2\big)^*$ and 
$\Delta m_{\bar{\imath}\bar{\jmath}}^2 = \big(\Delta m_{ij}^2\big)^*$.
The diagonal entries of the scalar mass matrix therefore obtain
the following compact form,
\begin{align}
\Delta m_{\bar{\imath}i}^2 = 
\frac{\left|W_{i k\vphantom{\bar{k}}} \Omega_{\bar{k}} - 2\, W_i\right|^2}{\varrho\,\Omega} \,.
\end{align}
In addition to the masses encoded in the scalar potential, the complex
scalar fields acquire further, effective masses from their nonminimal
coupling to $R_J$ in the Jordan-frame action (see Eq.~\eqref{eq:SbosJ}),
\begin{align}
\left(m_{ab}^R\right)^2 = \zeta R_J\,
\frac{\partial^2 \Omega}{\partial z_a \partial z_b} \,, \quad \zeta = \frac{1}{6} \,.
\end{align}
All scalar fields with a canonical kinetic function, thus,
receive a universal gravity-induced mass $m_R$,
\begin{align}
m_R^2 = \zeta R_J \,.
\end{align}


In summary, the entries of the total effective mass matrix in the Jordan frame, $M_J^2$,
read as follows,
\begin{align}
\label{eq:MJ2}
\left(M_J^2\right)_{ab} = 
\left(m_{ab}^0\right)^2 + \Delta m_{ab}^2 + \left(m_{ab}^R\right)^2 \,.
\end{align}
In our model, all scalar fields are canonically normalized by construction.
The eigenvalues of the matrix $M_J^2$ therefore directly correspond to the physical scalar
mass eigenvalues in the Jordan frame.
The situation is more complicated in the Einstein frame, where the scalar fields
parametrize the target space of a nonlinear sigma model (see Eq.~\eqref{eq:Sbos}). 
There, the scalar fields are \textit{a priori} not canonically normalized
which makes it more difficult to find the physical mass eigenvalues.
Without reference to the Jordan frame, the computation of the scalar mass spectrum
in the Einstein frame requires two steps.
First, one has to perform a field transformation that renders all fields canonically
normalized.
Then, one has to calculate the mass eigenvalues of these canonically normalized fields
as usual.
Our result in Eq.~\eqref{eq:MJ2},
however, allows us to bypass this complicated procedure.
Instead, we can simply make use of the universal scaling behavior
of physical mass scales when transforming back and forth between the Jordan frame
and the Einstein frame.
According to this scaling behavior, we know that the total effective scalar mass matrix in
the Einstein frame, $M^2$, must obtain the following form,
\begin{align}
M_{ab}^2 =  \mathcal{C}^2\left(M_J^2\right)_{ab} = \mathcal{C}^2\left[
\left(m_{ab}^0\right)^2 + \Delta m_{ab}^2 + \left(m_{ab}^R\right)^2\right] \,.
\end{align}
It would be interesting to check the validity of this result by means of
an explicit calculation in the Einstein frame.
Such a task is, however, beyond the scope of this paper and left for future work.


More details on the mass parameters for all fields with nonzero
spin (i.e., the fermions, vector boson, and gravitino in our model)
can be found in the literature.
The relevant expressions in the Jordan frame are spelled out in~\cite{Ferrara:2010yw},
while the standard Einstein-frame results are listed, e.g., in~\cite{Wess:1992cp}.


\subsection{Slow-roll parameters}
\label{app:slowroll_parameters}


Finally, let us discuss the relation between the inflationary slow-roll parameters
in the Einstein frame, $\varepsilon$ and $\eta$,
and their counterparts in the Jordan frame, $\varepsilon_J$ and $\eta_J$.
The results derived in this section will enable us to use our
results for the scalar potential in the Jordan frame (see Sec.~\ref{sec:model})
as input for a standard slow-roll analysis of
the inflationary dynamics in the Einstein frame (see Sec.~\ref{sec:inflation}).


Let us consider the action of
the complex inflaton field $s$ in the Einstein frame (see Eq.~\eqref{eq:Sbos}),
\begin{align}
\label{eq:Sinf}
S_{\rm inf} = - \int d^4 x\, \sqrt{-g} \left[
\mathcal{N}^2\,\partial_\mu s^* \partial^\mu s + V\left(s\right)\right] \,, \quad
\mathcal{N} = \mathcal{K}_{s^*s}^{1/2} \,.
\end{align}
As can be seen from this action, the inflaton field is not canonically
normalized in the Einstein frame.
This is made explicit by the noncanonical normalization factor of
the inflaton kinetic term, $\mathcal{N} \neq 1$.
However, in our slow-roll analysis, we will have
to work with the canonically normalized field $\hat{s}$.
The field $\hat{s}$ can be constructed as a function of the field $s$
by solving the following differential equations,
\begin{align}
\label{eq:PDEs}
\frac{\partial\hat{\sigma}}{\partial \sigma} = \mathcal{N}\left(\sigma,\tau\right) \,, \quad 
\frac{\partial\hat{\tau}}{\partial\tau} = \mathcal{N}\left(\sigma,\tau\right) \,, \quad 
s = \frac{1}{\sqrt{2}}\left(\sigma + i\tau\right) \,, \quad
\hat{s} = \frac{1}{\sqrt{2}}\left(\hat{\sigma} + i\hat{\tau}\right) \,.
\end{align}
In terms of the canonically normalized field $\hat{s}$, the
action in Eq.~\eqref{eq:Sinf} obtains its standard form,
\begin{align}
S_{\rm inf} = - \int d^4 x\, \sqrt{-g} \left[
\partial_\mu \hat{s}^* \partial^\mu \hat{s} + V\left(\hat{s}\right)\right] \,, \quad
V\left(\hat{s}\right) \equiv V\left(s\left(\hat{s}\right)\right) \,.
\end{align}
This action is the starting point of our standard slow-roll analysis.
The slow-roll parameters in the Einstein frame,
$\varepsilon$ and $\eta$, are defined in terms of the usual partial derivatives
of the scalar potential,
\begin{align}
\varepsilon = \frac{M_{\rm Pl}^2}{2} \left(\frac{V'}{V}\right)^2 \,, \quad
\eta = M_{\rm Pl}^2\,\frac{V''}{V} \,, \quad 
V' = \frac{\partial V}{\partial\hat{\sigma}} \,, \quad
V'' = \frac{\partial^2V}{\partial\hat{\sigma}^2} \,.
\end{align}
Here, we assumed that inflaton occurs along the real component $\hat{\sigma}$
of the complex inflaton field $\hat{s}$.
In the next step, we rewrite these expressions making use of
the chain rule and the relations in Eq.~\eqref{eq:PDEs},
\begin{align}
V' = \frac{\partial \sigma}{\partial\hat{\sigma}} 
\frac{\partial V}{\partial\sigma} = \frac{V_{\sigma}}{\mathcal{N}} \,, \quad
V'' = \frac{\partial \sigma}{\partial\hat{\sigma}} 
\frac{\partial V'}{\partial\sigma} =
\frac{1}{\mathcal{N}}\left(\frac{V_{\sigma\sigma}}{\mathcal{N}} -
\frac{\mathcal{N}_\sigma}{\mathcal{N}}\frac{V_\sigma}{\mathcal{N}}\right) \,,
\end{align}
where $V_\sigma = \partial V/\partial\sigma$,
$V_{\sigma\sigma} = \partial^2V/\partial\sigma^2$, and
$\mathcal{N}_\sigma = \partial\mathcal{N}/\partial\sigma$.
With these definitions and relations, we find
\begin{align}
\label{eq:eetilde}
\varepsilon = \frac{\tilde{\varepsilon}}{\mathcal{N}^2} \,, \quad
\eta = \frac{\tilde{\eta} - 2 \left(\nu\tilde{\varepsilon}\right)^{1/2}}{\mathcal{N}^2} \,,
\end{align}
where we assumed that $V_\sigma > 0$ and $\mathcal{N}_\sigma > 0$.
The parameters $\tilde{\varepsilon}$ and $\tilde{\eta}$ represent what
can be referred to as the naive slow-roll parameters in the
Einstein frame, i.e., the slow-roll parameters that one would obtain
if one ignored the noncanonical normalization factor $\mathcal{N}$
in Eq.~\eqref{eq:Sinf}.
Meanwhile, the factor $\nu$ is an auxiliary slow-roll parameter that accounts
for the field dependence of the factor $\mathcal{N}$,
\begin{align}
\tilde{\varepsilon} = \frac{M_{\rm Pl}^2}{2} \left(\frac{V_\sigma}{V}\right)^2 \,, \quad
\tilde{\eta} = M_{\rm Pl}^2\,\frac{V_{\sigma\sigma}}{V} \,, \quad 
\nu = \frac{M_{\rm Pl}^2}{2} \left(\frac{\mathcal{N}_\sigma}{\mathcal{N}}\right)^2 \,.
\end{align}


The expressions in Eq.~\eqref{eq:eetilde} are now well suited to
establish a connection to the Jordan frame. 
The naive slow-roll parameters $\tilde{\varepsilon}$ and $\tilde{\eta}$
can be readily related to the Jordan-frame slow-roll parameters $\varepsilon_J$
and $\eta_J$ by employing the relations for the partial derivatives of
the scalar potential in Eq.~\eqref{eq:VijVJij},
\begin{align}
\label{eq:eetildeJ}
\tilde{\varepsilon} =
\left(\varepsilon_J^{1/2} -2\, \xi_J^{1/2}\right)^2  \,, \quad 
\tilde{\eta} = \eta_J +
12\,\xi_J - 8\left(\varepsilon_J\,\xi_J\right)^{1/2} - 2\,\zeta_J \,,
\end{align}
where we assumed again a positive potential gradient, $V_\sigma^J > 0$.
The Jordan-frame slow-roll parameters $\varepsilon_J$ and $\eta_J$
are defined in terms of the usual partial derivatives of the Jordan-frame
scalar potential,
\begin{align}
\varepsilon_J = \frac{M_{\rm Pl}^2}{2} \left(\frac{V_\sigma^J}{V^J}\right)^2 \,, \quad
\eta_J = M_{\rm Pl}^2\,\frac{V_{\sigma\sigma}^J}{V^J} \qquad \textrm{with} \qquad
V_\sigma^J = \frac{\partial V^J}{\partial\sigma} \,, \quad
V_{\sigma\sigma}^J = \frac{\partial^2V^J}{\partial\sigma^2} \,.
\end{align}
In Eq.~\eqref{eq:eetildeJ}, we also introduced the auxiliary slow-roll
parameters $\xi_J$ and $\zeta_J$ which account for the field dependence
of the frame function $\Omega$.
These slow-roll parameters are defined as follows,
\begin{align}
\xi_J^{1/2} = \frac{M_{\rm Pl}}{\sqrt{2}}\frac{\Omega_\sigma}{\Omega} \,, \quad
\zeta_J = M_{\rm Pl}^2\,\frac{\Omega_{\sigma\sigma}}{\Omega} \qquad \textrm{with} \qquad
\Omega_\sigma = \frac{\partial\Omega}{\partial\sigma} \,, \quad
\Omega_{\sigma\sigma} = \frac{\partial^2\Omega}{\partial\sigma^2} \,.
\end{align}
Unlike in the Einstein frame, we do not have to
distinguish between naive and actual slow-roll parameters in the Jordan frame.
This is because, in our model, all scalar fields are canonically
normalized in the Jordan frame by construction.
In the language of Eq.~\eqref{eq:eetilde}, this can be rephrased by saying that the
normalization factor of the inflaton kinetic term in the Jordan frame
is simply trivial, $\mathcal{N}_J = 1$.


Combining our results in Eqs.~\eqref{eq:eetilde} and \eqref{eq:eetildeJ},
we finally obtain the following relations,
\begin{align}
\label{eq:eeEJ}
\varepsilon & = \frac{1}{\mathcal{N}^2} \left(\varepsilon_J^{1/2} -2\, \xi_J^{1/2}\right)^2 \,,
\\ \nonumber
\eta & = \frac{1}{\mathcal{N}^2} \left[ \eta_J +
12\,\xi_J - 8\left(\varepsilon_J\,\xi_J\right)^{1/2} - 2\,\zeta_J 
- 2\,\nu^{1/2}\left(\varepsilon_J^{1/2} -2\, \xi_J^{1/2}\right)\right] \,.
\end{align}
This is an important result that holds in any model with an Einstein-frame 
action as in Eq.~\eqref{eq:Sinf}.
Eq.~\eqref{eq:eeEJ} is the starting point for our computation
of the Einstein-frame slow-roll parameters in Sec.~\ref{sec:inflation}.
We emphasize that computing $\varepsilon$ and $\eta$ according
to Eq.~\eqref{eq:eeEJ} is considerably easier than a brute-force calculation
in the Einstein frame.
In the Einstein frame, we would have to deal with a complicated
K\"ahler potential, a more complicated scalar potential, and
a noncanonically normalized inflaton field.
Eq.~\eqref{eq:eeEJ} allows us to circumvent these complications and
determine the parameters $\varepsilon$ and $\eta$ simply based on the derivatives
of the Jordan-frame scalar potential $V^J$ and the frame function $\Omega$.



\bibliographystyle{JHEP}
\bibliography{arxiv_2}{}

\providecommand{\href}[2]{#2}\begingroup\raggedright\begin{thebibliography}{100}

\bibitem{Aaboud:2017bac}
{\bf ATLAS} Collaboration, M.~Aaboud et~al., {\it {Search for squarks and
  gluinos in events with an isolated lepton, jets, and missing transverse
  momentum at $\sqrt{s}=13$ TeV with the ATLAS detector}},  {\em Phys. Rev.}
  {\bf D96} (2017), no.~11 112010, [\href{http://arxiv.org/abs/1708.08232}{{\tt
  arXiv:1708.08232}}].

\bibitem{Sirunyan:2017cwe}
{\bf CMS} Collaboration, A.~M. Sirunyan et~al., {\it {Search for supersymmetry
  in multijet events with missing transverse momentum in proton-proton
  collisions at 13 TeV}},  {\em Phys. Rev.} {\bf D96} (2017), no.~3 032003,
  [\href{http://arxiv.org/abs/1704.07781}{{\tt arXiv:1704.07781}}].

\bibitem{Aad:2012tfa}
{\bf ATLAS} Collaboration, G.~Aad et~al., {\it {Observation of a new particle
  in the search for the Standard Model Higgs boson with the ATLAS detector at
  the LHC}},  {\em Phys. Lett.} {\bf B716} (2012) 1--29,
  [\href{http://arxiv.org/abs/1207.7214}{{\tt arXiv:1207.7214}}].

\bibitem{Chatrchyan:2012xdj}
{\bf CMS} Collaboration, S.~Chatrchyan et~al., {\it {Observation of a new boson
  at a mass of 125 GeV with the CMS experiment at the LHC}},  {\em Phys. Lett.}
  {\bf B716} (2012) 30--61, [\href{http://arxiv.org/abs/1207.7235}{{\tt
  arXiv:1207.7235}}].

\bibitem{Aad:2015zhl}
{\bf ATLAS, CMS} Collaboration, G.~Aad et~al., {\it {Combined Measurement of
  the Higgs Boson Mass in $pp$ Collisions at $\sqrt{s}=7$ and 8 TeV with the
  ATLAS and CMS Experiments}},  {\em Phys. Rev. Lett.} {\bf 114} (2015) 191803,
  [\href{http://arxiv.org/abs/1503.07589}{{\tt arXiv:1503.07589}}].

\bibitem{Okada:1990vk}
Y.~Okada, M.~Yamaguchi, and T.~Yanagida, {\it {Upper bound of the lightest
  Higgs boson mass in the minimal supersymmetric standard model}},  {\em Prog.
  Theor. Phys.} {\bf 85} (1991) 1--6.

\bibitem{Okada:1990gg}
Y.~Okada, M.~Yamaguchi, and T.~Yanagida, {\it {Renormalization group analysis
  on the Higgs mass in the softly broken supersymmetric standard model}},  {\em
  Phys. Lett.} {\bf B262} (1991) 54--58.

\bibitem{Ellis:1990nz}
J.~R. Ellis, G.~Ridolfi, and F.~Zwirner, {\it {Radiative corrections to the
  masses of supersymmetric Higgs bosons}},  {\em Phys. Lett.} {\bf B257} (1991)
  83--91.

\bibitem{Ellis:1991zd}
J.~R. Ellis, G.~Ridolfi, and F.~Zwirner, {\it {On radiative corrections to
  supersymmetric Higgs boson masses and their implications for LEP searches}},
  {\em Phys. Lett.} {\bf B262} (1991) 477--484.

\bibitem{Haber:1990aw}
H.~E. Haber and R.~Hempfling, {\it {Can the mass of the lightest Higgs boson of
  the minimal supersymmetric model be larger than m(Z)?}},  {\em Phys. Rev.
  Lett.} {\bf 66} (1991) 1815--1818.

\bibitem{Giudice:1998xp}
G.~F. Giudice, M.~A. Luty, H.~Murayama, and R.~Rattazzi, {\it {Gaugino mass
  without singlets}},  {\em JHEP} {\bf 12} (1998) 027,
  [\href{http://arxiv.org/abs/hep-ph/9810442}{{\tt hep-ph/9810442}}].

\bibitem{Wells:2003tf}
J.~D. Wells, {\it {Implications of supersymmetry breaking with a little
  hierarchy between gauginos and scalars}},  in {\em {11th International
  Conference on Supersymmetry and the Unification of Fundamental Interactions
  (SUSY 2003), Tucson, Arizona, June 5-10, 2003}}, 2003.
\newblock \href{http://arxiv.org/abs/hep-ph/0306127}{{\tt hep-ph/0306127}}.

\bibitem{Wells:2004di}
J.~D. Wells, {\it {PeV-scale supersymmetry}},  {\em Phys. Rev.} {\bf D71}
  (2005) 015013, [\href{http://arxiv.org/abs/hep-ph/0411041}{{\tt
  hep-ph/0411041}}].

\bibitem{Hall:2009nd}
L.~J. Hall and Y.~Nomura, {\it {A Finely-Predicted Higgs Boson Mass from A
  Finely-Tuned Weak Scale}},  {\em JHEP} {\bf 03} (2010) 076,
  [\href{http://arxiv.org/abs/0910.2235}{{\tt arXiv:0910.2235}}].

\bibitem{ArkaniHamed:2004fb}
N.~Arkani-Hamed and S.~Dimopoulos, {\it {Supersymmetric unification without low
  energy supersymmetry and signatures for fine-tuning at the LHC}},  {\em JHEP}
  {\bf 06} (2005) 073, [\href{http://arxiv.org/abs/hep-th/0405159}{{\tt
  hep-th/0405159}}].

\bibitem{Giudice:2004tc}
G.~F. Giudice and A.~Romanino, {\it {Split supersymmetry}},  {\em Nucl. Phys.}
  {\bf B699} (2004) 65--89, [\href{http://arxiv.org/abs/hep-ph/0406088}{{\tt
  hep-ph/0406088}}]. [Erratum: \textit{Nucl. Phys.} \textbf{B706} (2005)
  487-487].

\bibitem{ArkaniHamed:2004yi}
N.~Arkani-Hamed, S.~Dimopoulos, G.~F. Giudice, and A.~Romanino, {\it {Aspects
  of split supersymmetry}},  {\em Nucl. Phys.} {\bf B709} (2005) 3--46,
  [\href{http://arxiv.org/abs/hep-ph/0409232}{{\tt hep-ph/0409232}}].

\bibitem{Arvanitaki:2012ps}
A.~Arvanitaki, N.~Craig, S.~Dimopoulos, and G.~Villadoro, {\it {Mini-Split}},
  {\em JHEP} {\bf 02} (2013) 126, [\href{http://arxiv.org/abs/1210.0555}{{\tt
  arXiv:1210.0555}}].

\bibitem{ArkaniHamed:2012gw}
N.~Arkani-Hamed, A.~Gupta, D.~E. Kaplan, N.~Weiner, and T.~Zorawski, {\it
  {Simply Unnatural Supersymmetry}},
  \href{http://arxiv.org/abs/1212.6971}{{\tt arXiv:1212.6971}}.

\bibitem{Hall:2011jd}
L.~J. Hall and Y.~Nomura, {\it {Spread Supersymmetry}},  {\em JHEP} {\bf 01}
  (2012) 082, [\href{http://arxiv.org/abs/1111.4519}{{\tt arXiv:1111.4519}}].

\bibitem{Hall:2012zp}
L.~J. Hall, Y.~Nomura, and S.~Shirai, {\it {Spread Supersymmetry with Wino LSP:
  Gluino and Dark Matter Signals}},  {\em JHEP} {\bf 01} (2013) 036,
  [\href{http://arxiv.org/abs/1210.2395}{{\tt arXiv:1210.2395}}].

\bibitem{Ibe:2006de}
M.~Ibe, T.~Moroi, and T.~T. Yanagida, {\it {Possible Signals of Wino LSP at the
  Large Hadron Collider}},  {\em Phys. Lett.} {\bf B644} (2007) 355--360,
  [\href{http://arxiv.org/abs/hep-ph/0610277}{{\tt hep-ph/0610277}}].

\bibitem{Ibe:2011aa}
M.~Ibe and T.~T. Yanagida, {\it {The Lightest Higgs Boson Mass in Pure Gravity
  Mediation Model}},  {\em Phys. Lett.} {\bf B709} (2012) 374--380,
  [\href{http://arxiv.org/abs/1112.2462}{{\tt arXiv:1112.2462}}].

\bibitem{Ibe:2012hu}
M.~Ibe, S.~Matsumoto, and T.~T. Yanagida, {\it {Pure Gravity Mediation with
  $\mathit{m_{3/2} = 10\textrm{--}100\,\textrm{TeV}}$}},  {\em Phys. Rev.} {\bf
  D85} (2012) 095011, [\href{http://arxiv.org/abs/1202.2253}{{\tt
  arXiv:1202.2253}}].

\bibitem{Buchmuller:2015jna}
W.~Buchmüller, M.~Dierigl, F.~Ruehle, and J.~Schweizer, {\it {Split
  symmetries}},  {\em Phys. Lett.} {\bf B750} (2015) 615--619,
  [\href{http://arxiv.org/abs/1507.06819}{{\tt arXiv:1507.06819}}].

\bibitem{Pagels:1981ke}
H.~Pagels and J.~R. Primack, {\it {Supersymmetry, Cosmology and New TeV
  Physics}},  {\em Phys. Rev. Lett.} {\bf 48} (1982) 223.

\bibitem{Weinberg:1982zq}
S.~Weinberg, {\it {Cosmological Constraints on the Scale of Supersymmetry
  Breaking}},  {\em Phys. Rev. Lett.} {\bf 48} (1982) 1303.

\bibitem{Khlopov:1984pf}
M.~{\relax Yu}. Khlopov and A.~D. Linde, {\it {Is It Easy to Save the
  Gravitino?}},  {\em Phys. Lett.} {\bf 138B} (1984) 265--268.

\bibitem{Ellis:1984er}
J.~R. Ellis, D.~V. Nanopoulos, and S.~Sarkar, {\it {The Cosmology of Decaying
  Gravitinos}},  {\em Nucl. Phys.} {\bf B259} (1985) 175--188.

\bibitem{Gabbiani:1996hi}
F.~Gabbiani, E.~Gabrielli, A.~Masiero, and L.~Silvestrini, {\it {A Complete
  analysis of FCNC and CP constraints in general SUSY extensions of the
  standard model}},  {\em Nucl. Phys.} {\bf B477} (1996) 321--352,
  [\href{http://arxiv.org/abs/hep-ph/9604387}{{\tt hep-ph/9604387}}].

\bibitem{Starobinsky:1980te}
A.~A. Starobinsky, {\it {A New Type of Isotropic Cosmological Models Without
  Singularity}},  {\em Phys. Lett.} {\bf 91B} (1980) 99--102.

\bibitem{Guth:1980zm}
A.~H. Guth, {\it {The Inflationary Universe: A Possible Solution to the Horizon
  and Flatness Problems}},  {\em Phys. Rev.} {\bf D23} (1981) 347--356.

\bibitem{Linde:1981mu}
A.~D. Linde, {\it {A New Inflationary Universe Scenario: A Possible Solution of
  the Horizon, Flatness, Homogeneity, Isotropy and Primordial Monopole
  Problems}},  {\em Phys. Lett.} {\bf 108B} (1982) 389--393.

\bibitem{Albrecht:1982wi}
A.~Albrecht and P.~J. Steinhardt, {\it {Cosmology for Grand Unified Theories
  with Radiatively Induced Symmetry Breaking}},  {\em Phys. Rev. Lett.} {\bf
  48} (1982) 1220--1223.

\bibitem{Lyth:1998xn}
D.~H. Lyth and A.~Riotto, {\it {Particle physics models of inflation and the
  cosmological density perturbation}},  {\em Phys. Rept.} {\bf 314} (1999)
  1--146, [\href{http://arxiv.org/abs/hep-ph/9807278}{{\tt hep-ph/9807278}}].

\bibitem{Linde:2005ht}
A.~D. Linde, {\it {Particle physics and inflationary cosmology}},  {\em
  Contemp. Concepts Phys.} {\bf 5} (1990) 1--362,
  [\href{http://arxiv.org/abs/hep-th/0503203}{{\tt hep-th/0503203}}].

\bibitem{Randall:1994fr}
L.~Randall and S.~D. Thomas, {\it {Solving the cosmological moduli problem with
  weak scale inflation}},  {\em Nucl. Phys.} {\bf B449} (1995) 229--247,
  [\href{http://arxiv.org/abs/hep-ph/9407248}{{\tt hep-ph/9407248}}].

\bibitem{Riotto:1997iv}
A.~Riotto, {\it {Inflation and the nature of supersymmetry breaking}},  {\em
  Nucl. Phys.} {\bf B515} (1998) 413--435,
  [\href{http://arxiv.org/abs/hep-ph/9707330}{{\tt hep-ph/9707330}}].

\bibitem{Buchmuller:2000zm}
W.~Buchmüller, L.~Covi, and D.~Delepine, {\it {Inflation and supersymmetry
  breaking}},  {\em Phys. Lett.} {\bf B491} (2000) 183--189,
  [\href{http://arxiv.org/abs/hep-ph/0006168}{{\tt hep-ph/0006168}}].

\bibitem{Hardeman:2010fh}
S.~Hardeman, J.~M. Oberreuter, G.~A. Palma, K.~Schalm, and T.~van~der Aalst,
  {\it {The everpresent eta-problem: knowledge of all hidden sectors
  required}},  {\em JHEP} {\bf 04} (2011) 009,
  [\href{http://arxiv.org/abs/1012.5966}{{\tt arXiv:1012.5966}}].

\bibitem{Izawa:1997jc}
K.~I. Izawa, {\it {Supersymmetry-breaking models of inflation}},  {\em Prog.
  Theor. Phys.} {\bf 99} (1998) 157--160,
  [\href{http://arxiv.org/abs/hep-ph/9708315}{{\tt hep-ph/9708315}}].

\bibitem{Schmitz:2016kyr}
K.~Schmitz and T.~T. Yanagida, {\it {Polonyi Inflation: Dynamical supersymmetry
  breaking and late-time R symmetry breaking as the origin of cosmic
  inflation}},  {\em Phys. Rev.} {\bf D94} (2016), no.~7 074021,
  [\href{http://arxiv.org/abs/1604.04911}{{\tt arXiv:1604.04911}}].

\bibitem{Domcke:2017xvu}
V.~Domcke and K.~Schmitz, {\it {Unified model of D-term inflation}},  {\em
  Phys. Rev.} {\bf D95} (2017), no.~7 075020,
  [\href{http://arxiv.org/abs/1702.02173}{{\tt arXiv:1702.02173}}].

\bibitem{Harigaya:2017jny}
K.~Harigaya and K.~Schmitz, {\it {Unified Model of Chaotic Inflation and
  Dynamical Supersymmetry Breaking}},  {\em Phys. Lett.} {\bf B773} (2017)
  320--324, [\href{http://arxiv.org/abs/1707.03646}{{\tt arXiv:1707.03646}}].

\bibitem{Dimopoulos:1997fv}
S.~Dimopoulos, G.~R. Dvali, and R.~Rattazzi, {\it {Dynamical inflation and
  unification scale on quantum moduli spaces}},  {\em Phys. Lett.} {\bf B410}
  (1997) 119--124, [\href{http://arxiv.org/abs/hep-ph/9705348}{{\tt
  hep-ph/9705348}}].

\bibitem{Izawa:1997df}
K.~I. Izawa, M.~Kawasaki, and T.~Yanagida, {\it {Dynamical tuning of the
  initial condition for new inflation in supergravity}},  {\em Phys. Lett.}
  {\bf B411} (1997) 249--255, [\href{http://arxiv.org/abs/hep-ph/9707201}{{\tt
  hep-ph/9707201}}].

\bibitem{Hamaguchi:2008uy}
K.~Hamaguchi, K.-I. Izawa, and H.~Nakajima, {\it {Supersymmetric Inflation of
  Dynamical Origin}},  {\em Phys. Lett.} {\bf B662} (2008) 208--212,
  [\href{http://arxiv.org/abs/0801.2204}{{\tt arXiv:0801.2204}}].

\bibitem{Harigaya:2012pg}
K.~Harigaya, M.~Ibe, K.~Schmitz, and T.~T. Yanagida, {\it {Chaotic Inflation
  with a Fractional Power-Law Potential in Strongly Coupled Gauge Theories}},
  {\em Phys. Lett.} {\bf B720} (2013) 125--129,
  [\href{http://arxiv.org/abs/1211.6241}{{\tt arXiv:1211.6241}}].

\bibitem{Harigaya:2014sua}
K.~Harigaya, M.~Ibe, K.~Schmitz, and T.~T. Yanagida, {\it {Dynamical Chaotic
  Inflation in the Light of BICEP2}},  {\em Phys. Lett.} {\bf B733} (2014)
  283--287, [\href{http://arxiv.org/abs/1403.4536}{{\tt arXiv:1403.4536}}].

\bibitem{Harigaya:2014wta}
K.~Harigaya, M.~Ibe, K.~Schmitz, and T.~T. Yanagida, {\it {Dynamical fractional
  chaotic inflation}},  {\em Phys. Rev.} {\bf D90} (2014), no.~12 123524,
  [\href{http://arxiv.org/abs/1407.3084}{{\tt arXiv:1407.3084}}].

\bibitem{Argurio:2017joe}
R.~Argurio, D.~Coone, L.~Heurtier, and A.~Mariotti, {\it {Sgoldstino-less
  inflation and low energy SUSY breaking}},  {\em JCAP} {\bf 1707} (2017),
  no.~07 047, [\href{http://arxiv.org/abs/1705.06788}{{\tt arXiv:1705.06788}}].

\bibitem{Antoniadis:2017gjr}
I.~Antoniadis, A.~Chatrabhuti, H.~Isono, and R.~Knoops, {\it {Inflation from
  Supersymmetry Breaking}},  {\em Eur. Phys. J.} {\bf C77} (2017), no.~11 724,
  [\href{http://arxiv.org/abs/1706.04133}{{\tt arXiv:1706.04133}}].

\bibitem{Dudas:2017kfz}
E.~Dudas, T.~Gherghetta, Y.~Mambrini, and K.~A. Olive, {\it {Inflation and
  High-Scale Supersymmetry with an EeV Gravitino}},  {\em Phys. Rev.} {\bf D96}
  (2017), no.~11 115032, [\href{http://arxiv.org/abs/1710.07341}{{\tt
  arXiv:1710.07341}}].

\bibitem{Binetruy:1996xj}
P.~Binetruy and G.~R. Dvali, {\it {D term inflation}},  {\em Phys. Lett.} {\bf
  B388} (1996) 241--246, [\href{http://arxiv.org/abs/hep-ph/9606342}{{\tt
  hep-ph/9606342}}].

\bibitem{Halyo:1996pp}
E.~Halyo, {\it {Hybrid inflation from supergravity D terms}},  {\em Phys.
  Lett.} {\bf B387} (1996) 43--47,
  [\href{http://arxiv.org/abs/hep-ph/9606423}{{\tt hep-ph/9606423}}].

\bibitem{Evans:2017bjs}
J.~L. Evans, T.~Gherghetta, N.~Nagata, and M.~Peloso, {\it {Low-scale D-term
  inflation and the relaxion mechanism}},  {\em Phys. Rev.} {\bf D95} (2017),
  no.~11 115027, [\href{http://arxiv.org/abs/1704.03695}{{\tt
  arXiv:1704.03695}}].

\bibitem{Higaki:2012iq}
T.~Higaki, K.~S. Jeong, and F.~Takahashi, {\it {Hybrid inflation in high-scale
  supersymmetry}},  {\em JHEP} {\bf 12} (2012) 111,
  [\href{http://arxiv.org/abs/1211.0994}{{\tt arXiv:1211.0994}}].

\bibitem{Pallis:2014xva}
C.~Pallis and Q.~Shafi, {\it {From Hybrid to Quadratic Inflation With
  High-Scale Supersymmetry Breaking}},  {\em Phys. Lett.} {\bf B736} (2014)
  261--266, [\href{http://arxiv.org/abs/1405.7645}{{\tt arXiv:1405.7645}}].

\bibitem{Kadota:2017dbz}
K.~Kadota, T.~Kobayashi, and K.~Sumita, {\it {A viable D-term hybrid inflation
  model}},  {\em JCAP} {\bf 1711} (2017), no.~11 033,
  [\href{http://arxiv.org/abs/1707.00813}{{\tt arXiv:1707.00813}}].

\bibitem{Linde:1991km}
A.~D. Linde, {\it {Axions in inflationary cosmology}},  {\em Phys. Lett.} {\bf
  B259} (1991) 38--47.

\bibitem{Linde:1993cn}
A.~D. Linde, {\it {Hybrid inflation}},  {\em Phys. Rev.} {\bf D49} (1994)
  748--754, [\href{http://arxiv.org/abs/astro-ph/9307002}{{\tt
  astro-ph/9307002}}].

\bibitem{Buchmuller:2010yy}
W.~Buchmüller, K.~Schmitz, and G.~Vertongen, {\it {Matter and Dark Matter from
  False Vacuum Decay}},  {\em Phys. Lett.} {\bf B693} (2010) 421--425,
  [\href{http://arxiv.org/abs/1008.2355}{{\tt arXiv:1008.2355}}].

\bibitem{Buchmuller:2011mw}
W.~Buchmüller, K.~Schmitz, and G.~Vertongen, {\it {Entropy, Baryon Asymmetry
  and Dark Matter from Heavy Neutrino Decays}},  {\em Nucl. Phys.} {\bf B851}
  (2011) 481--532, [\href{http://arxiv.org/abs/1104.2750}{{\tt
  arXiv:1104.2750}}].

\bibitem{Buchmuller:2012wn}
W.~Buchmüller, V.~Domcke, and K.~Schmitz, {\it {Spontaneous B-L Breaking as
  the Origin of the Hot Early Universe}},  {\em Nucl. Phys.} {\bf B862} (2012)
  587--632, [\href{http://arxiv.org/abs/1202.6679}{{\tt arXiv:1202.6679}}].

\bibitem{Buchmuller:2012bt}
W.~Buchmüller, V.~Domcke, and K.~Schmitz, {\it {WIMP Dark Matter from
  Gravitino Decays and Leptogenesis}},  {\em Phys. Lett.} {\bf B713} (2012)
  63--67, [\href{http://arxiv.org/abs/1203.0285}{{\tt arXiv:1203.0285}}].

\bibitem{Schmitz:2012kaa}
K.~Schmitz, {\em {The B-L Phase Transition: Implications for Cosmology and
  Neutrinos}}.
\newblock PhD thesis, Hamburg U., 2012.
\newblock \href{http://arxiv.org/abs/1307.3887}{{\tt arXiv:1307.3887}}.

\bibitem{Buchmuller:2013lra}
W.~Buchmüller, V.~Domcke, K.~Kamada, and K.~Schmitz, {\it {The Gravitational
  Wave Spectrum from Cosmological $B-L$ Breaking}},  {\em JCAP} {\bf 1310}
  (2013) 003, [\href{http://arxiv.org/abs/1305.3392}{{\tt arXiv:1305.3392}}].

\bibitem{Buchmuller:2013dja}
W.~Buchmüller, V.~Domcke, K.~Kamada, and K.~Schmitz, {\it {A Minimal
  Supersymmetric Model of Particle Physics and the Early Universe}},  in {\em
  {Cosmology and Particle Physics beyond Standard Models}}
  (L.~\'Alvarez-Gaum\'e, G.~S. Djordjevi\'c, and D.~Stojkovi\'c, eds.),
  CERN-Proceedings-2014-001, pp.~47--77.
\newblock CERN, Geneva, 2014.
\newblock \href{http://arxiv.org/abs/1309.7788}{{\tt arXiv:1309.7788}}.

\bibitem{Domcke:2013pma}
V.~Domcke, {\em {Matter, Dark Matter and Gravitational Waves from a GUT-Scale
  U(1) Phase Transition}}.
\newblock PhD thesis, Hamburg U., 2013.

\bibitem{Fukugita:1986hr}
M.~Fukugita and T.~Yanagida, {\it {Baryogenesis Without Grand Unification}},
  {\em Phys. Lett.} {\bf B174} (1986) 45--47.

\bibitem{Minkowski:1977sc}
P.~Minkowski, {\it {$\mu \to e\gamma$ at a Rate of One Out of $10^{9}$ Muon
  Decays?}},  {\em Phys. Lett.} {\bf 67B} (1977) 421--428.

\bibitem{Yanagida:1979as}
T.~Yanagida, {\it {Horizontal Gauge Symmetry and Masses of Neutrinos}},  {\em
  Conf. Proc.} {\bf C7902131} (1979) 95--98.

\bibitem{Yanagida:1980xy}
T.~Yanagida, {\it {Horizontal Symmetry and Masses of Neutrinos}},  {\em Prog.
  Theor. Phys.} {\bf 64} (1980) 1103.

\bibitem{GellMann:1980vs}
M.~Gell-Mann, P.~Ramond, and R.~Slansky, {\it {Complex Spinors and Unified
  Theories}},  {\em Conf. Proc.} {\bf C790927} (1979) 315--321,
  [\href{http://arxiv.org/abs/1306.4669}{{\tt arXiv:1306.4669}}].

\bibitem{Mohapatra:1979ia}
R.~N. Mohapatra and G.~Senjanovic, {\it {Neutrino Mass and Spontaneous Parity
  Violation}},  {\em Phys. Rev. Lett.} {\bf 44} (1980) 912.

\bibitem{Ade:2015lrj}
{\bf Planck} Collaboration, P.~A.~R. Ade et~al., {\it {Planck 2015 results. XX.
  Constraints on inflation}},  {\em Astron. Astrophys.} {\bf 594} (2016) A20,
  [\href{http://arxiv.org/abs/1502.02114}{{\tt arXiv:1502.02114}}].

\bibitem{Copeland:1994vg}
E.~J. Copeland, A.~R. Liddle, D.~H. Lyth, E.~D. Stewart, and D.~Wands, {\it
  {False vacuum inflation with Einstein gravity}},  {\em Phys. Rev.} {\bf D49}
  (1994) 6410--6433, [\href{http://arxiv.org/abs/astro-ph/9401011}{{\tt
  astro-ph/9401011}}].

\bibitem{Dvali:1994ms}
G.~R. Dvali, Q.~Shafi, and R.~K. Schaefer, {\it {Large scale structure and
  supersymmetric inflation without fine tuning}},  {\em Phys. Rev. Lett.} {\bf
  73} (1994) 1886--1889, [\href{http://arxiv.org/abs/hep-ph/9406319}{{\tt
  hep-ph/9406319}}].

\bibitem{ORaifeartaigh:1975nky}
L.~O'Raifeartaigh, {\it {Spontaneous Symmetry Breaking for Chiral Scalar
  Superfields}},  {\em Nucl. Phys.} {\bf B96} (1975) 331--352.

\bibitem{Fayet:1974jb}
P.~Fayet and J.~Iliopoulos, {\it {Spontaneously Broken Supergauge Symmetries
  and Goldstone Spinors}},  {\em Phys. Lett.} {\bf 51B} (1974) 461--464.

\bibitem{Panagiotakopoulos:1997qd}
C.~Panagiotakopoulos, {\it {Blue perturbation spectra from hybrid inflation
  with canonical supergravity}},  {\em Phys. Rev.} {\bf D55} (1997)
  R7335--R7339, [\href{http://arxiv.org/abs/hep-ph/9702433}{{\tt
  hep-ph/9702433}}].

\bibitem{Linde:1997sj}
A.~D. Linde and A.~Riotto, {\it {Hybrid inflation in supergravity}},  {\em
  Phys. Rev.} {\bf D56} (1997) R1841--R1844,
  [\href{http://arxiv.org/abs/hep-ph/9703209}{{\tt hep-ph/9703209}}].

\bibitem{Rehman:2009nq}
M.~U. Rehman, Q.~Shafi, and J.~R. Wickman, {\it {Supersymmetric Hybrid
  Inflation Redux}},  {\em Phys. Lett.} {\bf B683} (2010) 191--195,
  [\href{http://arxiv.org/abs/0908.3896}{{\tt arXiv:0908.3896}}].

\bibitem{Rehman:2009yj}
M.~U. Rehman, Q.~Shafi, and J.~R. Wickman, {\it {Minimal Supersymmetric Hybrid
  Inflation, Flipped SU(5) and Proton Decay}},  {\em Phys. Lett.} {\bf B688}
  (2010) 75--81, [\href{http://arxiv.org/abs/0912.4737}{{\tt
  arXiv:0912.4737}}].

\bibitem{Nakayama:2010xf}
K.~Nakayama, F.~Takahashi, and T.~T. Yanagida, {\it {Constraint on the
  gravitino mass in hybrid inflation}},  {\em JCAP} {\bf 1012} (2010) 010,
  [\href{http://arxiv.org/abs/1007.5152}{{\tt arXiv:1007.5152}}].

\bibitem{Buchmuller:2014epa}
W.~Buchmüller, V.~Domcke, K.~Kamada, and K.~Schmitz, {\it {Hybrid Inflation in
  the Complex Plane}},  {\em JCAP} {\bf 1407} (2014) 054,
  [\href{http://arxiv.org/abs/1404.1832}{{\tt arXiv:1404.1832}}].

\bibitem{Komargodski:2009pc}
Z.~Komargodski and N.~Seiberg, {\it {Comments on the Fayet-Iliopoulos Term in
  Field Theory and Supergravity}},  {\em JHEP} {\bf 06} (2009) 007,
  [\href{http://arxiv.org/abs/0904.1159}{{\tt arXiv:0904.1159}}].

\bibitem{Dienes:2009td}
K.~R. Dienes and B.~Thomas, {\it {On the Inconsistency of Fayet-Iliopoulos
  Terms in Supergravity Theories}},  {\em Phys. Rev.} {\bf D81} (2010) 065023,
  [\href{http://arxiv.org/abs/0911.0677}{{\tt arXiv:0911.0677}}].

\bibitem{Banks:2010zn}
T.~Banks and N.~Seiberg, {\it {Symmetries and Strings in Field Theory and
  Gravity}},  {\em Phys. Rev.} {\bf D83} (2011) 084019,
  [\href{http://arxiv.org/abs/1011.5120}{{\tt arXiv:1011.5120}}].

\bibitem{Cribiori:2017laj}
N.~Cribiori, F.~Farakos, M.~Tournoy, and A.~van Proeyen, {\it {Fayet-Iliopoulos
  terms in supergravity without gauged R-symmetry}},  {\em JHEP} {\bf 04}
  (2018) 032, [\href{http://arxiv.org/abs/1712.08601}{{\tt arXiv:1712.08601}}].

\bibitem{Kuzenko:2018jlz}
S.~M. Kuzenko, {\it {Taking a vector supermultiplet apart: Alternative
  Fayet–Iliopoulos-type terms}},  {\em Phys. Lett.} {\bf B781} (2018)
  723--727, [\href{http://arxiv.org/abs/1801.04794}{{\tt arXiv:1801.04794}}].

\bibitem{Dine:1987xk}
M.~Dine, N.~Seiberg, and E.~Witten, {\it {Fayet-Iliopoulos Terms in String
  Theory}},  {\em Nucl. Phys.} {\bf B289} (1987) 589--598.

\bibitem{Atick:1987gy}
J.~J. Atick, L.~J. Dixon, and A.~Sen, {\it {String Calculation of
  Fayet-Iliopoulos d Terms in Arbitrary Supersymmetric Compactifications}},
  {\em Nucl. Phys.} {\bf B292} (1987) 109--149.

\bibitem{Green:1984sg}
M.~B. Green and J.~H. Schwarz, {\it {Anomaly Cancellation in Supersymmetric
  D=10 Gauge Theory and Superstring Theory}},  {\em Phys. Lett.} {\bf 149B}
  (1984) 117--122.

\bibitem{Komargodski:2010rb}
Z.~Komargodski and N.~Seiberg, {\it {Comments on Supercurrent Multiplets,
  Supersymmetric Field Theories and Supergravity}},  {\em JHEP} {\bf 07} (2010)
  017, [\href{http://arxiv.org/abs/1002.2228}{{\tt arXiv:1002.2228}}].

\bibitem{Binetruy:2004hh}
P.~Binetruy, G.~Dvali, R.~Kallosh, and A.~Van~Proeyen, {\it {Fayet-Iliopoulos
  terms in supergravity and cosmology}},  {\em Class. Quant. Grav.} {\bf 21}
  (2004) 3137--3170, [\href{http://arxiv.org/abs/hep-th/0402046}{{\tt
  hep-th/0402046}}].

\bibitem{Coughlan:1983ci}
G.~D. Coughlan, W.~Fischler, E.~W. Kolb, S.~Raby, and G.~G. Ross, {\it
  {Cosmological Problems for the Polonyi Potential}},  {\em Phys. Lett.} {\bf
  131B} (1983) 59--64.

\bibitem{Ellis:1986zt}
J.~R. Ellis, D.~V. Nanopoulos, and M.~Quiros, {\it {On the Axion, Dilaton,
  Polonyi, Gravitino and Shadow Matter Problems in Supergravity and Superstring
  Models}},  {\em Phys. Lett.} {\bf B174} (1986) 176--182.

\bibitem{Domcke:2014zqa}
V.~Domcke, K.~Schmitz, and T.~T. Yanagida, {\it {Dynamical D-Terms in
  Supergravity}},  {\em Nucl. Phys.} {\bf B891} (2015) 230--258,
  [\href{http://arxiv.org/abs/1410.4641}{{\tt arXiv:1410.4641}}].

\bibitem{Kibble:1976sj}
T.~W.~B. Kibble, {\it {Topology of Cosmic Domains and Strings}},  {\em J.
  Phys.} {\bf A9} (1976) 1387--1398.

\bibitem{Jeannerot:2003qv}
R.~Jeannerot, J.~Rocher, and M.~Sakellariadou, {\it {How generic is cosmic
  string formation in SUSY GUTs}},  {\em Phys. Rev.} {\bf D68} (2003) 103514,
  [\href{http://arxiv.org/abs/hep-ph/0308134}{{\tt hep-ph/0308134}}].

\bibitem{Vilenkin:1984ib}
A.~Vilenkin, {\it {Cosmic Strings and Domain Walls}},  {\em Phys. Rept.} {\bf
  121} (1985) 263--315.

\bibitem{Hindmarsh:1994re}
M.~B. Hindmarsh and T.~W.~B. Kibble, {\it {Cosmic strings}},  {\em Rept. Prog.
  Phys.} {\bf 58} (1995) 477--562,
  [\href{http://arxiv.org/abs/hep-ph/9411342}{{\tt hep-ph/9411342}}].

\bibitem{Ade:2013xla}
{\bf Planck} Collaboration, P.~A.~R. Ade et~al., {\it {Planck 2013 results.
  XXV. Searches for cosmic strings and other topological defects}},  {\em
  Astron. Astrophys.} {\bf 571} (2014) A25,
  [\href{http://arxiv.org/abs/1303.5085}{{\tt arXiv:1303.5085}}].

\bibitem{Ade:2015xua}
{\bf Planck} Collaboration, P.~A.~R. Ade et~al., {\it {Planck 2015 results.
  XIII. Cosmological parameters}},  {\em Astron. Astrophys.} {\bf 594} (2016)
  A13, [\href{http://arxiv.org/abs/1502.01589}{{\tt arXiv:1502.01589}}].

\bibitem{Sanidas:2012ee}
S.~A. Sanidas, R.~A. Battye, and B.~W. Stappers, {\it {Constraints on cosmic
  string tension imposed by the limit on the stochastic gravitational wave
  background from the European Pulsar Timing Array}},  {\em Phys. Rev.} {\bf
  D85} (2012) 122003, [\href{http://arxiv.org/abs/1201.2419}{{\tt
  arXiv:1201.2419}}].

\bibitem{Blanco-Pillado:2017rnf}
J.~J. Blanco-Pillado, K.~D. Olum, and X.~Siemens, {\it {New limits on cosmic
  strings from gravitational wave observation}},  {\em Phys. Lett.} {\bf B778}
  (2018) 392--396, [\href{http://arxiv.org/abs/1709.02434}{{\tt
  arXiv:1709.02434}}].

\bibitem{Ringeval:2017eww}
C.~Ringeval and T.~Suyama, {\it {Stochastic gravitational waves from cosmic
  string loops in scaling}},  {\em JCAP} {\bf 1712} (2017), no.~12 027,
  [\href{http://arxiv.org/abs/1709.03845}{{\tt arXiv:1709.03845}}].

\bibitem{Mota:2014uka}
H.~F. Santana~Mota and M.~Hindmarsh, {\it {Big-Bang Nucleosynthesis and
  Gamma-Ray Constraints on Cosmic Strings with a large Higgs condensate}},
  {\em Phys. Rev.} {\bf D91} (2015), no.~4 043001,
  [\href{http://arxiv.org/abs/1407.3599}{{\tt arXiv:1407.3599}}].

\bibitem{Charnock:2016nzm}
T.~Charnock, A.~Avgoustidis, E.~J. Copeland, and A.~Moss, {\it {CMB constraints
  on cosmic strings and superstrings}},  {\em Phys. Rev.} {\bf D93} (2016),
  no.~12 123503, [\href{http://arxiv.org/abs/1603.01275}{{\tt
  arXiv:1603.01275}}].

\bibitem{Lizarraga:2016onn}
J.~Lizarraga, J.~Urrestilla, D.~Daverio, M.~Hindmarsh, and M.~Kunz, {\it {New
  CMB constraints for Abelian Higgs cosmic strings}},  {\em JCAP} {\bf 1610}
  (2016), no.~10 042, [\href{http://arxiv.org/abs/1609.03386}{{\tt
  arXiv:1609.03386}}].

\bibitem{Battye:2010hg}
R.~Battye, B.~Garbrecht, and A.~Moss, {\it {Tight constraints on F- and D-term
  hybrid inflation scenarios}},  {\em Phys. Rev.} {\bf D81} (2010) 123512,
  [\href{http://arxiv.org/abs/1001.0769}{{\tt arXiv:1001.0769}}].

\bibitem{Babu:2015xba}
K.~S. Babu, K.~Schmitz, and T.~T. Yanagida, {\it {Pure gravity mediation and
  spontaneous $B$$-$$L$ breaking from strong dynamics}},  {\em Nucl. Phys.}
  {\bf B905} (2016) 73--95, [\href{http://arxiv.org/abs/1507.04467}{{\tt
  arXiv:1507.04467}}].

\bibitem{Nilles:1983ge}
H.~P. Nilles, {\it {Supersymmetry, Supergravity and Particle Physics}},  {\em
  Phys. Rept.} {\bf 110} (1984) 1--162.

\bibitem{Ferrara:2010yw}
S.~Ferrara, R.~Kallosh, A.~Linde, A.~Marrani, and A.~Van~Proeyen, {\it {Jordan
  Frame Supergravity and Inflation in NMSSM}},  {\em Phys. Rev.} {\bf D82}
  (2010) 045003, [\href{http://arxiv.org/abs/1004.0712}{{\tt
  arXiv:1004.0712}}].

\bibitem{Ferrara:2010in}
S.~Ferrara, R.~Kallosh, A.~Linde, A.~Marrani, and A.~Van~Proeyen, {\it
  {Superconformal Symmetry, NMSSM, and Inflation}},  {\em Phys. Rev.} {\bf D83}
  (2011) 025008, [\href{http://arxiv.org/abs/1008.2942}{{\tt
  arXiv:1008.2942}}].

\bibitem{Postma:2014vaa}
M.~Postma and M.~Volponi, {\it {Equivalence of the Einstein and Jordan
  frames}},  {\em Phys. Rev.} {\bf D90} (2014), no.~10 103516,
  [\href{http://arxiv.org/abs/1407.6874}{{\tt arXiv:1407.6874}}].

\bibitem{Kamenshchik:2014waa}
A.~{\relax Yu}. Kamenshchik and C.~F. Steinwachs, {\it {Question of quantum
  equivalence between Jordan frame and Einstein frame}},  {\em Phys. Rev.} {\bf
  D91} (2015), no.~8 084033, [\href{http://arxiv.org/abs/1408.5769}{{\tt
  arXiv:1408.5769}}].

\bibitem{Stelle:1978ye}
K.~S. Stelle and P.~C. West, {\it {Minimal Auxiliary Fields for Supergravity}},
   {\em Phys. Lett.} {\bf 74B} (1978) 330--332.

\bibitem{Ferrara:1978em}
S.~Ferrara and P.~van Nieuwenhuizen, {\it {The Auxiliary Fields of
  Supergravity}},  {\em Phys. Lett.} {\bf 74B} (1978) 333.

\bibitem{Cremmer:1982en}
E.~Cremmer, S.~Ferrara, L.~Girardello, and A.~Van~Proeyen, {\it {Yang-Mills
  Theories with Local Supersymmetry: Lagrangian, Transformation Laws and
  SuperHiggs Effect}},  {\em Nucl. Phys.} {\bf B212} (1983) 413.

\bibitem{Kugo:1982mr}
T.~Kugo and S.~Uehara, {\it {Improved Superconformal Gauge Conditions in the
  $N=1$ Supergravity {Yang-Mills} Matter System}},  {\em Nucl. Phys.} {\bf
  B222} (1983) 125--138.

\bibitem{Buchmuller:2012ex}
W.~Buchmüller, V.~Domcke, and K.~Schmitz, {\it {Superconformal D-Term
  Inflation}},  {\em JCAP} {\bf 1304} (2013) 019,
  [\href{http://arxiv.org/abs/1210.4105}{{\tt arXiv:1210.4105}}].

\bibitem{Buchmuller:2013zfa}
W.~Buchmüller, V.~Domcke, and K.~Kamada, {\it {The Starobinsky Model from
  Superconformal D-Term Inflation}},  {\em Phys. Lett.} {\bf B726} (2013)
  467--470, [\href{http://arxiv.org/abs/1306.3471}{{\tt arXiv:1306.3471}}].

\bibitem{Randall:1998uk}
L.~Randall and R.~Sundrum, {\it {Out of this world supersymmetry breaking}},
  {\em Nucl. Phys.} {\bf B557} (1999) 79--118,
  [\href{http://arxiv.org/abs/hep-th/9810155}{{\tt hep-th/9810155}}].

\bibitem{Luty:2001jh}
M.~A. Luty and R.~Sundrum, {\it {Supersymmetry breaking and composite extra
  dimensions}},  {\em Phys. Rev.} {\bf D65} (2002) 066004,
  [\href{http://arxiv.org/abs/hep-th/0105137}{{\tt hep-th/0105137}}].

\bibitem{Luty:2001zv}
M.~Luty and R.~Sundrum, {\it {Anomaly mediated supersymmetry breaking in
  four-dimensions, naturally}},  {\em Phys. Rev.} {\bf D67} (2003) 045007,
  [\href{http://arxiv.org/abs/hep-th/0111231}{{\tt hep-th/0111231}}].

\bibitem{Ibe:2005pj}
M.~Ibe, K.~I. Izawa, Y.~Nakayama, Y.~Shinbara, and T.~Yanagida, {\it
  {Conformally sequestered SUSY breaking in vector-like gauge theories}},  {\em
  Phys. Rev.} {\bf D73} (2006) 015004,
  [\href{http://arxiv.org/abs/hep-ph/0506023}{{\tt hep-ph/0506023}}].

\bibitem{Ibe:2005qv}
M.~Ibe, K.~I. Izawa, Y.~Nakayama, Y.~Shinbara, and T.~Yanagida, {\it {More on
  conformally sequestered SUSY breaking}},  {\em Phys. Rev.} {\bf D73} (2006)
  035012, [\href{http://arxiv.org/abs/hep-ph/0509229}{{\tt hep-ph/0509229}}].

\bibitem{Cremmer:1983bf}
E.~Cremmer, S.~Ferrara, C.~Kounnas, and D.~V. Nanopoulos, {\it {Naturally
  Vanishing Cosmological Constant in N=1 Supergravity}},  {\em Phys. Lett.}
  {\bf 133B} (1983) 61.

\bibitem{Ellis:1983sf}
J.~R. Ellis, A.~B. Lahanas, D.~V. Nanopoulos, and K.~Tamvakis, {\it {No-Scale
  Supersymmetric Standard Model}},  {\em Phys. Lett.} {\bf 134B} (1984) 429.

\bibitem{Lahanas:1986uc}
A.~B. Lahanas and D.~V. Nanopoulos, {\it {The Road to No Scale Supergravity}},
  {\em Phys. Rept.} {\bf 145} (1987) 1.

\bibitem{Witten:1985xb}
E.~Witten, {\it {Dimensional Reduction of Superstring Models}},  {\em Phys.
  Lett.} {\bf 155B} (1985) 151.

\bibitem{Stewart:1994ts}
E.~D. Stewart, {\it {Inflation, supergravity and superstrings}},  {\em Phys.
  Rev.} {\bf D51} (1995) 6847--6853,
  [\href{http://arxiv.org/abs/hep-ph/9405389}{{\tt hep-ph/9405389}}].

\bibitem{Dumitrescu:2010ca}
T.~T. Dumitrescu, Z.~Komargodski, and M.~Sudano, {\it {Global Symmetries and
  D-Terms in Supersymmetric Field Theories}},  {\em JHEP} {\bf 11} (2010) 052,
  [\href{http://arxiv.org/abs/1007.5352}{{\tt arXiv:1007.5352}}].

\bibitem{Kawamura:2010mb}
Y.~Kawamura, {\it {Limitation on Magnitude of $D$-components}},  {\em Prog.
  Theor. Phys.} {\bf 125} (2011) 509--520,
  [\href{http://arxiv.org/abs/1008.5223}{{\tt arXiv:1008.5223}}].

\bibitem{Kawasaki:2000yn}
M.~Kawasaki, M.~Yamaguchi, and T.~Yanagida, {\it {Natural chaotic inflation in
  supergravity}},  {\em Phys. Rev. Lett.} {\bf 85} (2000) 3572--3575,
  [\href{http://arxiv.org/abs/hep-ph/0004243}{{\tt hep-ph/0004243}}].

\bibitem{Grisaru:1979wc}
M.~T. Grisaru, W.~Siegel, and M.~Rocek, {\it {Improved Methods for
  Supergraphs}},  {\em Nucl. Phys.} {\bf B159} (1979) 429.

\bibitem{Buchbinder:1994iw}
I.~L. Buchbinder, S.~Kuzenko, and Z.~Yarevskaya, {\it {Supersymmetric effective
  potential: Superfield approach}},  {\em Nucl. Phys.} {\bf B411} (1994)
  665--692.

\bibitem{Grisaru:1996ve}
M.~T. Grisaru, M.~Rocek, and R.~von Unge, {\it {Effective Kahler potentials}},
  {\em Phys. Lett.} {\bf B383} (1996) 415--421,
  [\href{http://arxiv.org/abs/hep-th/9605149}{{\tt hep-th/9605149}}].

\bibitem{Brignole:2000kg}
A.~Brignole, {\it {One loop Kahler potential in non renormalizable theories}},
  {\em Nucl. Phys.} {\bf B579} (2000) 101--116,
  [\href{http://arxiv.org/abs/hep-th/0001121}{{\tt hep-th/0001121}}].

\bibitem{Nibbelink:2006si}
S.~Groot~Nibbelink and T.~S. Nyawelo, {\it {Effective action of softly broken
  supersymmetric theories}},  {\em Phys. Rev.} {\bf D75} (2007) 045002,
  [\href{http://arxiv.org/abs/hep-th/0612092}{{\tt hep-th/0612092}}].

\bibitem{Seto:2005qg}
O.~Seto and J.~Yokoyama, {\it {Hiding cosmic strings in supergravity D-term
  inflation}},  {\em Phys. Rev.} {\bf D73} (2006) 023508,
  [\href{http://arxiv.org/abs/hep-ph/0508172}{{\tt hep-ph/0508172}}].

\bibitem{Lin:2006xta}
C.-M. Lin and J.~McDonald, {\it {Supergravity modification of D-term hybrid
  inflation: Solving the cosmic string and spectral index problems via a
  right-handed sneutrino}},  {\em Phys. Rev.} {\bf D74} (2006) 063510,
  [\href{http://arxiv.org/abs/hep-ph/0604245}{{\tt hep-ph/0604245}}].

\bibitem{Rocher:2006nh}
J.~Rocher and M.~Sakellariadou, {\it {D-term inflation in non-minimal
  supergravity}},  {\em JCAP} {\bf 0611} (2006) 001,
  [\href{http://arxiv.org/abs/hep-th/0607226}{{\tt hep-th/0607226}}].

\bibitem{Buchmuller:2013uta}
W.~Buchmüller, V.~Domcke, and C.~Wieck, {\it {No-scale D-term inflation with
  stabilized moduli}},  {\em Phys. Lett.} {\bf B730} (2014) 155--160,
  [\href{http://arxiv.org/abs/1309.3122}{{\tt arXiv:1309.3122}}].

\bibitem{Izawa:1996pk}
K.-I. Izawa and T.~Yanagida, {\it {Dynamical supersymmetry breaking in
  vector-like gauge theories}},  {\em Prog. Theor. Phys.} {\bf 95} (1996)
  829--830, [\href{http://arxiv.org/abs/hep-th/9602180}{{\tt hep-th/9602180}}].

\bibitem{Intriligator:1996pu}
K.~A. Intriligator and S.~D. Thomas, {\it {Dynamical supersymmetry breaking on
  quantum moduli spaces}},  {\em Nucl. Phys.} {\bf B473} (1996) 121--142,
  [\href{http://arxiv.org/abs/hep-th/9603158}{{\tt hep-th/9603158}}].

\bibitem{Manohar:1983md}
A.~Manohar and H.~Georgi, {\it {Chiral Quarks and the Nonrelativistic Quark
  Model}},  {\em Nucl. Phys.} {\bf B234} (1984) 189--212.

\bibitem{Georgi:1986kr}
H.~Georgi and L.~Randall, {\it {Flavor Conserving CP Violation in Invisible
  Axion Models}},  {\em Nucl. Phys.} {\bf B276} (1986) 241--252.

\bibitem{Luty:1997fk}
M.~A. Luty, {\it {Naive dimensional analysis and supersymmetry}},  {\em Phys.
  Rev.} {\bf D57} (1998) 1531--1538,
  [\href{http://arxiv.org/abs/hep-ph/9706235}{{\tt hep-ph/9706235}}].

\bibitem{Cohen:1997rt}
A.~G. Cohen, D.~B. Kaplan, and A.~E. Nelson, {\it {Counting 4 pis in strongly
  coupled supersymmetry}},  {\em Phys. Lett.} {\bf B412} (1997) 301--308,
  [\href{http://arxiv.org/abs/hep-ph/9706275}{{\tt hep-ph/9706275}}].

\bibitem{Seiberg:1994bz}
N.~Seiberg, {\it {Exact results on the space of vacua of four-dimensional SUSY
  gauge theories}},  {\em Phys. Rev.} {\bf D49} (1994) 6857--6863,
  [\href{http://arxiv.org/abs/hep-th/9402044}{{\tt hep-th/9402044}}].

\bibitem{Chacko:1998si}
Z.~Chacko, M.~A. Luty, and E.~Ponton, {\it {Calculable dynamical supersymmetry
  breaking on deformed moduli spaces}},  {\em JHEP} {\bf 12} (1998) 016,
  [\href{http://arxiv.org/abs/hep-th/9810253}{{\tt hep-th/9810253}}].

\bibitem{Harigaya:2015soa}
K.~Harigaya, M.~Ibe, K.~Schmitz, and T.~T. Yanagida, {\it {Peccei-Quinn
  Symmetry from Dynamical Supersymmetry Breaking}},  {\em Phys. Rev.} {\bf D92}
  (2015), no.~7 075003, [\href{http://arxiv.org/abs/1505.07388}{{\tt
  arXiv:1505.07388}}].

\bibitem{Polonyi:1977pj}
J.~Polonyi, {\it {Generalization of the Massive Scalar Multiplet Coupling to
  the Supergravity}},  Tech. Rep. KFKI-77-93, Hungary Central Research
  Institute, 1977.

\bibitem{Coleman:1973jx}
S.~R. Coleman and E.~J. Weinberg, {\it {Radiative Corrections as the Origin of
  Spontaneous Symmetry Breaking}},  {\em Phys. Rev.} {\bf D7} (1973)
  1888--1910.

\bibitem{Martin:2001vx}
S.~P. Martin, {\it {Two loop effective potential for a general renormalizable
  theory and softly broken supersymmetry}},  {\em Phys. Rev.} {\bf D65} (2002)
  116003, [\href{http://arxiv.org/abs/hep-ph/0111209}{{\tt hep-ph/0111209}}].

\bibitem{Intriligator:2010be}
K.~Intriligator and M.~Sudano, {\it {General Gauge Mediation with Gauge
  Messengers}},  {\em JHEP} {\bf 06} (2010) 047,
  [\href{http://arxiv.org/abs/1001.5443}{{\tt arXiv:1001.5443}}].

\bibitem{Chiba:2008ia}
T.~Chiba and M.~Yamaguchi, {\it {Extended Slow-Roll Conditions and Rapid-Roll
  Conditions}},  {\em JCAP} {\bf 0810} (2008) 021,
  [\href{http://arxiv.org/abs/0807.4965}{{\tt arXiv:0807.4965}}].

\bibitem{Buchmuller:2014rfa}
W.~Buchmüller, V.~Domcke, and K.~Schmitz, {\it {The Chaotic Regime of D-Term
  Inflation}},  {\em JCAP} {\bf 1411} (2014), no.~11 006,
  [\href{http://arxiv.org/abs/1406.6300}{{\tt arXiv:1406.6300}}].

\bibitem{Buchmuller:2014dda}
W.~Buchmüller and K.~Ishiwata, {\it {Grand Unification and Subcritical Hybrid
  Inflation}},  {\em Phys. Rev.} {\bf D91} (2015), no.~8 081302,
  [\href{http://arxiv.org/abs/1412.3764}{{\tt arXiv:1412.3764}}].

\bibitem{Watari:2000jh}
T.~Watari and T.~Yanagida, {\it {$\mathcal{N}=2$ supersymmetry in a hybrid
  inflation model}},  {\em Phys. Lett.} {\bf B499} (2001) 297--304,
  [\href{http://arxiv.org/abs/hep-ph/0011389}{{\tt hep-ph/0011389}}].

\bibitem{Martin:1997ns}
S.~P. Martin, {\it {A Supersymmetry primer}},  {\em Adv. Ser. Direct. High
  Energy Phys.} {\bf 21} (2010) 1--153,
  [\href{http://arxiv.org/abs/hep-ph/9709356}{{\tt hep-ph/9709356}}].
  [\textit{Adv. Ser. Direct. High Energy Phys.} \textbf{18} (1998) 1--98].

\bibitem{Olive:1989jg}
K.~A. Olive and M.~Srednicki, {\it {New Limits on Parameters of the
  Supersymmetric Standard Model from Cosmology}},  {\em Phys. Lett.} {\bf B230}
  (1989) 78--82.

\bibitem{Griest:1989zh}
K.~Griest, M.~Kamionkowski, and M.~S. Turner, {\it {Supersymmetric Dark Matter
  Above the W Mass}},  {\em Phys. Rev.} {\bf D41} (1990) 3565--3582.

\bibitem{Harigaya:2013asa}
K.~Harigaya, M.~Ibe, and T.~T. Yanagida, {\it {A Closer Look at Gaugino Masses
  in Pure Gravity Mediation Model/Minimal Split SUSY Model}},  {\em JHEP} {\bf
  12} (2013) 016, [\href{http://arxiv.org/abs/1310.0643}{{\tt
  arXiv:1310.0643}}].

\bibitem{Jeong:2012en}
K.~S. Jeong and F.~Takahashi, {\it {A Gravitino-rich Universe}},  {\em JHEP}
  {\bf 01} (2013) 173, [\href{http://arxiv.org/abs/1210.4077}{{\tt
  arXiv:1210.4077}}].

\bibitem{Gherghetta:1999sw}
T.~Gherghetta, G.~F. Giudice, and J.~D. Wells, {\it {Phenomenological
  consequences of supersymmetry with anomaly induced masses}},  {\em Nucl.
  Phys.} {\bf B559} (1999) 27--47,
  [\href{http://arxiv.org/abs/hep-ph/9904378}{{\tt hep-ph/9904378}}].

\bibitem{Ibe:2004tg}
M.~Ibe, R.~Kitano, H.~Murayama, and T.~Yanagida, {\it {Viable supersymmetry and
  leptogenesis with anomaly mediation}},  {\em Phys. Rev.} {\bf D70} (2004)
  075012, [\href{http://arxiv.org/abs/hep-ph/0403198}{{\tt hep-ph/0403198}}].

\bibitem{Bolz:2000fu}
M.~Bolz, A.~Brandenburg, and W.~Buchmüller, {\it {Thermal production of
  gravitinos}},  {\em Nucl. Phys.} {\bf B606} (2001) 518--544,
  [\href{http://arxiv.org/abs/hep-ph/0012052}{{\tt hep-ph/0012052}}]. [Erratum:
  \textit{Nucl. Phys.} \textbf{B790} (2008) 336-337].

\bibitem{Kawasaki:2006gs}
M.~Kawasaki, F.~Takahashi, and T.~T. Yanagida, {\it {Gravitino overproduction
  in inflaton decay}},  {\em Phys. Lett.} {\bf B638} (2006) 8--12,
  [\href{http://arxiv.org/abs/hep-ph/0603265}{{\tt hep-ph/0603265}}].

\bibitem{Hisano:2006nn}
J.~Hisano, S.~Matsumoto, M.~Nagai, O.~Saito, and M.~Senami, {\it
  {Non-perturbative effect on thermal relic abundance of dark matter}},  {\em
  Phys. Lett.} {\bf B646} (2007) 34--38,
  [\href{http://arxiv.org/abs/hep-ph/0610249}{{\tt hep-ph/0610249}}].

\bibitem{Cirelli:2007xd}
M.~Cirelli, A.~Strumia, and M.~Tamburini, {\it {Cosmology and Astrophysics of
  Minimal Dark Matter}},  {\em Nucl. Phys.} {\bf B787} (2007) 152--175,
  [\href{http://arxiv.org/abs/0706.4071}{{\tt arXiv:0706.4071}}].

\bibitem{Barbier:2004ez}
R.~Barbier et~al., {\it {R-parity violating supersymmetry}},  {\em Phys. Rept.}
  {\bf 420} (2005) 1--202, [\href{http://arxiv.org/abs/hep-ph/0406039}{{\tt
  hep-ph/0406039}}].

\bibitem{Mukhanov:1990me}
V.~F. Mukhanov, H.~A. Feldman, and R.~H. Brandenberger, {\it {Theory of
  cosmological perturbations. Part 1. Classical perturbations. Part 2. Quantum
  theory of perturbations. Part 3. Extensions}},  {\em Phys. Rept.} {\bf 215}
  (1992) 203--333.

\bibitem{Vilenkin:1981zs}
A.~Vilenkin, {\it {Gravitational Field of Vacuum Domain Walls and Strings}},
  {\em Phys. Rev.} {\bf D23} (1981) 852--857.

\bibitem{Gelmini:1988sf}
G.~B. Gelmini, M.~Gleiser, and E.~W. Kolb, {\it {Cosmology of Biased Discrete
  Symmetry Breaking}},  {\em Phys. Rev.} {\bf D39} (1989) 1558.

\bibitem{Larsson:1996sp}
S.~E. Larsson, S.~Sarkar, and P.~L. White, {\it {Evading the cosmological
  domain wall problem}},  {\em Phys. Rev.} {\bf D55} (1997) 5129--5135,
  [\href{http://arxiv.org/abs/hep-ph/9608319}{{\tt hep-ph/9608319}}].

\bibitem{Sikivie:1982qv}
P.~Sikivie, {\it {Of Axions, Domain Walls and the Early Universe}},  {\em Phys.
  Rev. Lett.} {\bf 48} (1982) 1156--1159.

\bibitem{Chang:1998tb}
S.~Chang, C.~Hagmann, and P.~Sikivie, {\it {Studies of the motion and decay of
  axion walls bounded by strings}},  {\em Phys. Rev.} {\bf D59} (1998) 023505,
  [\href{http://arxiv.org/abs/hep-ph/9807374}{{\tt hep-ph/9807374}}].

\bibitem{Choi:2009jt}
K.-S. Choi, H.~P. Nilles, S.~Ramos-Sanchez, and P.~K.~S. Vaudrevange, {\it
  {Accions}},  {\em Phys. Lett.} {\bf B675} (2009) 381--386,
  [\href{http://arxiv.org/abs/0902.3070}{{\tt arXiv:0902.3070}}].

\bibitem{Dias:2014osa}
A.~G. Dias, A.~C.~B. Machado, C.~C. Nishi, A.~Ringwald, and P.~Vaudrevange,
  {\it {The Quest for an Intermediate-Scale Accidental Axion and Further
  ALPs}},  {\em JHEP} {\bf 06} (2014) 037,
  [\href{http://arxiv.org/abs/1403.5760}{{\tt arXiv:1403.5760}}].

\bibitem{Ringwald:2015dsf}
A.~Ringwald and K.~Saikawa, {\it {Axion dark matter in the post-inflationary
  Peccei-Quinn symmetry breaking scenario}},  {\em Phys. Rev.} {\bf D93}
  (2016), no.~8 085031, [\href{http://arxiv.org/abs/1512.06436}{{\tt
  arXiv:1512.06436}}]. [Addendum: \textit{Phys. Rev.} \textbf{D94} (2016) no.4
  049908].

\bibitem{Nakayama:2016gxi}
K.~Nakayama, F.~Takahashi, and N.~Yokozaki, {\it {Gravitational waves from
  domain walls and their implications}},  {\em Phys. Lett.} {\bf B770} (2017)
  500--506, [\href{http://arxiv.org/abs/1612.08327}{{\tt arXiv:1612.08327}}].

\bibitem{Saikawa:2017hiv}
K.~Saikawa, {\it {A review of gravitational waves from cosmic domain walls}},
  {\em Universe} {\bf 3} (2017), no.~2 40,
  [\href{http://arxiv.org/abs/1703.02576}{{\tt arXiv:1703.02576}}].

\bibitem{Ellis:2017erg}
S.~A.~R. Ellis and J.~D. Wells, {\it {High-scale supersymmetry, the Higgs boson
  mass, and gauge unification}},  {\em Phys. Rev.} {\bf D96} (2017), no.~5
  055024, [\href{http://arxiv.org/abs/1706.00013}{{\tt arXiv:1706.00013}}].

\bibitem{Ellis:2015jwa}
S.~A.~R. Ellis and J.~D. Wells, {\it {Visualizing gauge unification with
  high-scale thresholds}},  {\em Phys. Rev.} {\bf D91} (2015), no.~7 075016,
  [\href{http://arxiv.org/abs/1502.01362}{{\tt arXiv:1502.01362}}].

\bibitem{Wess:1992cp}
J.~Wess and J.~Bagger, {\em {Supersymmetry and supergravity}}.
\newblock Princeton Series in Physics. Princeton University Press, Princeton,
  New Jersey, second~ed., 1992.
\newblock 259 pages.

\end{thebibliography}\endgroup


\end{document}